\documentclass[12pt]{article}

\usepackage{amsmath,amssymb,bm,graphicx,subfigure,makecell}
\usepackage{lineno,booktabs,url,color}
\usepackage{cite}
\usepackage{hyperref,verbatim}
\usepackage{xcolor}

\usepackage{multirow}

\def\tL{{\tilde L}}

\evensidemargin - .5truecm
\allowdisplaybreaks[1]

\begin{document}

\begin{titlepage}
\nopagebreak

\renewcommand{\thefootnote}{\fnsymbol{footnote}}
\vskip 2cm

\vspace*{1cm}
\begin{center}
{\Large \bf
Transverse-momentum resummation effects on angular\vspace{.2em}\\ coefficients in \boldmath{$Z$} and \boldmath{$W$} boson hadroproduction
}
\end{center}

\par \vspace{1.5mm}
\begin{center}
    {\bf Stefano Camarda${}^{(a)}$},
    {\bf Giancarlo Ferrera${}^{(b)}$},\\
    {\bf Lorenzo Rossi${}^{(b)}$}
    and
        {\bf Gabriele Francesco Sala${}^{(b)}$}
\vspace{5mm}

${}^{(a)}$
CERN, CH-1211 Geneva, Switzerland

${}^{(b)}$
Dipartimento di Fisica, Universit\`a di Milano and\\ INFN, Sezione di Milano,
I-20133 Milan, Italy\\\vspace{1mm}
\vspace{.5cm}

\begin{quote}
\pretolerance 10000
\begin{abstract}
\noindent

We present a comprehensive analysis of the angular coefficients of $Z$ and $W$ boson production at hadron colliders in different kinematical ranges, using data from the ATLAS, LHCb, and CMS Collaborations at the LHC, including the recent CMS measurement at $\sqrt{s}=13$~TeV, as well as CDF data at the Tevatron.
We provide theoretical predictions obtained by consistently combining the
resummation of logarithmically enhanced QCD corrections at small transverse
momentum $q_T$ up to next-to-next-to-leading logarithmic accuracy (NNLL)
with the fixed-order calculation through $\mathcal{O}(\alpha_S^2)$.
The fixed-order helicity cross sections entering the finite component are
evaluated using the new complete analytic NLO calculation implemented in
the {\ttfamily DYTurbo} program. This implementation makes it possible to obtain fast
and numerically stable predictions for all the angular coefficients and to
perform an extensive study of their theoretical uncertainties.
We quantify the impact of transverse-momentum resummation on the angular coefficients
and we estimate the associated theoretical uncertainties, including scale variations, parton distribution functions, non-perturbative effects, the treatment of the transition region and the recoil prescription used to generate the lepton angular distribution of the resummed component.
Within the estimated theoretical uncertainties, resummation improves
the description of several measurements, particularly for $A_2$, while
its impact on $A_0$ depends more strongly on the dataset and on the
theoretical choices.
The Lam--Tung combination $A_0-A_2$ is essentially unaffected by resummation.

\vskip .4cm
\end{abstract}

 \vfill\end{quote}
\end{center}

\begin{flushleft}
November 2026
\vspace*{-1cm}
\end{flushleft}
\end{titlepage}

\renewcommand{\thefootnote}{\fnsymbol{footnote}}

\newpage

\section{Introduction}
    The production of $Z/\gamma^*$ and $W$ bosons with subsequent leptonic decays constitutes a cornerstone process in the physics programme of hadron colliders, such as the LHC and the Tevatron. Owing to the very large production cross section and the cleanliness of the experimental signature, this process has enabled high-precision measurements of boson production cross sections in a wide range of multi-differential observables\,\cite{Aad:2014xaa,Aad:2015auj,ATLAS:2019zci,ATLAS:2023lsr,Chatrchyan:2011wt,Khachatryan:2016nbe,CMS:2019raw,Aaij:2015gna,Aaij:2015zlq,Aaij:2016mgv,Affolder:1999jh,Aaltonen:2012fi,Abbott:1999wk,Abazov:2007ac,Abazov:2010kn}. These measurements are typically performed inclusively with respect to the kinematics of the gauge-boson decay and have been extensively exploited for phenomenological studies, notably for the determination of parton distribution functions, as well as for the extraction of fundamental parameters of the Standard Model, such as the $W$-boson mass $m_W$\,\cite{ATLAS:2024erm,LHCb:2021bjt,CMS:2024lrd,D0:2013jba,CDF:2022hxs,Rottoli:2023xdc}, the effective Weinberg angle\,\cite{CMS:2011utm,CDF:2005qwt,D0:2017ekd} and, more recently, the strong coupling constant $\alpha_S$\,\cite{ATLAS:2023lhg,Camarda:2022qdg}.

Beyond inclusive observables, additional and complementary tests of the QCD dynamics governing gauge-boson production can be obtained by studying the fully differential kinematic distributions of the final-state leptons\,\cite{CDF:2011ksg,CMS:2015cyj,ATLAS:2016rnf,LHCb:2022tbc,ATLAS:2025mlt,CDF:2005qwt,CMS:2026amb}. In this framework, the reconstructed lepton kinematics provide direct sensitivity to the polarisation state of the intermediate gauge boson\,\cite{CMS:2011kaj,ATLAS:2012au}, thereby offering a detailed probe of the underlying production mechanism. The angular structure of the process can be parametrised in terms of a set of eight frame-dependent angular coefficients $A_i$ ($i=0,\ldots,7$)\,\cite{Collins:1977iv,Mirkes:1992hu,Mirkes:1994eb,Mirkes:1994dp,Lam:1978pu,Lam:1978zr}, which depend on the invariant mass, transverse momentum, and rapidity of the lepton pair, and fully characterise the production of the intermediate gauge boson.

A precise determination of the angular coefficients plays an important role in the measurement of the $W$-boson mass $m_W$ at hadron colliders. Although the extraction of $m_W$ relies primarily on other kinematic observables, such as the transverse mass and the lepton transverse momentum, an accurate modelling of the QCD dynamics and of the polarisation of the $W$ boson is essential to control the Monte Carlo predictions used to describe the kinematic distributions of the leptons originating from $W$-boson decays\,\cite{Pellen:2022fom}.

On the theoretical side, the angular coefficients have been computed within perturbative QCD to high accuracy\,\cite{Collins:1977iv,Mirkes:1992hu,Mirkes:1994eb,Mirkes:1994dp,Lam:1978pu,Lam:1978zr,Gauld:2017tww}, and fixed-order predictions are known to provide a reliable description of these observables over a broad kinematic range\,\cite{Lambertsen:2016wgj,Gauld:2017tww}. At the same time, studies of inclusive and fiducial Drell--Yan cross sections have shown that the inclusion of soft gluon resummation is crucial for achieving theoretical control in the region of small transverse momentum of the vector boson\,\cite{Dokshitzer:1978yd,Parisi:1979se,Collins:1984kg,Catani:2010pd,Becher:2010tm,Ebert:2016gcn,Scimemi:2017etj,Bizon:2018foh,Bizon:2019zgf,Becher:2019bnm,Alioli:2021qbf,Ebert:2020dfc,Billis:2024dqq,Camarda:2025lbt,Bozzi:2005wk,Bozzi:2010xn,Bacchetta:2024qre,Bacchetta:2025ara}. This observation naturally motivates the extension of resummed calculations also to the angular observables. The behaviour of the angular coefficients at small transverse momentum in the presence of soft-gluon resummation, including the fate of the Lam--Tung relation\,\cite{Lam:1978pu,Lam:1978zr}, has been investigated in Refs.\,\cite{Boer:2006eq,Berger:2007jw,Berger:2007si}, while the role of power corrections in resummed predictions for leptonic observables, including the angular coefficients, has been discussed in Ref.\,\cite{Ebert:2020dfc}.

In this work we follow this approach employing the {\ttfamily DYTurbo} program\,\cite{Camarda:2019zyx}, which provides resummed predictions matched to fixed order and is based on the calculation presented in Ref.\,\cite{Catani:2015vma}, which includes the leptonic decay of the vector boson and the full dependence on the kinematics of the final--state leptons. Building on this setup, we study the impact of resummation on the angular coefficients and provide improved predictions across the full kinematic phase space,
comparing resummed and fixed-order predictions with the available measurements. Particular attention is devoted to the mechanism through which resummation affects the angular coefficients and to the theoretical uncertainties of the predictions.

The paper is organised as follows. In Sec.~\ref{s:formalism} we present an overview of the theoretical framework employed. In Sec.~\ref{sec:sudakov} we discuss the structure of the angular coefficients in the resummed calculation and the mechanisms through which resummation affects them. The comparisons with experimental data are presented in Sec.~\ref{s:results}, while in Sec.~\ref{s:uncertainties} we quantify the origin of the resummation effects and the theoretical uncertainties of the predictions. Our conclusions are summarised in Sec.~\ref{s:conclusions}.

\section{Transverse-momentum resummation formalism}
\label{s:formalism}

In this section we briefly outline the main elements of our formalism, which combines the standard perturbative QCD description of vector--boson production with the resummation of logarithmically enhanced contributions, more details can be found in Refs.\,\cite{Bozzi:2003jy,Bozzi:2005wk,Bozzi:2010xn,Catani:2013tia,Catani:2015vma}. We consider the process
\begin{equation}
   h_1 + h_2 \to V + X \to l_3 + l_4 + X,
\end{equation}
where the vector boson $V$ is produced in the collision of the incoming hadrons $h_1$ and $h_2$, and subsequently decays into the final--state leptons $l_3$ and $l_4$.

The fully differential cross section in the lepton kinematics is characterised by six independent variables. In our formulation, we express the cross section in terms of the transverse momentum $\mathbf{q}_T$ of the vector boson (with magnitude $q_T = \sqrt{\mathbf{q}_T^{\,2}}$), its rapidity $y$, and the invariant mass $M$ of the dilepton system. The angular distribution of the final--state leptons is described by the polar and azimuthal angles $\boldsymbol{\Omega} = \{\theta,\phi\}$ of one lepton, usually defined in the so--called Collins--Soper rest frame of the vector boson\,\cite{Collins:1977iv}.

The corresponding hadronic differential cross section is given by
\begin{align}
\label{e:dxshadr}
\frac{d\sigma_{h_1 h_2 \to l_3 l_4}}{d^2\mathbf{q}_T\, dM^2\, dy\, d\boldsymbol{\Omega}}
(\mathbf{q}_T, M, y, s, \boldsymbol{\Omega})
&= \sum_{a_1,a_2}
\int_0^1 dx_1 \int_0^1 dx_2\,
f_{a_1/h_1}(x_1,\mu_F^{2})\,
f_{a_2/h_2}(x_2,\mu_F^{2}) \notag \\
&\quad \times
\frac{d\hat{\sigma}_{a_1 a_2 \to l_3 l_4}}{d^2\mathbf{q}_T\, dM^2\, d\hat{y}\, d\boldsymbol{\Omega}}
\left(\mathbf{q}_T, M, \hat{y}, \hat{s}, \boldsymbol{\Omega};
\, \alpha_S(\mu_R^{2}), \mu_R^{2}, \mu_F^{2}\right)\,,
\end{align}
where $f_{a/h}(x,\mu_F^2)$, with $a = q_f, \bar q_f, g$, are the parton distribution functions of the incoming hadron $h$ evaluated at the factorisation scale $\mu_F$, and $d\hat{\sigma}_{a_1 a_2 \to l_3 l_4}$ denotes the corresponding partonic differential cross section. The quantity $\hat{s} = x_1 x_2 s$ is the squared partonic centre--of--mass energy, while $\hat{y} = y - \ln(\sqrt{x_1/x_2})$ is the rapidity of the vector boson in the partonic centre--of--mass frame. The parameter $\mu_R$ denotes the renormalisation scale.

The partonic cross section is decomposed as
\begin{equation}
\label{partXS2}
    d\hat\sigma_{a_1 a_2 \to l_3 l_4}
    = d\hat\sigma^{(\mathrm{res.})}_{a_1 a_2 \to l_3 l_4}
    + d\hat\sigma^{(\mathrm{fin.})}_{a_1 a_2 \to l_3 l_4},
\end{equation}
where the first term is the resummed component, which contains and resums all logarithmically enhanced contributions of the type
$\alpha_S^n\, (M^2/q_T^2)\ln^m(M^2/q_T^2)$ (with $0 \le m \le 2n-1$), while the second term is the finite component computed at fixed order in perturbation theory.

In the formalism adopted in this work, the resummation is performed in the impact-parameter space $\mathbf{b}$, conjugate to $\mathbf{q}_T$, where the kinematic constraint of
transverse-momentum conservation takes a factorised form\,\mbox{
\cite{Parisi:1979se}};
alternative formulations directly in transverse-momentum space have also been developed\,\mbox{
\cite{Monni:2016ktx,Ebert:2016gcn}}. The resummed component is obtained from its $b$-space counterpart through a Bessel transformation,
\begin{equation}
{\label{e:bessel}
\frac{d\hat\sigma^{(\rm res.)}_{a_1a_2\to l_3l_4}}{d^2\mathbf{q}_T\, dM^2\, d\hat{y}\, d\boldsymbol{\Omega}}
=\!\!\! \sum_{b_1,b_2=q_f,\bar{q}_{f'}}\!\!\!\frac{d\hat\sigma_{b_1b_2\to l_3l_4}^{(0)}}{d\boldsymbol{\Omega}} \frac{1}{\hat s}\int_0^{\infty}\frac{db}{2\pi}\; b\, J_0(b q_T)\;
\mathcal{W}_{a_1a_2,b_1b_2\to V}\left(b, M, \hat y, \hat s;\, \alpha_S, \mu_R^2, \mu_F^2\right),
}\end{equation}
where $J_0$ is the zeroth-order Bessel function. The function $\mathcal{W}$ is written in the following factorised form\,\cite{Bozzi:2005wk,Catani:2015vma}\,\footnote{The exponential structure is most conveniently expressed in
double Mellin space, with moments $N_1$ and $N_2$ conjugate to the two
longitudinal momentum fractions. For compactness, the partonic indices
$\{a_1a_2,b_1b_2\}$ are suppressed in Eq.~\eqref{resum}. This scalar
notation applies channel by channel, or directly to flavour non-singlet
channels; the complete calculation retains the full partonic and flavour
structure.}:
\begin{equation}
\label{resum}
\mathcal{W}_{V}^{N_1,N_2}\left(b, M;\, \alpha_S, \mu_R^2, \mu_F^2\right)
= \mathcal{H}_V^{N_1,N_2}\left(M;
\alpha_S,\mu_R,\mu_F,Q \right)
\times \exp\{{\mathcal{G}^{N_1,N_2}}(\alpha_S,\tL;\mu_R,Q)\}\,,
\end{equation}
where
we have introduced
the logarithmic expansion parameter $\tL\equiv \ln ({Q^2 b^2}/{b_0^2}+1)$
with $b_0=2e^{-\gamma_E}$ ($\gamma_E=0.5772...$ is the Euler-Mascheroni constant).
The scales $\mu_R$ and $\mu_F$ are the customary
renormalization and factorization scales,
while $Q$ is the resummation scale whose variations can be used
to estimate the effect of uncalculated logarithmic terms at higher orders\,\cite{Bozzi:2003jy}.

The factor $d{\hat \sigma}^{(0)}_{b_1b_2\to l_3l_4}$ in Eq.\,(\ref{e:bessel})
is the Born level differential cross section for the quark-antiquark annihilation
subprocess, which carries the full dependence on the lepton angles $\boldsymbol{\Omega}$, defined in the rest frame of the vector boson specified by the recoil prescription discussed below.
The functions ${\cal H}^{V}_{a_1a_2 \rightarrow q \bar{q}}$\,\cite{Catani:2000vq,Catani:2013tia,Catani:2012qa,Gehrmann:2012ze}
are (process dependent) coefficients which contain
the hard-collinear contributions
and have a standard fixed-order expansion in powers of $\alpha_S=\alpha_S(\mu_R^2)$.

The universal (process independent) form factor $\exp\{{\cal G}\}$
contains and resums in an exponentiated form all the $\alpha_S^n\tL^{m}$ (with $n,m\geq 1$) logarithmic terms
that order-by-order in $\alpha_S$ are logarithmically divergent
as $b \to \infty$ (i.e.\ $q_T\to 0$).

At N$^k$LL accuracy, the function ${\cal G}$ includes the logarithmic
terms from $\alpha_S^n\tL^{n+1}$ down to
$\alpha_S^n\tL^{n+1-k}$. The hard-collinear function ${\cal H}^{V}$
and the finite component are consistently included through
$\mathcal{O}(\alpha_S^k)$. In particular, at NNLL accuracy we retain
the functions $g^{(1)}$, $g^{(2)}$ and $g^{(3)}$, together with
${\cal H}^{V}$ and the finite component through
$\mathcal{O}(\alpha_S^2)$. Throughout this paper we denote this
accuracy as NNLL+NLO, following the perturbative counting of the
finite-$q_T$ helicity cross sections, whose leading contribution is
of $\mathcal{O}(\alpha_S)$. In the inclusive Drell--Yan counting of
Ref.~\cite{Catani:2015vma}, the same fixed-order truncation is
conventionally denoted as NNLL+NNLO.

The function ${\cal G}$ is singular when $\alpha_S\tL=\pi/\beta_0$ (where $\beta_0$ is the one-loop coefficient of the QCD $\beta$ function)
which corresponds to the region of transverse-momenta
of the order of the Landau pole of the QCD coupling
or $b^{-1}\sim \Lambda_{\mathrm{QCD}}$. This signals
that a truly non-perturbative region is approached and
perturbative results (including resummed ones) are not reliable. In this region, a model for NP QCD effects is required, which must include a regularization of the singular behavior of the function $\mathcal{G}$. As noted in Refs.~\,\cite{Qiu:2000hf,Catani:1996yz,Laenen:2000de,Lustermans:2019plv}, the specific form of this regularization remains somewhat arbitrary. In this work, we adopt the so-called $b_\star$ prescription, originally proposed in Ref.\,\cite{Collins:1984kg},
by freezing the $b$ dependence of
$\exp\{\mathcal{G}(\alpha_S,\widetilde L)\}$ before it reaches the
Landau singularity.

To achieve this, a parameter $b_{\text{max}}$ is introduced with the replacement
\begin{equation}
\label{e:bstar}
b \to b_\star = \frac{b}{\sqrt{1 + b^2/b_{\text{max}}^2}}\,.
\end{equation}
This prescription guarantees that the variable $b_\star$ saturates to $b_{\text{max}}$ at large values of $b$.
However, $b_{\star}$ also introduces spurious power corrections
that scale like
$(\Lambda_{\text{QCD}}/q_T)^k$\,\cite{Catani:1996yz,Kulesza:2002rh,Laenen:2000de,Kulesza:2003wn},
with $k>0$. In the region $q_T \simeq \Lambda_{\text{QCD}}$, these power
corrections become sizeable and have to be modelled by including in
Eq.~\eqref{resum} a non-perturbative form factor.
We use
\begin{equation}
\label{e:snp}
S_{\rm NP}(b)=\exp\left\{-g_j(b)-g_K(b)\,\ln\frac{M^2}{Q_0^2}\right\},
\end{equation}
with
\begin{equation}
g_j(b)=\frac{g_1 b^2}{\sqrt{1+\lambda b^2}}+{\rm sign}(q)\left(1-e^{-|q|b^4}\right),\qquad
g_K(b)=g_0\left\{1-\exp\left[-\frac{C_F\,\alpha_S(b_0^2/b_\star^2)\,b^2}{\pi\, g_0\, b_{\rm max}^2}\right]\right\},
\label{e:gjgk}
\end{equation}
as in Ref.\,\cite{Camarda:2025lbt}. We use the central values of the recent determination of Ref.\,\cite{Camarda:2025lbt}: $g_1=0.732$~GeV$^2$, $\lambda=0.92$~GeV$^2$, $q=-0.136$~GeV$^4$, $g_0=0.83$, $Q_0=1$~GeV and $b_{\rm max}=1.5$~GeV$^{-1}$. The dependence of the results on the non-perturbative parameters is studied in Sec.~\ref{s:uncertainties}.

The finite component $d\hat\sigma^{(\rm fin.)}$ in Eq.~\eqref{partXS2} is obtained by subtracting from the fixed-order cross section the expansion of the resummed component at the same perturbative order (the asymptotic term).
In the present calculation the $\mathcal O(\alpha_S^2)$
fixed-order term is evaluated using the new
complete analytic calculation of the unpolarised and helicity cross
sections at finite $q_T$, recently implemented in
{\ttfamily DYTurbo}\,\cite{Camarda:2026kdu}.
This analytic implementation replaces the substantially slower numerical
evaluation previously used for the helicity cross sections and makes the
large number of calculations required by the uncertainty analysis
numerically feasible.
At large $q_T$ the resummed and asymptotic terms cancel only up to contributions beyond the accuracy of the calculation. In order to recover the fixed-order result in this region, both terms are multiplied by a transition function, which we choose as $w(q_T)=1$ for $q_T<kM$ and $w(q_T)=\exp[-(q_T-kM)^2/(\delta M)^2]$ for $q_T\geq kM$, with $k=0.33$ and $\delta=0.25$. The dependence on the form of this function provides an estimate of the matching uncertainty in the transition region between the resummed and fixed-order regimes, which we discuss in Sec.~\ref{s:uncertainties}.

Equation~\eqref{e:bessel} is derived at leading power in $q_T/M$,
where the angular dependence is evaluated with Born-level partonic
kinematics. At finite $q_T$, momentum conservation requires a prescription
for assigning the transverse recoil to the two incoming Born-level
partons. Different recoil prescriptions leave the leading-power
logarithmically enhanced terms unchanged and differ by power-suppressed
contributions in $q_T/M$\,\cite{Catani:2015vma}.

The default prescription used in {\ttfamily DYTurbo} generates the
Born-level angular distribution in the Collins--Soper frame. Within the
general construction of Ref.~\cite{Catani:2015vma}, this corresponds to
sharing the transverse recoil symmetrically,
\begin{equation}
\mathbf{k}_{T1}=\mathbf{k}_{T2}=\frac{\mathbf{q}_T}{2}\,.
\end{equation}
The recoil prescription has no effect on the cross section integrated
over the lepton angles, but it modifies the power-suppressed angular
projections. Its dependence is therefore included in the uncertainty
assessment of the angular coefficients.
 This prescription accounts for the linear power corrections of leptonic origin to azimuthally symmetric observables, but not for the linear power corrections originating from the hadronic tensor, which affect the observables sensitive to the azimuthal dependence of the lepton distribution\,\cite{Ebert:2020dfc}. As discussed in the next section, the recoil prescription plays a central role for the angular coefficients.

\section{Resummation effects on angular coefficients}
\label{sec:sudakov}

Fixed-order perturbative calculations fail to provide a reliable description
of the $q_T$ distribution of the vector boson at small transverse momenta,
where large logarithmic contributions of the type
$\alpha_S^n \ln^m(M^2/q_T^2)$ spoil the convergence of the perturbative
expansion and resummation to all orders is required. Nevertheless,
fixed-order predictions for the angular coefficients $A_i$ are known to
provide a reasonable description of the data even at small values of
$q_T$\,\cite{Gauld:2017tww,Lambertsen:2016wgj}. The theoretical basis of this
behaviour, and in particular the behaviour of the Lam--Tung relation in the presence
of soft-gluon resummation, has been investigated in
Refs.\,\cite{Boer:2006eq,Berger:2007jw,Berger:2007si}. In this section we recall
the structure of the angular coefficients in the resummed calculation, and we
discuss the mechanisms through which resummation affects them in the
intermediate $q_T$ region.

The DY cross section, fully differential in the lepton
kinematics, can be decomposed by expanding its angular dependence as
\begin{align}
\frac{d\sigma}{d^2\mathbf{q}_T\,dy\,dM^2\,d\cos\theta\,d\phi}
&= \frac{3}{16 \pi} \frac{d\sigma}{d^2\mathbf{q}_T\,dy\,dM^2}
\Bigg\{
\left(1+\cos^2\theta\right)
+ \frac{A_0}{2}\left(1-3\cos^2\theta\right)
+ A_1\sin 2\theta\cos\phi \notag\\
&\quad
+ \frac{A_2}{2}\sin^2\theta\cos 2\phi
+ A_3\sin\theta\cos\phi
+ A_4\cos\theta
+ A_5\sin^2\theta\sin 2\phi \notag\\
&\quad
+ A_6\sin 2\theta\sin\phi
+ A_7\sin\theta\sin\phi
\Bigg\},
\label{e:Aicoeff}
\end{align}
where $A_i(q_T, y, M)$, with $i = 0,\ldots,7$, are the eight
frame-dependent angular coefficients. Each coefficient is the ratio of a
helicity cross section $d\sigma^{(i)}$ to the unpolarised cross section
$d\sigma^{(U+L)}\equiv d\sigma/(d^2\mathbf{q}_T\,dy\,dM^2)$,
\begin{equation}
  A_i(q_T, y, M)
  =
  \frac{d\sigma^{(i)}}{d^2\mathbf{q}_T\,dy\,dM^2}\,
  \left(\frac{d\sigma^{(U+L)}}{d^2\mathbf{q}_T\,dy\,dM^2}\right)^{-1},
\label{eq:Ai_ratio}
\end{equation}
and it can be extracted from the moments of the lepton angular distribution\,\cite{Mirkes:1994eb,ATLAS:2016rnf},
\begin{equation}
\begin{aligned}
A_0&=4-10\,\langle\cos^2\theta\rangle, & A_1&=5\,\langle\sin2\theta\cos\phi\rangle, & A_2&=10\,\langle\sin^2\theta\cos2\phi\rangle, & A_3&=4\,\langle\sin\theta\cos\phi\rangle,\\
A_4&=4\,\langle\cos\theta\rangle, & A_5&=5\,\langle\sin^2\theta\sin2\phi\rangle, & A_6&=5\,\langle\sin2\theta\sin\phi\rangle, & A_7&=4\,\langle\sin\theta\sin\phi\rangle,
\end{aligned}
\label{eq:moments}
\end{equation}
where $\langle f\rangle$ denotes the average of a function $f(\theta,\phi)$ over the
angular distribution of Eq.~\eqref{e:Aicoeff}.

In the resummed component of Eq.~\eqref{e:bessel}, the dependence on
the lepton angles is entirely carried by the Born-level factor
$d\hat{\sigma}^{(0)}_{q\bar q\to l_3l_4}$, while the QCD kernel is
independent of $\boldsymbol{\Omega}$. For each Born channel, i.e.\ for a given flavour and a given
assignment of the quark and the antiquark to the two incoming hadrons, the same leading-power kernel therefore multiplies all the
angular projections and cancels in any non-vanishing angular ratio.

At hadron level, complete cancellation of the leading-power kernel
requires the Born angular ratio to be independent of the quark flavour
and beam orientation. Otherwise, only the LL tower cancels universally:
beyond LL, collinear evolution and hard-collinear coefficients reweight
the different channels, leaving residual leading-power logarithms while
the angular coefficient remains finite as $q_T\to0$. This applies in
particular to $A_4$, whose Born ratio depends on the electroweak
couplings and on the quark direction. Its numerator and denominator
must therefore be resummed separately.

The finite component contains no leading-power
$q_T^{-2}$ logarithmic singularities. It is therefore power suppressed
relative to the unpolarised resummed contribution as $q_T\to0$ and
does not alter the finiteness argument.

In the Collins--Soper frame, and for the perturbative contributions
considered here, only $A_4$ has a non-vanishing Born projection at
strict leading power.
In perturbation theory, $A_0$ and
$A_2$ are suppressed by $q_T^2/M^2$, and $A_1$ and $A_3$ by at least
$q_T/M$\,\cite{Gauld:2017tww,Ebert:2020dfc}. Their first non-zero
contributions are therefore not controlled by leading-power
factorisation. Although these power-suppressed terms can be
logarithmically enhanced\,\cite{Boer:2006eq,Berger:2007jw}, they cannot
produce a power divergence as $q_T\to0$.

This power counting is frame dependent. A regular rotation can generate
additional leading-power angular projections, but cannot increase the
overall power singularity. The cancellation arguments given above
remain valid in any such frame.

The angular coefficients are obtained from the matched cross section
of Eq.~\eqref{partXS2}, whose resummed and finite components have
different angular structures. Resummation can therefore affect the
corresponding ratios.

The angular coefficients are defined in the Collins--Soper frame. With
our default recoil prescription, the Born-level distributions entering
the resummed and asymptotic components are evaluated in the same frame
and contain only the $(1+\cos^2\theta)$ and $\cos\theta$ harmonics.
These components therefore contribute only to
$d\sigma^{(U+L)}$ and $d\sigma^{(4)}$. At $q_T>0$ and at the accuracy
considered here, the helicity cross sections $d\sigma^{(i)}$ with
$i\neq4$ are thus entirely determined by the fixed-order component.
At fixed $q_T$, $y$ and $M$, one then has
\begin{equation}
A_i^{\rm NNLL+NLO}
=\frac{d\sigma^{(i)}_{\rm NLO}}{d\sigma^{(U+L)}_{\rm NNLL+NLO}}
=A_i^{\rm NLO}\;\frac{d\sigma^{(U+L)}_{\rm NLO}}{d\sigma^{(U+L)}_{\rm NNLL+NLO}}\,,
\qquad i\neq4\,.
\label{eq:Ai_res}
\end{equation}
Thus, for $i\neq4$, resummation affects the angular coefficients only
through the unpolarised denominator and produces a common rescaling.
In the intermediate region $q_T\sim20$--$50$~GeV, where the matched
unpolarised cross section exceeds the fixed-order result, the rescaling
factor ranges from about $0.8$ to $0.95$. It approaches unity at large
$q_T$, where the transition function switches off the difference
between the resummed and asymptotic components. This effect is
pointwise in $q_T$ and does not rely on bin averaging. Since $A_0$ and
$A_2$ receive the same rescaling, resummation does not generate an
additional Lam--Tung violation and only marginally affects
$A_0-A_2$\,\cite{Berger:2007jw,Berger:2007si}.

The bin-averaged angular coefficients are ratios of the corresponding
integrated helicity and unpolarised cross sections,
\begin{equation}
\langle A_i\rangle_{\rm bin}
=
\frac{\displaystyle\int_{\rm bin}dq_T\,
      \dfrac{d\sigma^{(i)}}{dq_T}}
     {\displaystyle\int_{\rm bin}dq_T\,
      \dfrac{d\sigma^{(U+L)}}{dq_T}}\,.
\label{eq:Ai_bin}
\end{equation}
Resummation can therefore also affect the coefficients by changing the
$q_T$ weights within each bin. In Sec.~\ref{s:uncertainties} we show
that, for $q_T\gtrsim20$~GeV, this bin-averaging effect accounts for
less than $5\%$ of the total shift, which is dominated by the pointwise
effect of Eq.~\eqref{eq:Ai_res}.

Equation~\eqref{eq:Ai_res} holds for our default recoil prescription.
Alternative prescriptions modify the Born parton momenta and hence the
polar axis of the Born angular distribution. Within the prescriptions
implemented in {\ttfamily DYTurbo}, this axis lies in the production
plane and is related to the Collins--Soper axis by a rotation
$\beta(q_T,y,M)$\,\cite{Catani:2015vma}. When the resulting
distribution is projected in the Collins--Soper frame, the unpolarised
cross section is unchanged, while the resummed minus asymptotic
contribution has angular coefficients
\begin{equation}
A_0=A_2=\sin^2\beta\,,\qquad
A_1=\pm\tfrac{1}{2}\sin2\beta\,,\qquad
A_3=A_4^{(0)}\sin\beta\,,\qquad
A_4=A_4^{(0)}\cos\beta\,,
\label{eq:rotation}
\end{equation}
where $A_4^{(0)}$ is the Born-level coefficient. Choosing the polar
axis along one of the incoming partons, as in the
Gottfried--Jackson construction, gives
$\sin^2\beta=q_T^2/(M^2+q_T^2)$. The recoil contributions are therefore
suppressed by $q_T^2/M^2$ for $A_0$ and $A_2$, and by $q_T/M$ for
$A_1$ and $A_3$, but can be numerically relevant in the intermediate
$q_T$ region. We include this prescription dependence in the
uncertainty assessment of Sec.~\ref{s:uncertainties}. Since the recoil
contributions to $A_0$ and $A_2$ are equal, $A_0-A_2$ is independent
of the prescription within this class.

Two further remarks are in order. First, the recoil prescription fixes only the kinematic power corrections associated with the Born-level resummed contribution; it does not resum the independent power-suppressed hadronic structure functions~\cite{Ebert:2020dfc}. These contributions are nevertheless included at fixed order through
$\mathcal{O}(\alpha_S^2)$. Second, the $\mathcal{O}(\alpha_S^3)$ corrections computed in Ref.~\cite{Gauld:2017tww} are sizeable for $A_2$ and $A_0-A_2$, but small for $A_0$, in the intermediate-$q_T$ region. They are unrelated to transverse-momentum resummation and are required to describe the Lam--Tung violation observed at large $q_T$.

\section{Numerical Predictions}
\label{s:results}
In the following, we present a comparison between experimental data and theoretical predictions for $Z$ and $W$ production followed by leptonic decays.
The predictions are obtained using the {\ttfamily DYTurbo} numerical program\,\cite{Camarda:2019zyx} at
NNLL+NLO accuracy with central scale choices, $\mu_R = \mu_F = Q = M$.
We compare the results against the fixed-order predictions at NLO in order to assess the impact of the resummation. We use the analytic implementation in {\ttfamily {DYTurbo}}
of the $\mathcal{O}(\alpha_S^2)$ helicity cross sections for vector-boson production at $q_T>0$\,\cite{Camarda:2026kdu}, which allows us to compute the unpolarised and helicity cross sections directly in the bins of the measurements, integrated over the full phase space of the leptons, since the measured coefficients are extrapolated to the full phase space by the experiments. In the first bin, which contains $q_T=0$, the fixed-order prediction is obtained with the $q_T$-subtraction method, with a cut $q_T^{\rm cut}/M=0.008$. The resummed component is evaluated with the Collins--Soper recoil prescription, the transition function and the non-perturbative form factor described in Sec.~\ref{s:formalism}. The numerical uncertainties of the predictions are below $1\%$ of the experimental uncertainties; the theoretical uncertainties are discussed in Sec.~\ref{s:uncertainties}. For $A_4$ the resummed component contributes also to the numerator, and both $d\sigma^{(4)}$ and $d\sigma^{(U+L)}$ are resummed. Since in $pp$ collisions the polar axis of the Collins--Soper frame is oriented along the longitudinal momentum of the boson, the projection on $d\sigma^{(4)}$ depends on the sign of the boson rapidity; the rapidity-integrated resummed contribution to $d\sigma^{(4)}$ and the corresponding counterterm are therefore computed with a rapidity-dependent (two-dimensional) Mellin inversion, which reproduces the unpolarised cross section of the standard calculation within $0.05\%$.
The cross section is computed by convolving the partonic expression in Eq.~\eqref{e:dxshadr} with PDFs from the NNPDF4.0 set\,\cite{NNPDF:2021njg}
using $\alpha_S(m_Z^2) = 0.118$ for the reference value of the QCD coupling.
As for the electroweak couplings, we adopt the $G_\mu$ scheme, where the corresponding input parameters are collected in Table~\ref{t_EW_inputs}.
\begin{table}[h!]
\centering
\begin{tabular}{|c|c|c|}
\hline
\textbf{Constant} & \textbf{Value} & \textbf{Unit} \\
\hline
$G_F$ & $1.1663787 \times 10^{-5}$ & GeV$^{-2}$ \\
$m_Z$ & $91.1876$ & GeV \\
$\Gamma_Z$ & $2.4950$ & GeV \\
$m_W$ & $80.385$ & GeV \\
$\Gamma_W$ & $2.091$ & GeV \\
\hline
\end{tabular}
\caption{Numerical values of the electroweak constants used as input for the $G_\mu$ scheme in our calculations.}
\label{t_EW_inputs}
\end{table}

The datasets included in our phenomenological analysis are summarised in Table~\ref{t:data}. They comprise measurements of $Z$-boson angular
distributions performed by the CDF\,\cite{CDF:2011ksg},
CMS\,\cite{CMS:2015cyj}, ATLAS\,\cite{ATLAS:2016rnf}, and
LHCb\,\cite{LHCb:2022tbc} collaborations, including the recent CMS measurement at $\sqrt{s}=13$~TeV\,\cite{CMS:2026amb}. These are complemented by the CDF\,\cite{CDF:2005qwt} and the most recent ATLAS measurements of $W$-boson production\,\cite{ATLAS:2025mlt}.

For the ATLAS $Z$-boson analysis, both inclusive and rapidity-binned distributions are available; however, the inclusive distribution is obtained through an extrapolation beyond the fiducial detector coverage and therefore carries additional reconstruction uncertainties. For this reason, we restrict ourselves to the central rapidity bin $|y|<1$. Since CMS provides measurements in several rapidity intervals, including a $|y|<1$ bin, this choice ensures a consistent and homogeneous comparison between the ATLAS and CMS datasets.

The LHCb data, which extend the rapidity coverage into the forward region, provide a complementary dataset that allows us to probe the theoretical predictions in a broader kinematic regime.

\begin{table}[h!]
\footnotesize
\begin{center}
\renewcommand{\tabcolsep}{0.4pc}
\renewcommand{\arraystretch}{1.2}
\begin{tabular}{|c|c|c|c|c|c|}
\hline
  Experiment & Exchanged boson & $N_{\text{dat}}$ & $\sqrt{s}$ [GeV] & $y$ cut & Ref. \\
  \hline
  \hline
  CDF & $Z$ & 5 & 1960 & Inclusive & \cite{CDF:2011ksg} \\
  \hline
    CMS & $Z$ & 7 & 8000 & $|y|<1$ & \cite{CMS:2015cyj} \\
  \hline
  ATLAS & $Z$ & 18 & 8000 & $|y|<1$ & \cite{ATLAS:2016rnf} \\
  \hline
    LHCb & $Z$ & 6 & 13000 & $y>2$ & \cite{LHCb:2022tbc} \\
  \hline
    CMS & $Z$ & 7 & 13000 & $|y|<1$ & \cite{CMS:2026amb} \\
  \hline
    CMS & $Z$ & 7 & 13000 & $1<|y|<2.4$ & \cite{CMS:2026amb} \\
  \hline
  CDF & $W^-$ & 4 & 1800 & Inclusive & \cite{CDF:2005qwt} \\
  \hline
      ATLAS & $W^+$ & 10 & 13000 & Inclusive & \cite{ATLAS:2025mlt} \\
  \hline
        ATLAS & $W^-$ & 10 & 13000 & Inclusive & \cite{ATLAS:2025mlt} \\
  \hline
\end{tabular}
\caption{Experimental data sets included in this analysis. Each row contains the boson exchanged, the number of datapoints included, the center-of-mass energy $\sqrt{s}$, the acceptance region in the
variable $y$, and the published reference.}
\label{t:data}
\end{center}
\end{table}

To assess the impact of the resummed contributions in a quantitative manner, and not only through a visual inspection of the distributions, we introduce a statistical estimator that allows us to evaluate their effect on a dataset-by-dataset basis. A purely visual comparison can in fact be misleading, particularly for ATLAS measurements, where strong (anti-)correlations exist among data points, and these are not easily visible in standard plots. For this reason, we adopt the standard $\chi^2$ function, defined as
\begin{equation}
\chi^2 =
\left( \vec{D}_d - \vec{P}_d \right)^T
V_d^{-1}
\left( \vec{D}_d - \vec{P}_d \right) \, ,
\end{equation}
where the index $d$ denotes the specific dataset. The vector $\vec{D}_d$ denotes the data points of the $d$-th dataset, $V_d$ is the corresponding covariance matrix, and $\vec{P}_d$ represents the theoretical predictions for that dataset. The covariance matrix is the sum of the experimental covariance matrix and of the diagonal matrix of the squared numerical uncertainties of the predictions. For the ATLAS $Z$-boson measurement we use the bin-to-bin correlations of the covariance matrix provided by the Collaboration\,\cite{ATLAS:2016rnf}, combined with the total uncertainties of the published coefficients, i.e.\ $V_{ij}=\rho_{ij}\sigma_i\sigma_j$. For the other measurements we add statistical and systematic uncertainties in quadrature and neglect bin-to-bin correlations. For the ATLAS $W$-boson measurement, statistical correlations between the bins and between the coefficients are provided\,\cite{ATLAS:2025mlt}: we include them in the covariance matrix, and we add the systematic uncertainties, for which no correlations are published, on the diagonal. The covariance of $A_0-A_2$ is obtained from those of $A_0$ and $A_2$ and from their statistical correlation. Since no parameters are fitted, $N_{\rm d.o.f.}$ is the number of data points included in the comparison. The values of $\chi^2/N_{d.o.f.}$ obtained with the experimental uncertainties only are collected for all the datasets in Table~\ref{t:chi2th} (columns NLO and CS), together with those that include the theoretical uncertainties of the predictions, which are discussed in Sec.~\ref{s:impact}.

Since not all experiments provide measurements for the complete set of angular coefficients $A_i$, and since in several cases the available measurements are either statistically compatible with zero or affected by relatively large uncertainties, in this work we concentrate on a subset of observables with robust experimental coverage and clear phenomenological relevance. Specifically, we focus on the coefficients $A_0$ and $A_2$, together with their difference $A_0 - A_2$, which vanishes at $\mathcal{O}(\alpha_S)$ according to the Lam--Tung relation and therefore provides a test of the QCD dynamics beyond the first order, complementary to the individual coefficients. In addition, we include the coefficients $A_3$ and $A_4$ for the $W$.

\subsection{\texorpdfstring{CDF $Z$-boson}{CDF Z-boson}}

The CDF collaboration was the first to measure and publish angular coefficients for
$Z$-boson production. The analysis was performed using a dilepton invariant-mass
selection in the range $66 < M < 116$~GeV, and the angular coefficients were extracted
as functions of the transverse momentum $q_T$ in a rapidity-inclusive setup. Owing to
the relatively limited statistical precision of the Tevatron dataset, the number of
available data points is small and the corresponding $q_T$ binning is rather coarse.
For this reason, we compare our theoretical predictions with the CDF measurements over
the full accessible $q_T$ range, without imposing additional kinematic restrictions.

\begin{figure}[h!]
\centering
\includegraphics[width=0.48\textwidth]{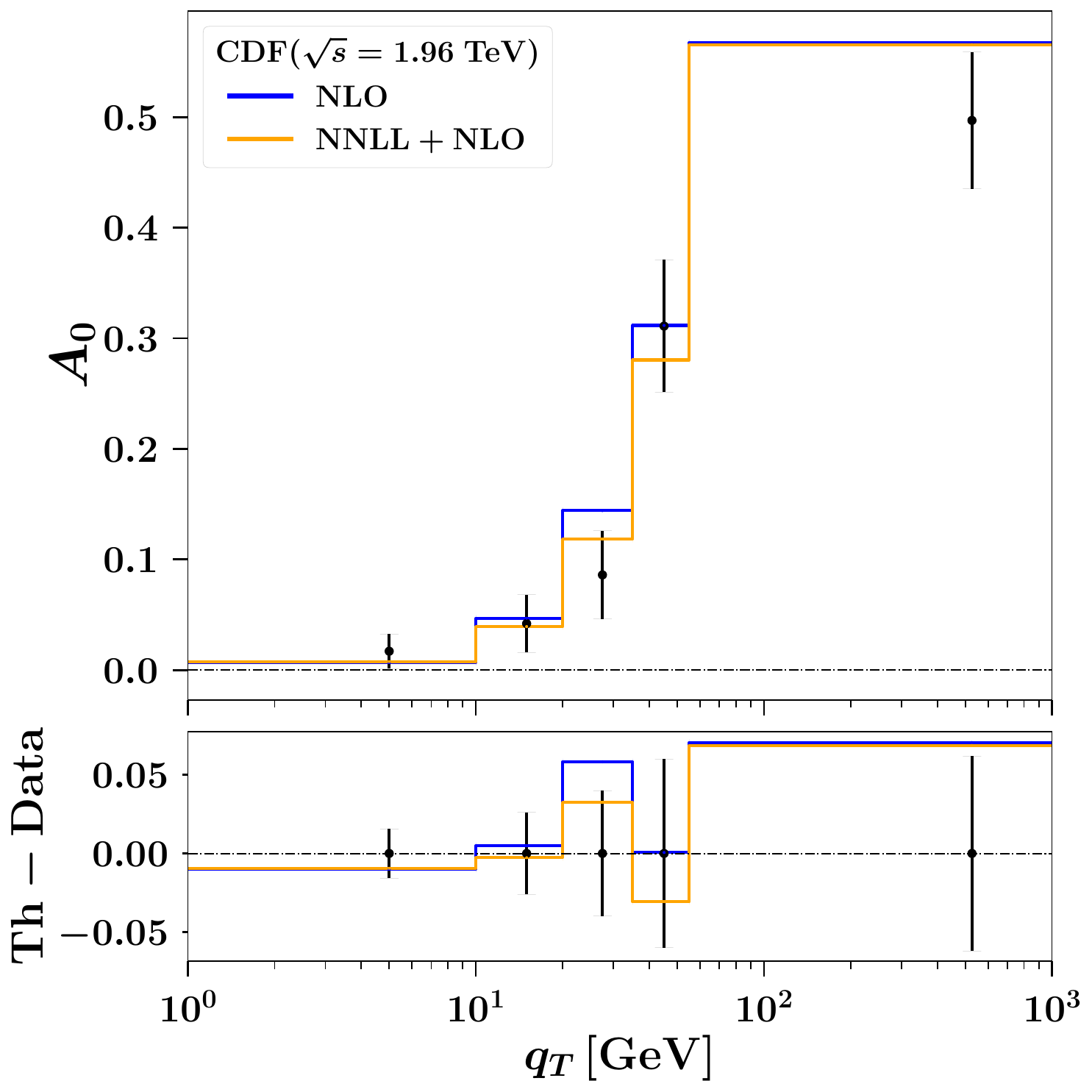}
\includegraphics[width=0.48\textwidth]{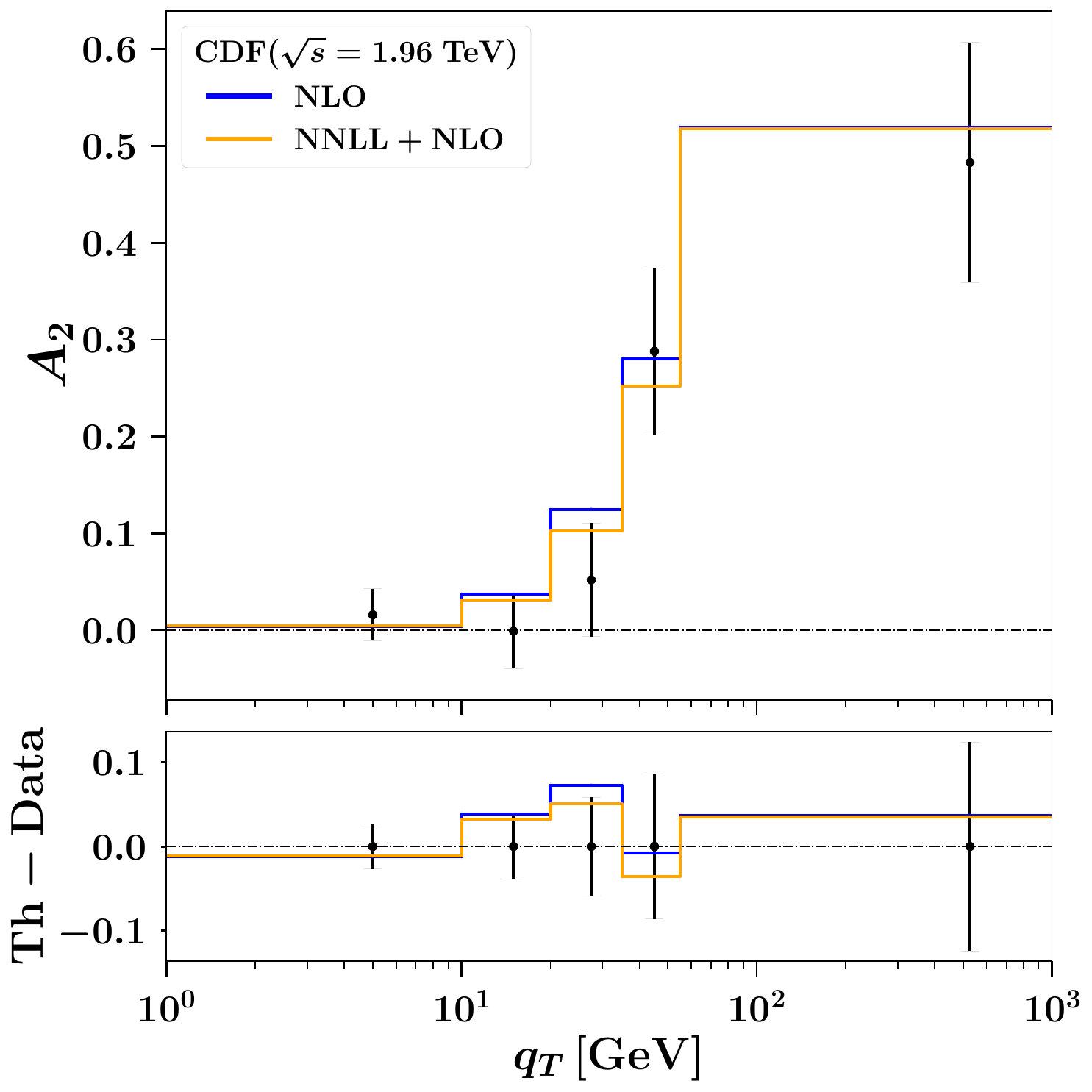}
\includegraphics[width=0.48\textwidth]{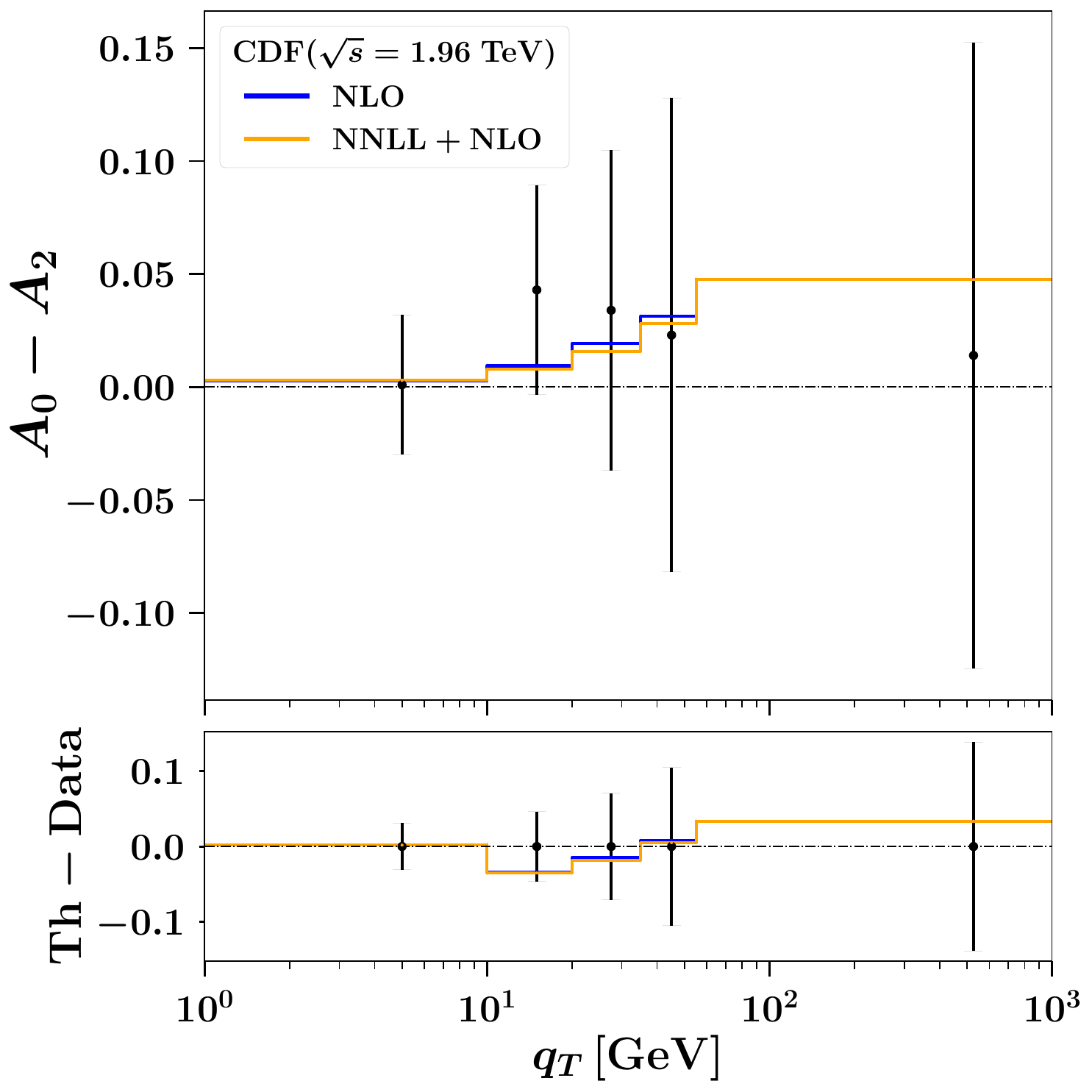}
\caption{Comparison between experimental data and theoretical predictions at NLO (blue) and NNLL+NLO (orange) for the angular coefficients $A_0$ (upper left), $A_2$ (upper right), and the Lam--Tung combination $A_0-A_2$ (lower) in $Z$-boson production at CDF, shown as functions of $q_T$. The lower panels display the differences between theory and data.}
\label{f:A_CDF}
\end{figure}

The $q_T$ distributions of the angular coefficients $A_0$, $A_2$, and their difference
$A_0 - A_2$ are shown in Fig.~\ref{f:A_CDF}. The experimental data are represented by
black points and are compared to theoretical predictions obtained at NLO accuracy
(blue) and at NNLL+NLO accuracy (orange). The lower panels display the differences
between the theoretical predictions and the experimental measurements, highlighting
the regions where higher-order effects are most relevant.

 The fixed-order NLO calculation already provides a reasonable description of the data, as observed in the literature. The inclusion of the resummed contribution {lowers the predictions for $A_0$ and $A_2$ } in the intermediate transverse-momentum region, $q_T \sim 20$--$50$~GeV,
where the NLO predictions lie above the data. The Lam--Tung combination is only marginally affected by resummation (see Sec.~\ref{sec:sudakov}). Given the limited precision of the CDF data, all the predictions are statistically acceptable (see Table~\ref{t:chi2th}); the stability of these changes against the theoretical uncertainties is assessed in Sec.~\ref{s:uncertainties}.

\subsection{\texorpdfstring{CMS $Z$-boson at 8 TeV}{CMS Z-boson at 8 TeV}}

The CMS collaboration presented the first measurement of angular coefficients for $Z$-boson production at the LHC. The analysis was performed using a dilepton invariant-mass window of $81 < M < 101$~GeV, and the angular coefficients were extracted as functions of the transverse momentum $q_T$ in two rapidity intervals, $|y| \in [0.0,\,1.0]$ and $|y| \in [1.0,\,2.1]$. In the following, we restrict our comparison to the central rapidity bin $|y|<1$, which provides the highest statistical precision and allows for a direct and consistent comparison with the ATLAS measurements.

\begin{figure}[h!]
\centering
\includegraphics[width=0.48\textwidth]{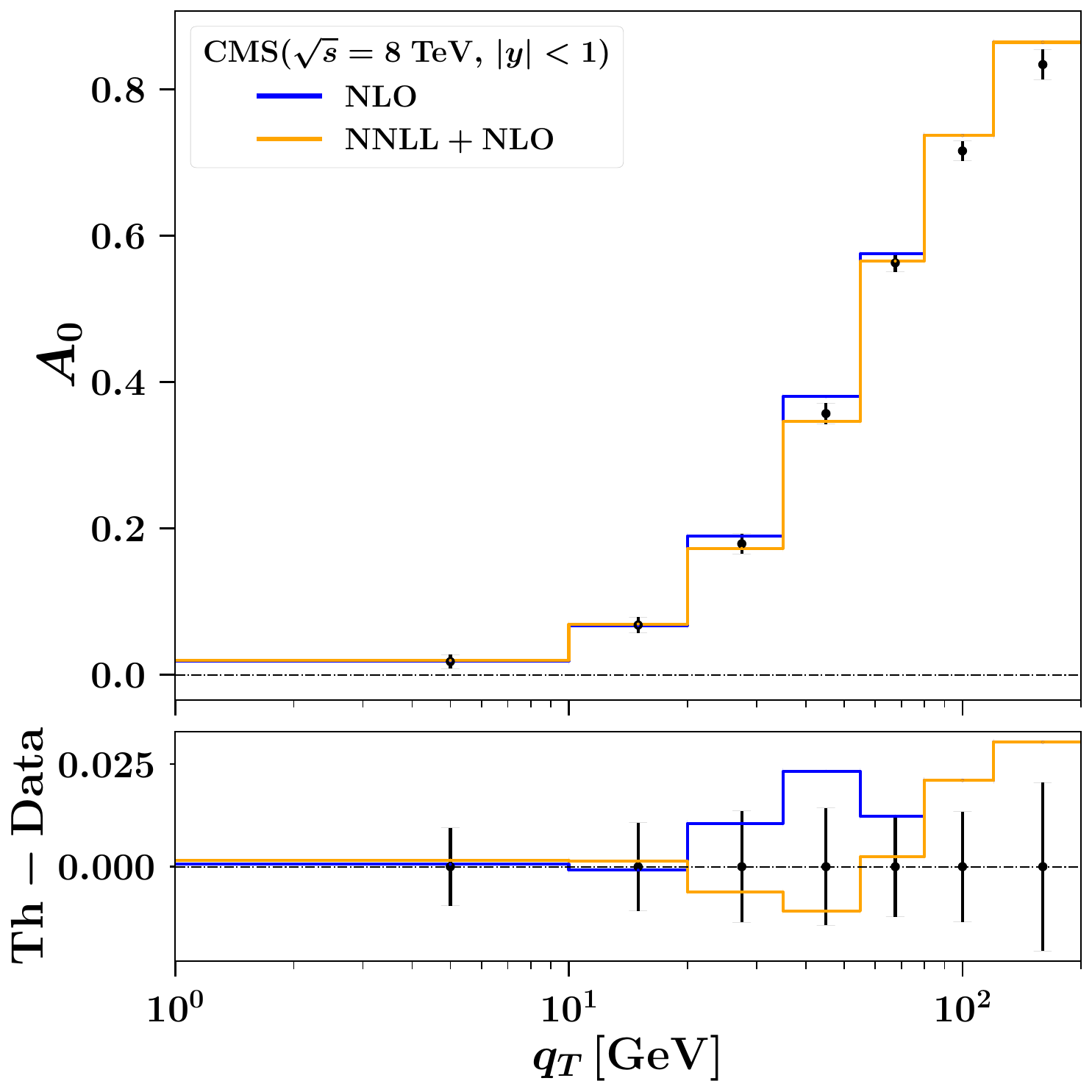}
\includegraphics[width=0.48\textwidth]{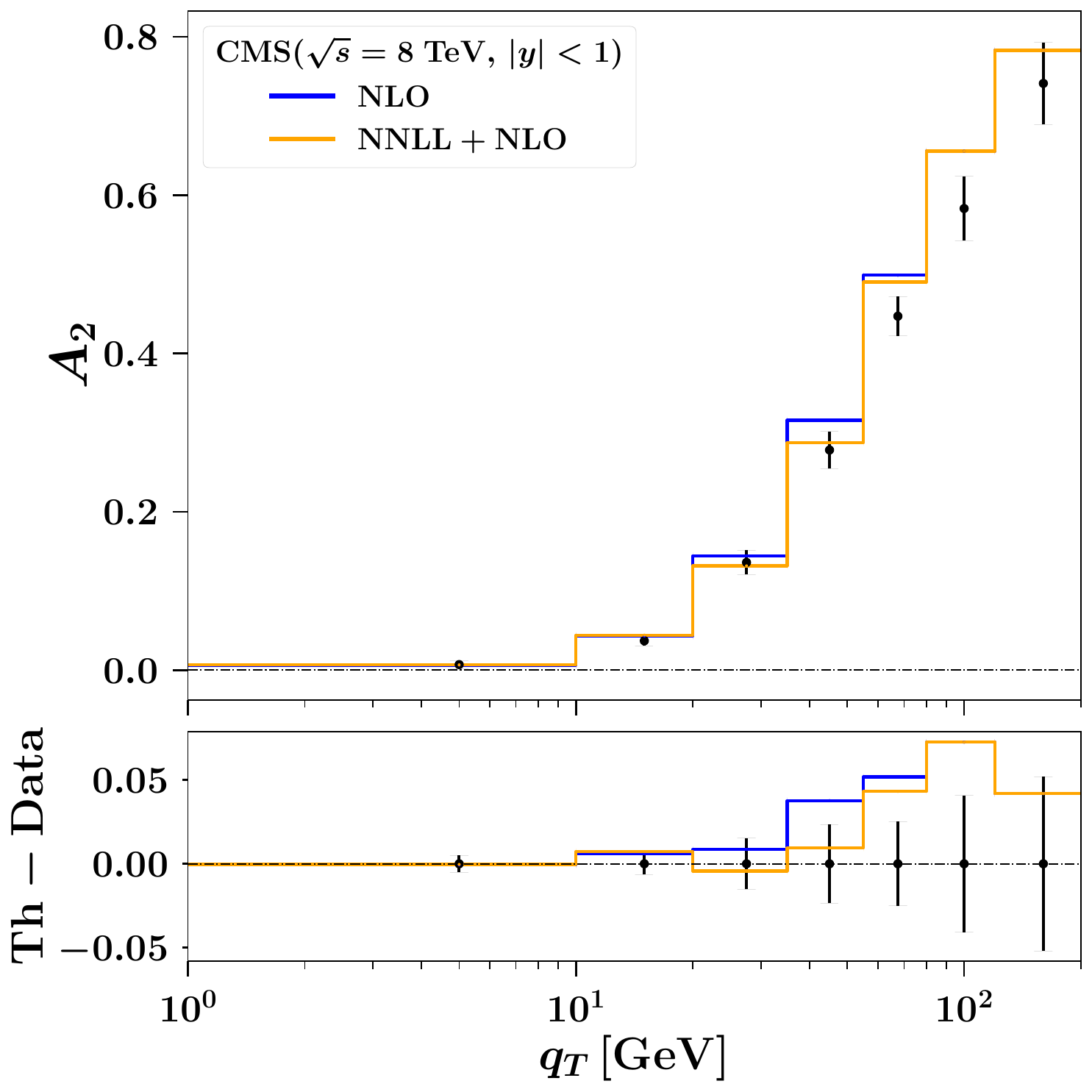}
\includegraphics[width=0.48\textwidth]{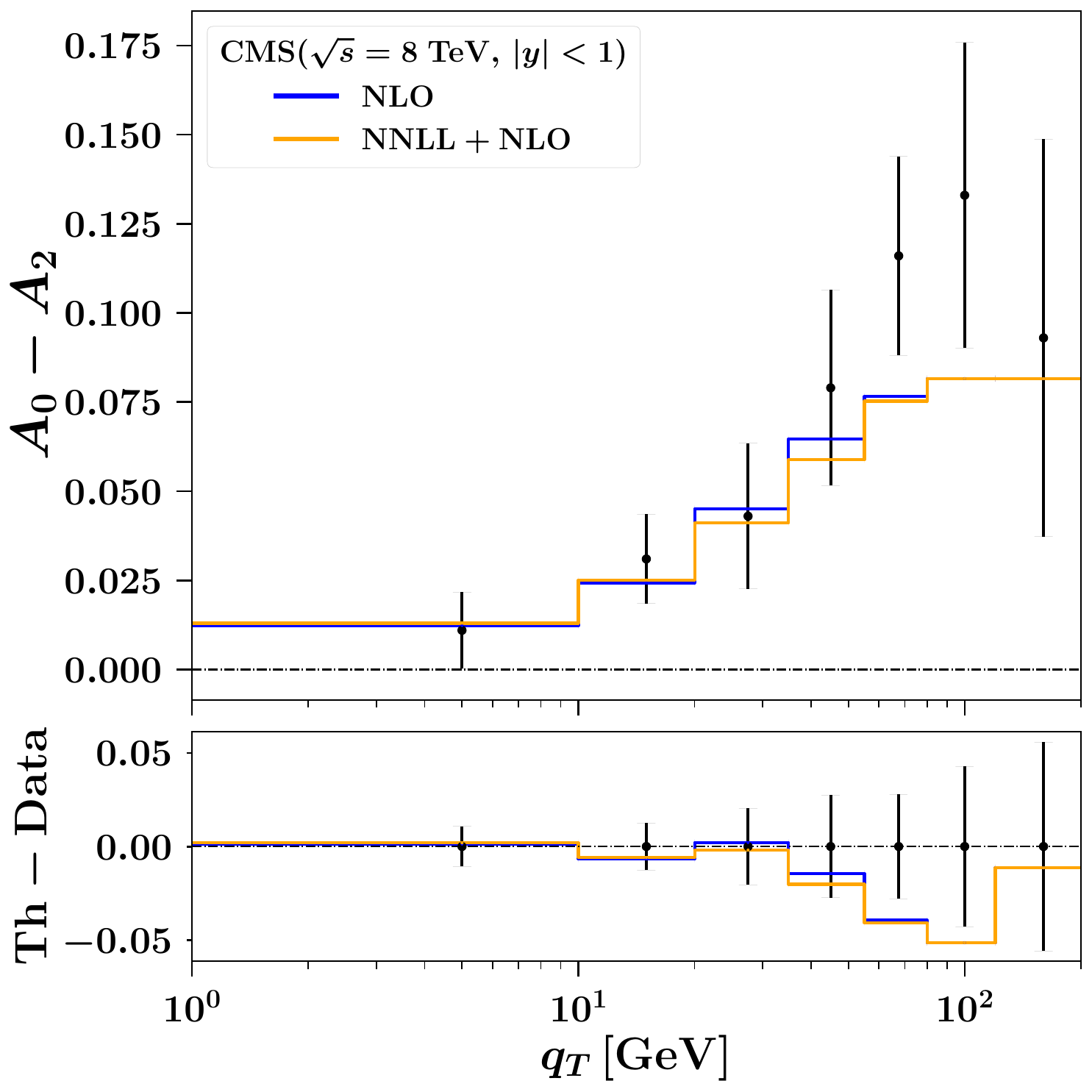}
 \caption{Comparison between experimental data and theoretical predictions at NLO (blue) and NNLL+NLO (orange) for the angular coefficients $A_0$ (upper left), $A_2$ (upper right), and $A_0-A_2$ (lower) in $Z$-boson production at CMS with $|y|<1$, shown as functions of $q_T$. The lower panels display the differences of the predictions with respect to the corresponding experimental results.}
\label{f:A_CMS}
\end{figure}

The CMS data in this rapidity interval are provided in seven $q_T$ bins spanning the range $q_T \in [0,\,200]$~GeV. Compared to the Tevatron measurements, the CMS dataset features a finer binning and significantly improved statistical precision, while still maintaining moderate bin widths. This makes the CMS measurements particularly well-suited for investigating the interplay between fixed-order and resummed
perturbative predictions over a broad kinematic range.

The $q_T$ distributions of the angular coefficients $A_0$, $A_2$, and their difference $A_0 - A_2$ are shown in Fig.~\ref{f:A_CMS}. The experimental data are displayed as black points and compared to theoretical predictions obtained at NLO accuracy (blue) and at NNLL+NLO accuracy (orange). The lower panels show the differences between the theoretical predictions and the experimental measurements, which provide a clearer visualisation of the impact of higher-order corrections.

In contrast to the CDF case, the increased precision of the CMS data leads to a more pronounced sensitivity to resummation effects, particularly in the intermediate transverse-momentum region, $q_T \sim 20$--$50$~GeV. While the fixed-order NLO predictions already capture the overall trends of the data, the inclusion of the resummed contribution results in a visibly improved agreement across multiple $q_T$ bins.

  For the Lam--Tung combination $A_0-A_2$ the change is small, as expected from the common rescaling of $A_0$ and $A_2$ discussed in Sec.~\ref{sec:sudakov}.

 \subsection{\texorpdfstring{ATLAS $Z$-boson}{ATLAS Z-boson}}

The ATLAS measurements have been performed within the invariant-mass range $80 < M < 100$~GeV, and the angular coefficients have been extracted as functions of $q_T$ in several rapidity intervals. As discussed above, we restrict our analysis to the central rapidity bin $|y|<1$, and we compare our predictions to the ATLAS data in the region $q_T \in [0,\,100]$~GeV. This choice is motivated by the fact that the impact of resummation effects becomes negligible beyond this transverse-momentum range, where fixed-order perturbation theory alone provides an accurate description.

Two features of the ATLAS dataset are worth noting before presenting the comparison. First, it has the finest $q_T$ binning among the datasets considered in this work, which results in visible point-to-point fluctuations between neighbouring bins. Second, in the first bin, $q_T \in [0,2.5]$~GeV, the $A_2$ coefficient is expected to receive a non-negligible, possibly dominant, contribution from the Boer--Mulders functions\,\cite{Boer:1997nt,Boer:1999mm}, which are transverse-momentum-dependent distributions describing transversely polarized partons inside unpolarized hadrons. Since these functions are not included in our theoretical framework, we exclude this bin from the analysis of $A_2$ and, for consistency, from the analysis of all other observables as well.

\begin{figure}[h!]
\centering
  \includegraphics[width=0.48\textwidth]{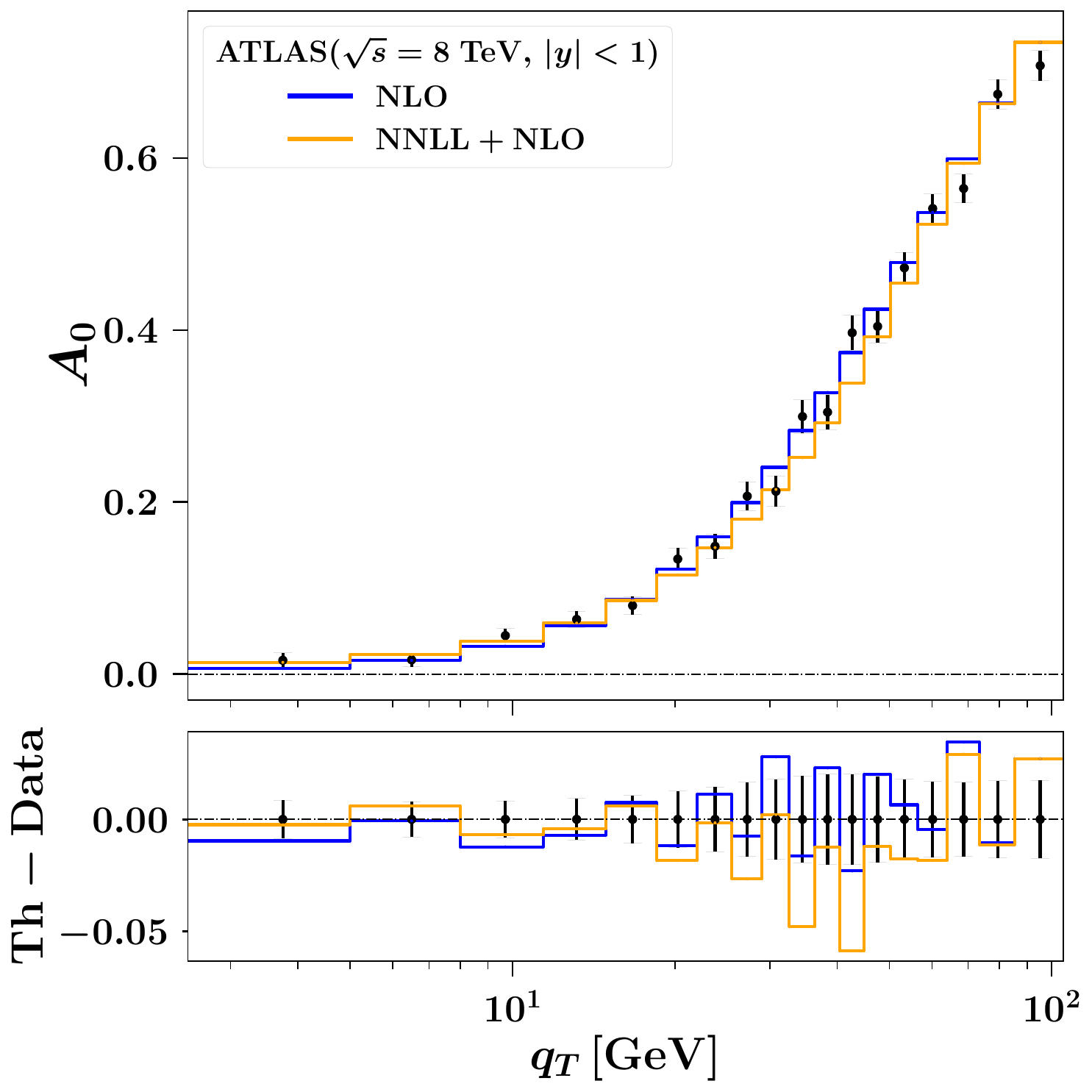}
\includegraphics[width=0.48\textwidth]{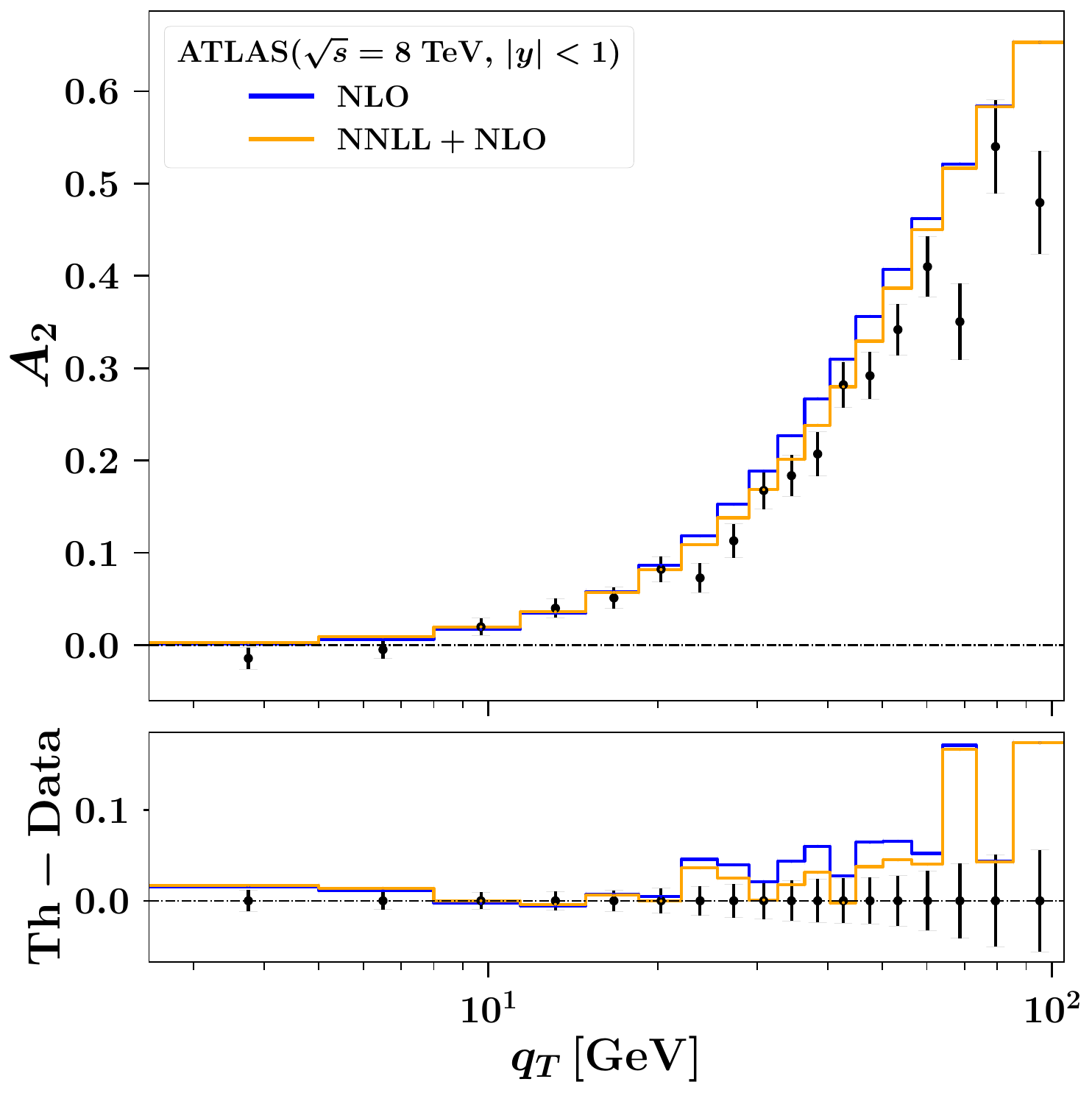}
\includegraphics[width=0.48\textwidth]{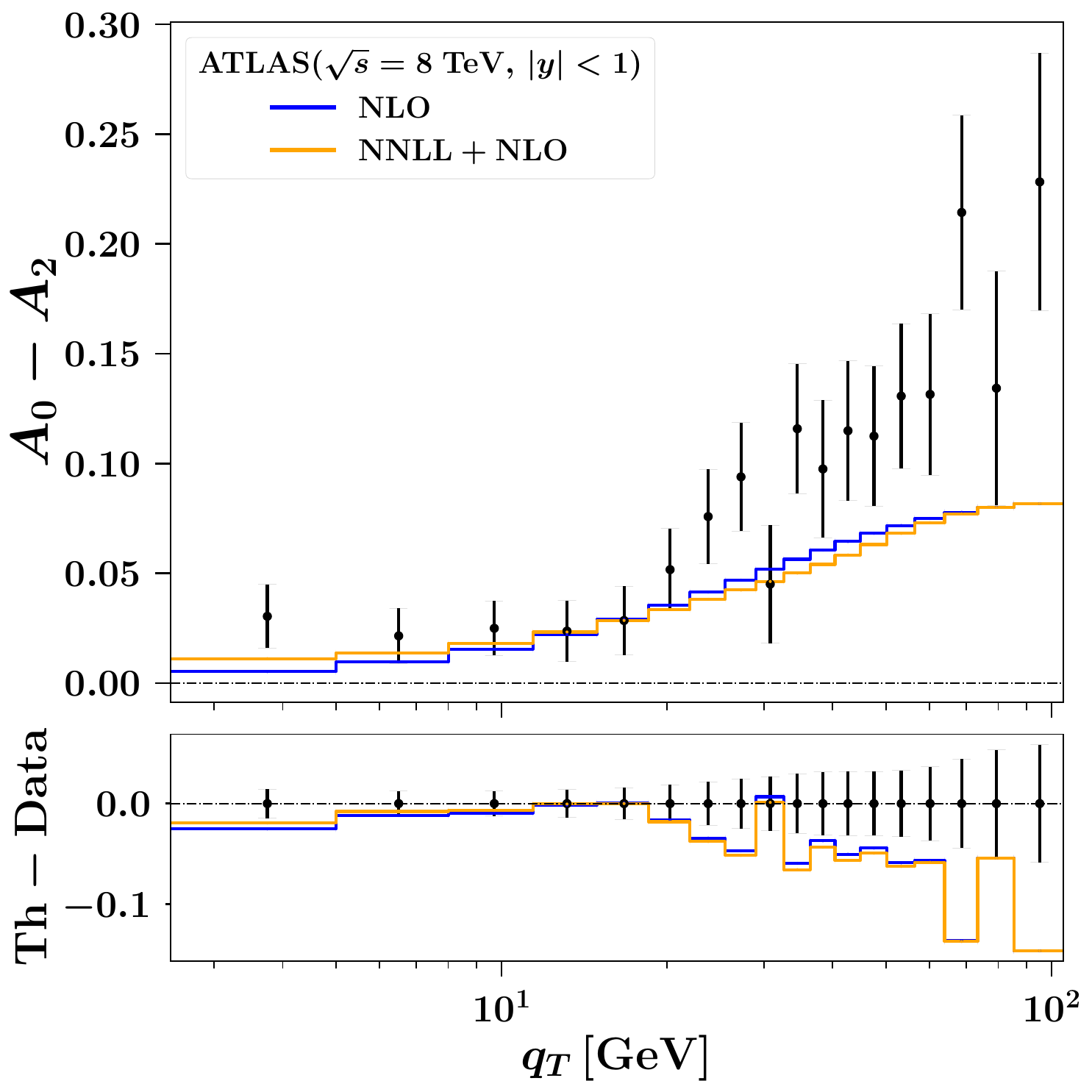}
 \caption{Comparison between experimental data and theoretical predictions at NLO (blue) and NNLL+NLO (orange) for the angular coefficients $A_0$ (upper left), $A_2$ (upper right), and $A_0-A_2$ (lower) in $Z$-boson production at ATLAS with $|y|<1$, shown as functions of $q_T$. The lower panels display the differences of the predictions with respect to the corresponding experimental results.}
\label{f:A_ATLAS}
\end{figure}

As already pointed out, the ATLAS measurements exhibit strong bin-by-bin correlations, so a purely visual inspection of Fig.~\ref{f:A_ATLAS} can be misleading. Nevertheless, a non-negligible improvement in the description of $A_2$ is visible in the intermediate $q_T$ ($q_T \sim 20$--$50$~GeV) region once the resummed contribution is included.

For $A_0$, instead, between 30 and 50~GeV the resummed prediction lies below the data. As shown in Sec.~\ref{s:uncertainties}, the sign of the change in the description of $A_0$ is not stable against the variation of the recoil prescription and of the scales, whereas the improvement for $A_2$ persists for all the variations considered. For $q_T\gtrsim20$~GeV the $A_0-A_2$ data lie above both predictions, in a region where $A_0-A_2$ is generated by the $\mathcal{O}(\alpha_S^2)$ corrections and is sensitive to the $\mathcal{O}(\alpha_S^3)$ ones\,\cite{Gauld:2017tww}. Finally, we note that the size of the shift induced by resummation is similar for the finely binned ATLAS data and for the CMS data in the same rapidity range (see Sec.~\ref{s:uncertainties}), as expected from the pointwise nature of the effect discussed in Sec.~\ref{sec:sudakov}.
\subsection{\texorpdfstring{LHCb $Z$-boson}{LHCb Z-boson}}

The LHCb measurement provides the first determination of the angular coefficients in the
forward-rapidity region ($y>2$), thereby offering information complementary to that
obtained by ATLAS and CMS, which are primarily sensitive to the central-rapidity region.
The analysis is performed in the dilepton invariant-mass window
$75 < M < 105$~GeV, and the angular coefficients are presented as functions of the
transverse momentum $q_T$.

\begin{figure}[h!]
\centering  \includegraphics[width=0.46\textwidth]{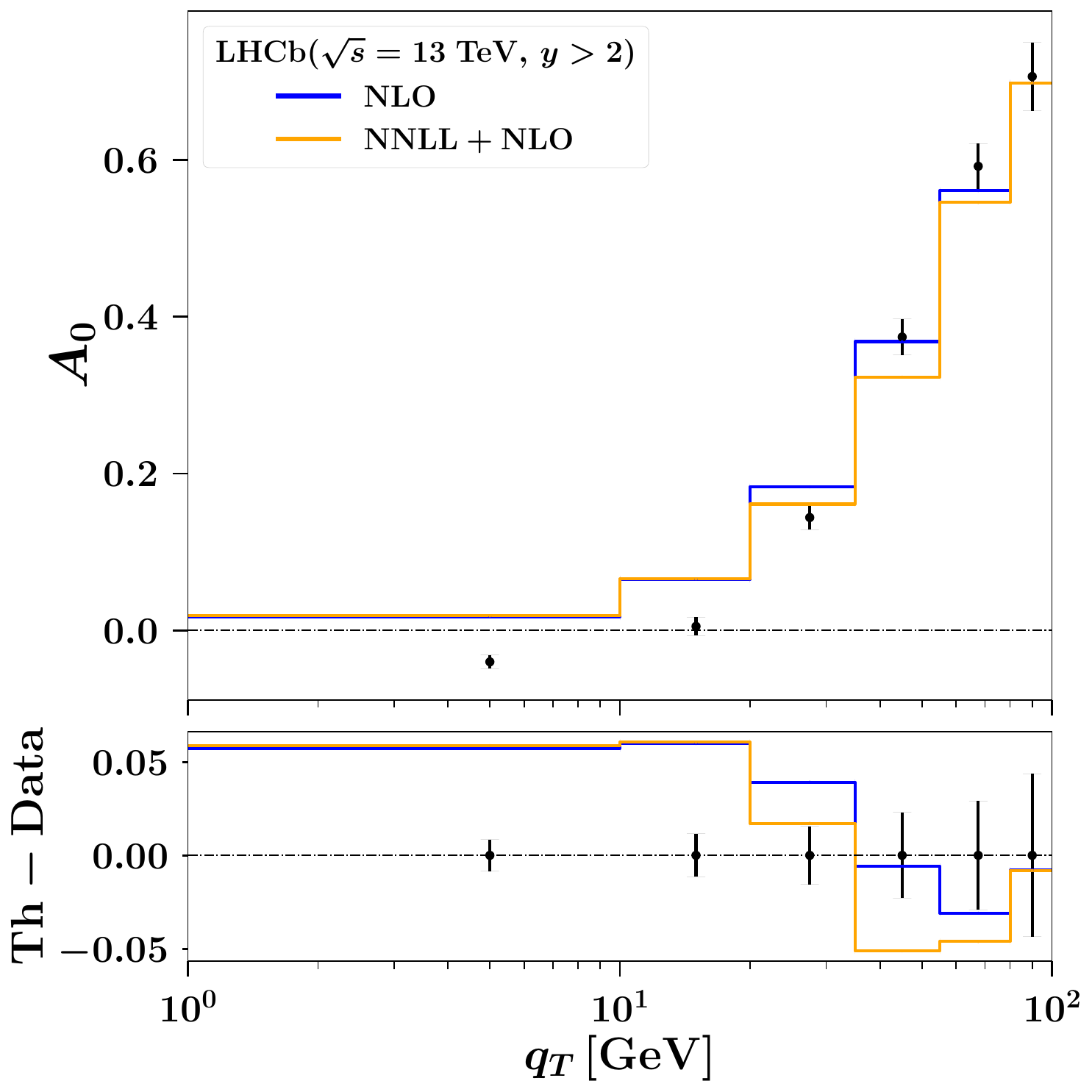}
\includegraphics[width=0.46\textwidth]{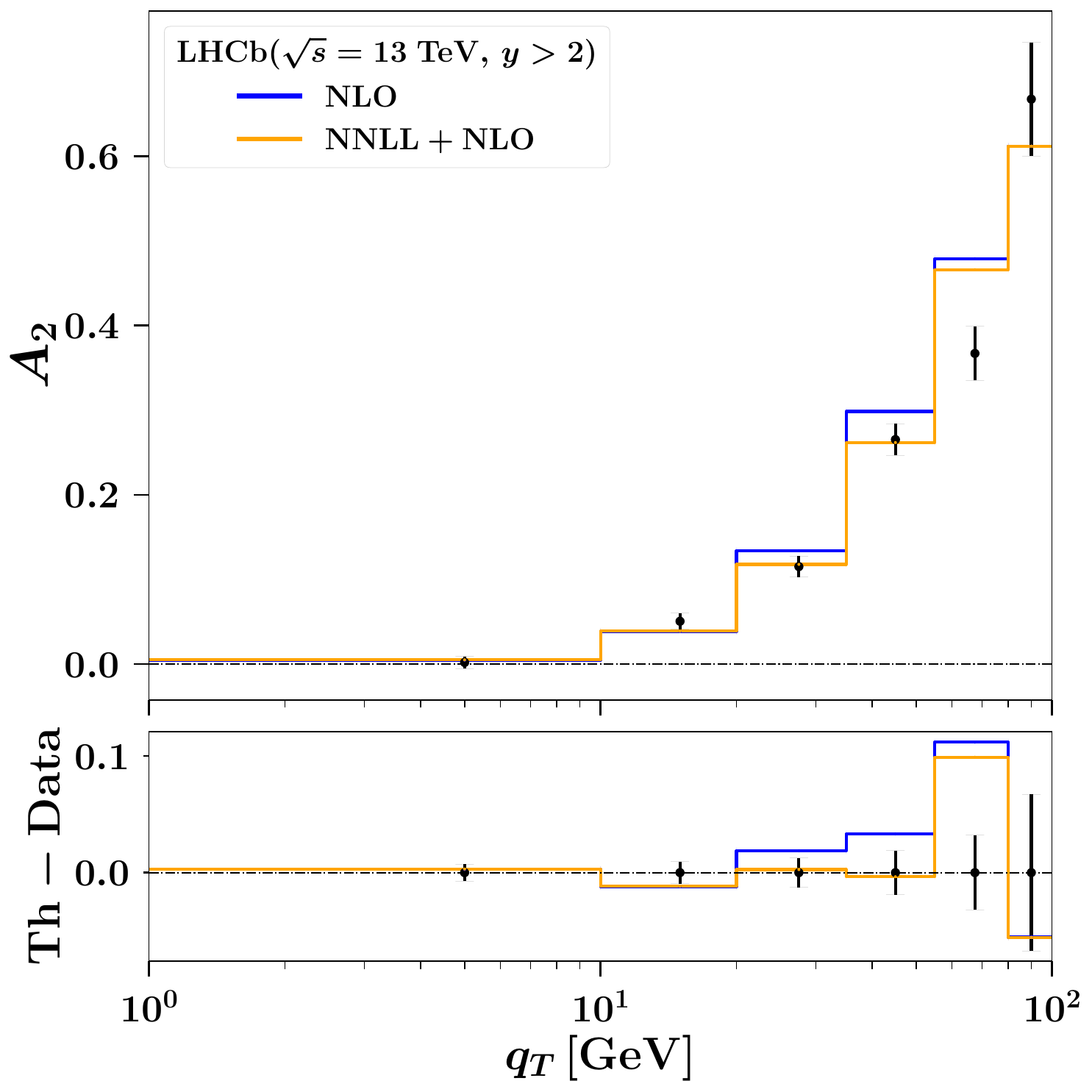}
\includegraphics[width=0.46\textwidth]{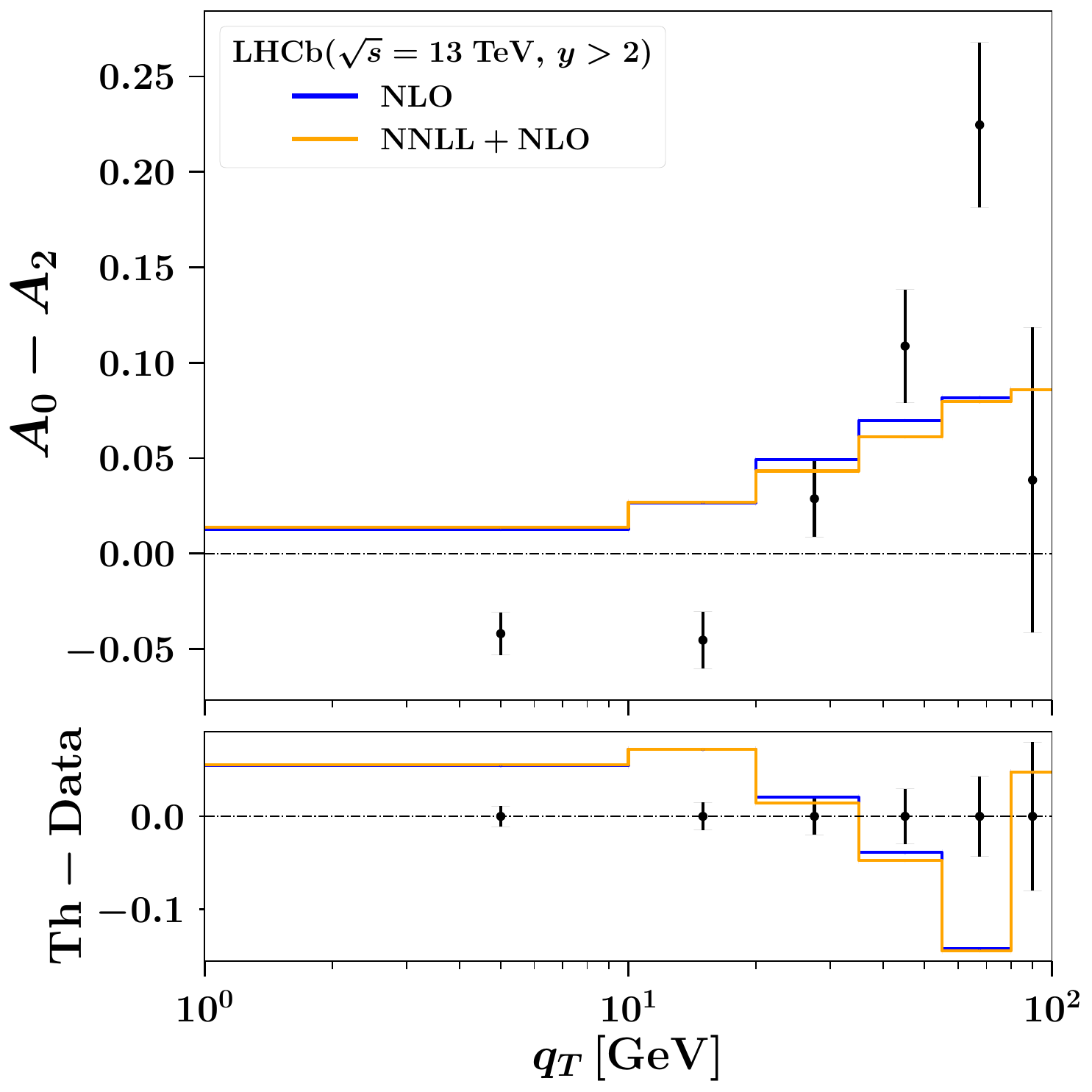}
 \caption{Comparison between experimental data and theoretical predictions at NLO (blue) and NNLL+NLO (orange) for the angular coefficients $A_0$ (upper left), $A_2$ (upper right), and $A_0-A_2$ (lower) in $Z$-boson production at LHCb with $y> 2$, shown as functions of $q_T$. The lower panels display the differences of the predictions with respect to the corresponding experimental results.}
\label{f:A_LHCb}
\end{figure}

The $q_T$ distributions of the angular coefficients $A_0$, $A_2$, and their difference
$A_0-A_2$ are shown in Fig.~\ref{f:A_LHCb}, where the experimental data are compared to
theoretical predictions at NLO and at NNLL+NLO accuracy. As already observed in the
CDF and CMS analyses, the inclusion of resummed contributions leads to a systematic
reduction of the theoretical predictions in the intermediate-$q_T$ region. The
qualitative trend induced by resummation therefore remains consistent also in the
forward-rapidity region probed by LHCb.

In this case, however, a quantitative assessment of the impact of the resummed
contribution is more challenging. In particular,
the first data point of the $A_0$
distribution is negative, i.e.\ outside the physical region $A_0\ge0$, and the second
one is compatible with zero: they lie $6.7$ and $5.2$ standard deviations below the
NLO prediction, respectively, a deviation much larger than the effect of resummation,
which is negligible in this region. These points dominate
the corresponding $\chi^2$ values and, as a result, also affect the $\chi^2$ of the
derived Lam--Tung combination $A_0-A_2$.

For this reason, the $\chi^2$ values associated with $A_0$ and $A_0-A_2$ should be interpreted with caution in the LHCb case; we therefore also report in Table~\ref{t:chi2th}, in brackets, the reduced $\chi^2$ obtained by
excluding the first two points. Without these points, as shown in Sec.~\ref{s:uncertainties}, the sign of the change in the description of $A_0$ is not stable against the theoretical uncertainties.

By contrast, the behaviour
of the $A_2$ coefficient is more stable and displays a clearer sensitivity to
resummation effects. In this case the comparison of the differential
distributions indicates improved agreement when resummed predictions are
employed. This suggests that, despite the reduced precision and the presence of
pathological points in some observables, the forward-rapidity LHCb data are compatible
with the same qualitative impact of resummation observed in the central region.

\subsection{\texorpdfstring{CMS $Z$-boson at 13 TeV}{CMS Z-boson at 13 TeV}}
\label{s:CMS13}

The CMS collaboration has recently measured the angular coefficients of $Z$-boson production in the dimuon channel at $\sqrt{s}=13$~TeV, with an integrated luminosity of $140~{\rm fb}^{-1}$\,\cite{CMS:2026amb}. The measurement is performed in the invariant-mass window $81<M<101$~GeV, as a function of $q_T$ in two rapidity intervals, $|y|<1$ and $1<|y|<2.4$; the latter extends the comparison to an intermediate region between the central ATLAS and CMS measurements at 8~TeV and the forward LHCb measurement. As for the measurement at 8~TeV, we consider the seven bins with $q_T<200$~GeV. These are the most precise $Z$-boson measurements considered in this work: in the intermediate region the uncertainties of $A_0$ and $A_2$ are below $0.01$. For $20<q_T<80$~GeV and $|y|<1$, the experimental uncertainties are smaller than those of the measurement at 8~TeV by a factor of $1.3$--$3.5$.

\begin{figure}[h!]
\centering
\includegraphics[width=0.48\textwidth]{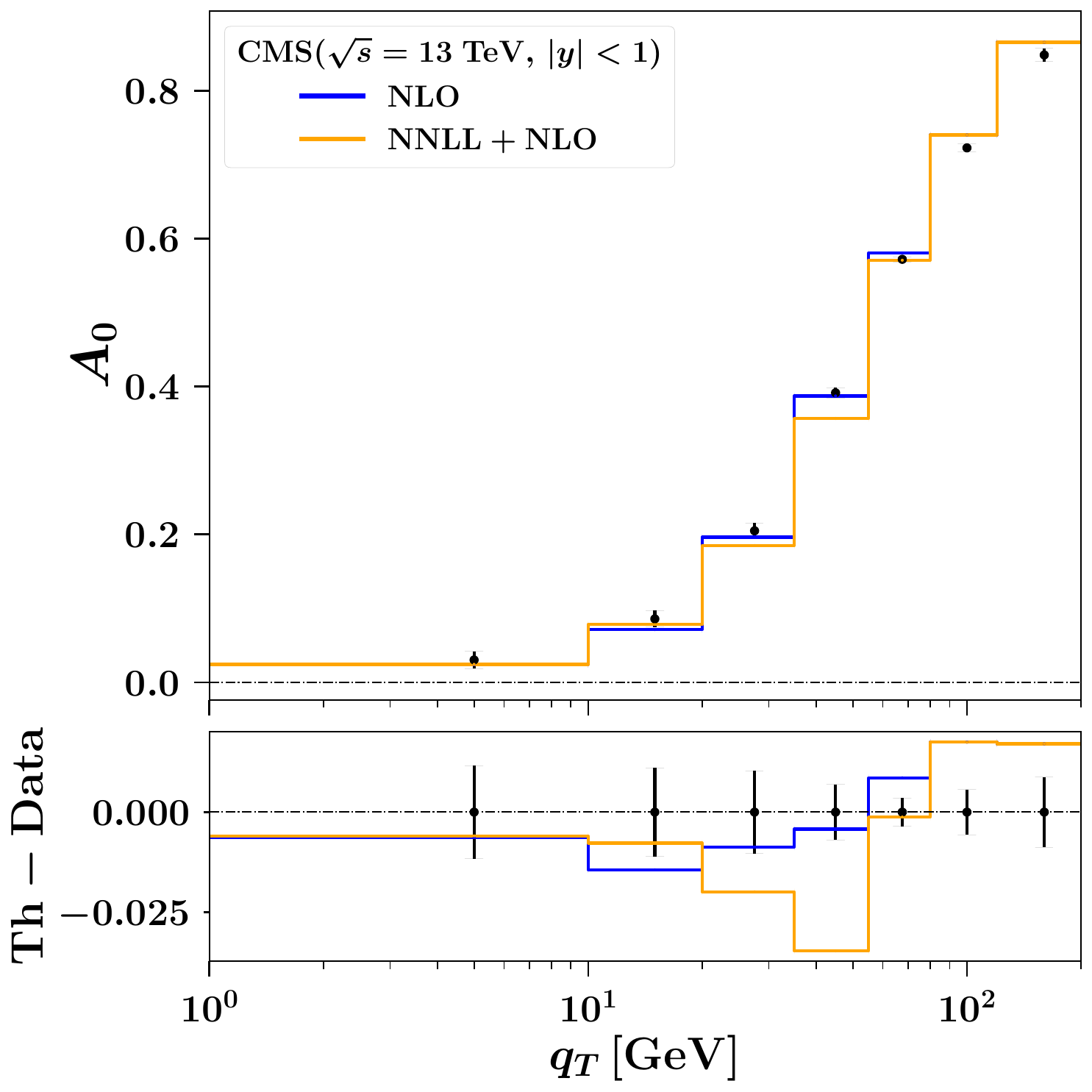}
\includegraphics[width=0.48\textwidth]{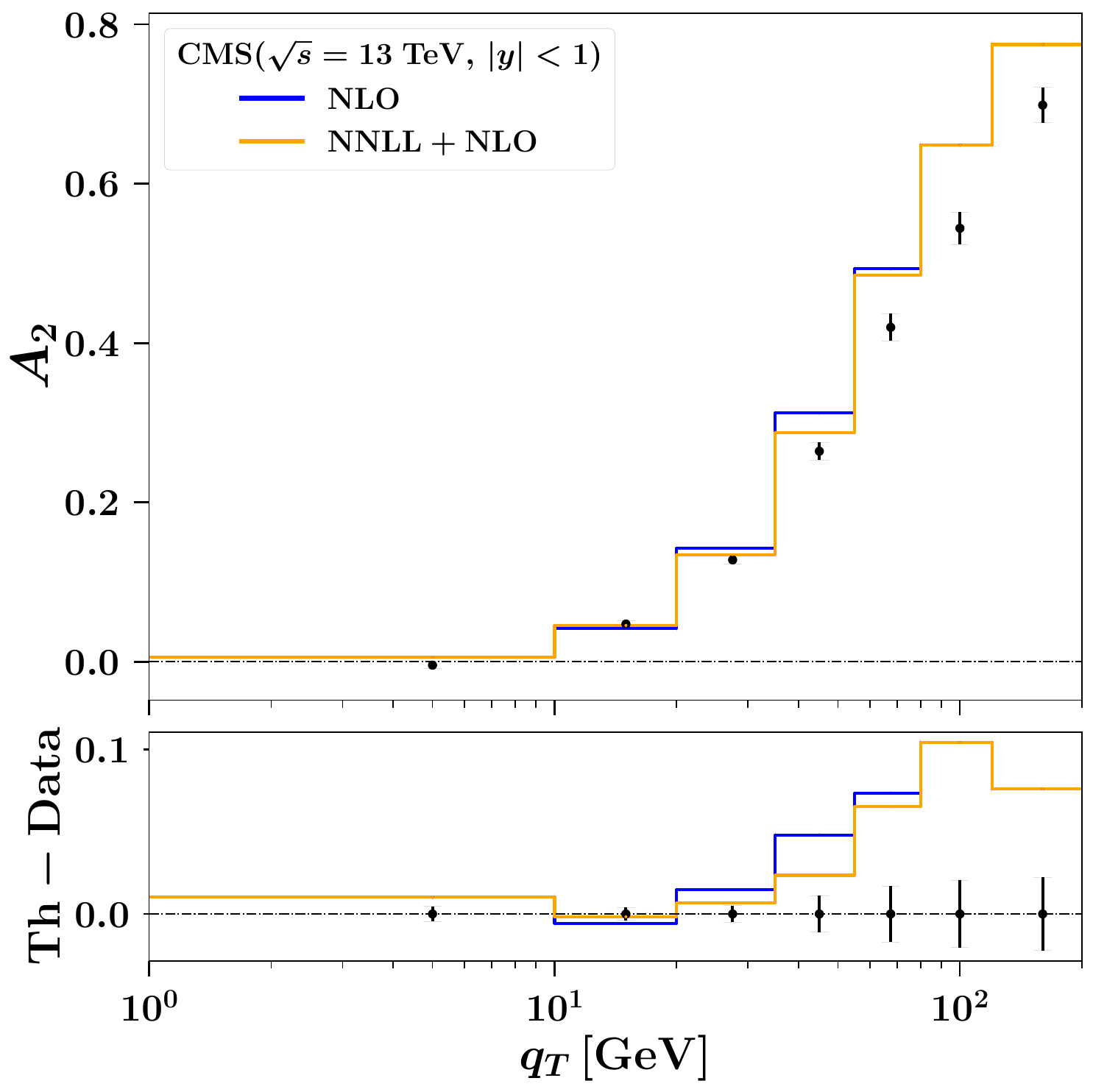}
\includegraphics[width=0.48\textwidth]{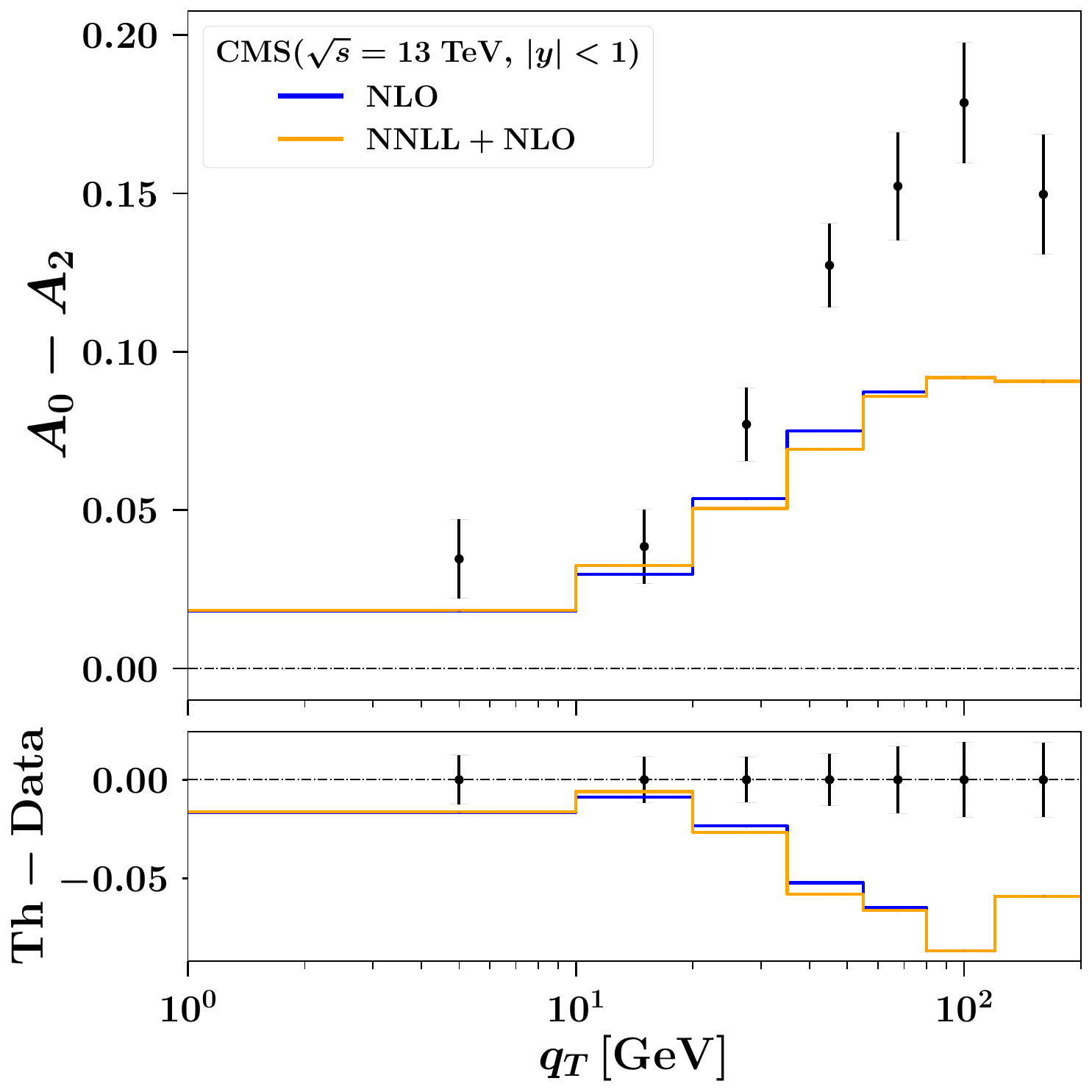}
\caption{Comparison between experimental data and theoretical predictions at NLO (blue) and NNLL+NLO (orange) for the angular coefficients $A_0$ (upper left), $A_2$ (upper right), and $A_0-A_2$ (lower) in $Z$-boson production at CMS at $\sqrt{s}=13$~TeV with $|y|<1$, shown as functions of $q_T$. The lower panels display the differences of the predictions with respect to the corresponding experimental results.}
\label{f:A_CMS13a}
\end{figure}

\begin{figure}[h!]
\centering
\includegraphics[width=0.45\textwidth]{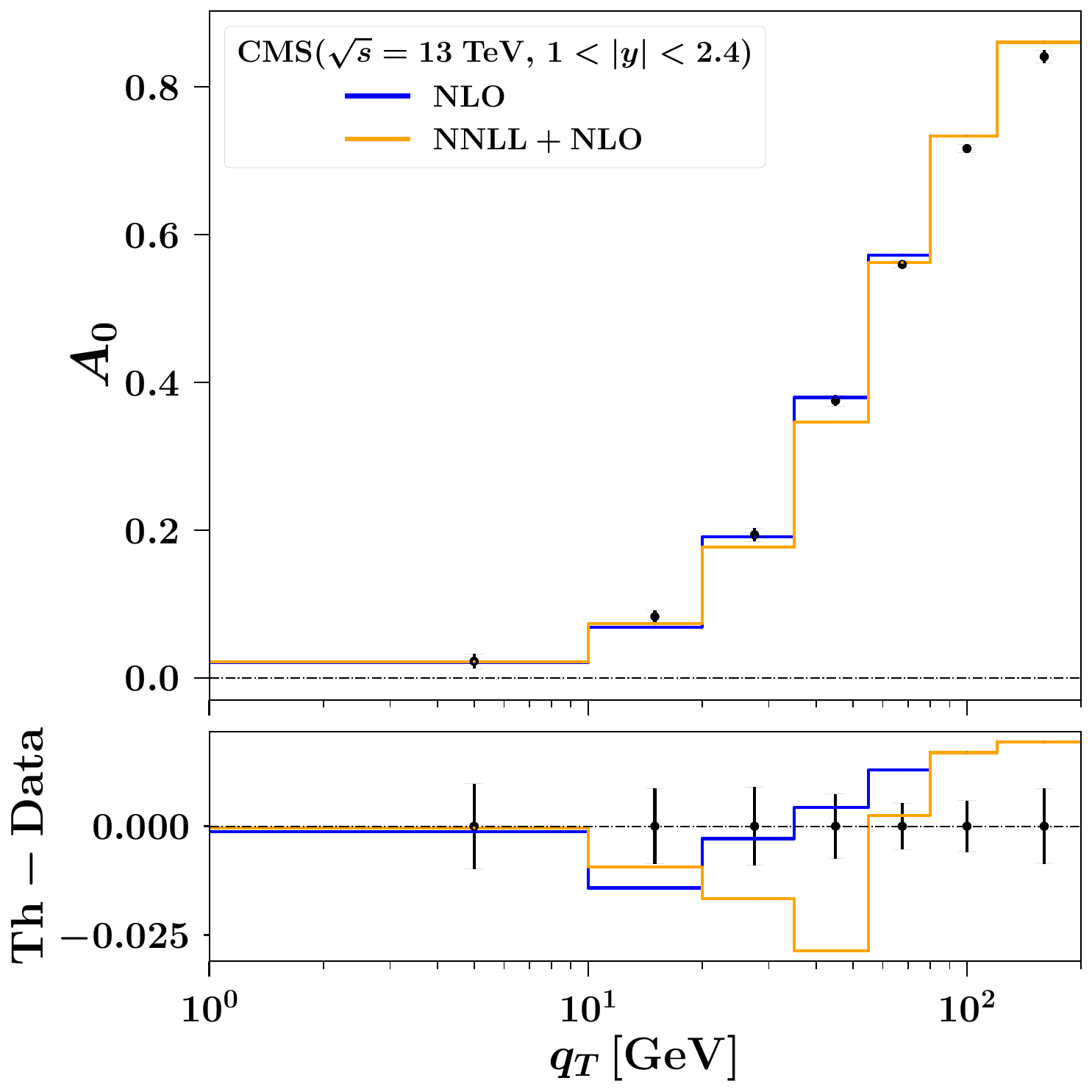}
\includegraphics[width=0.45\textwidth]{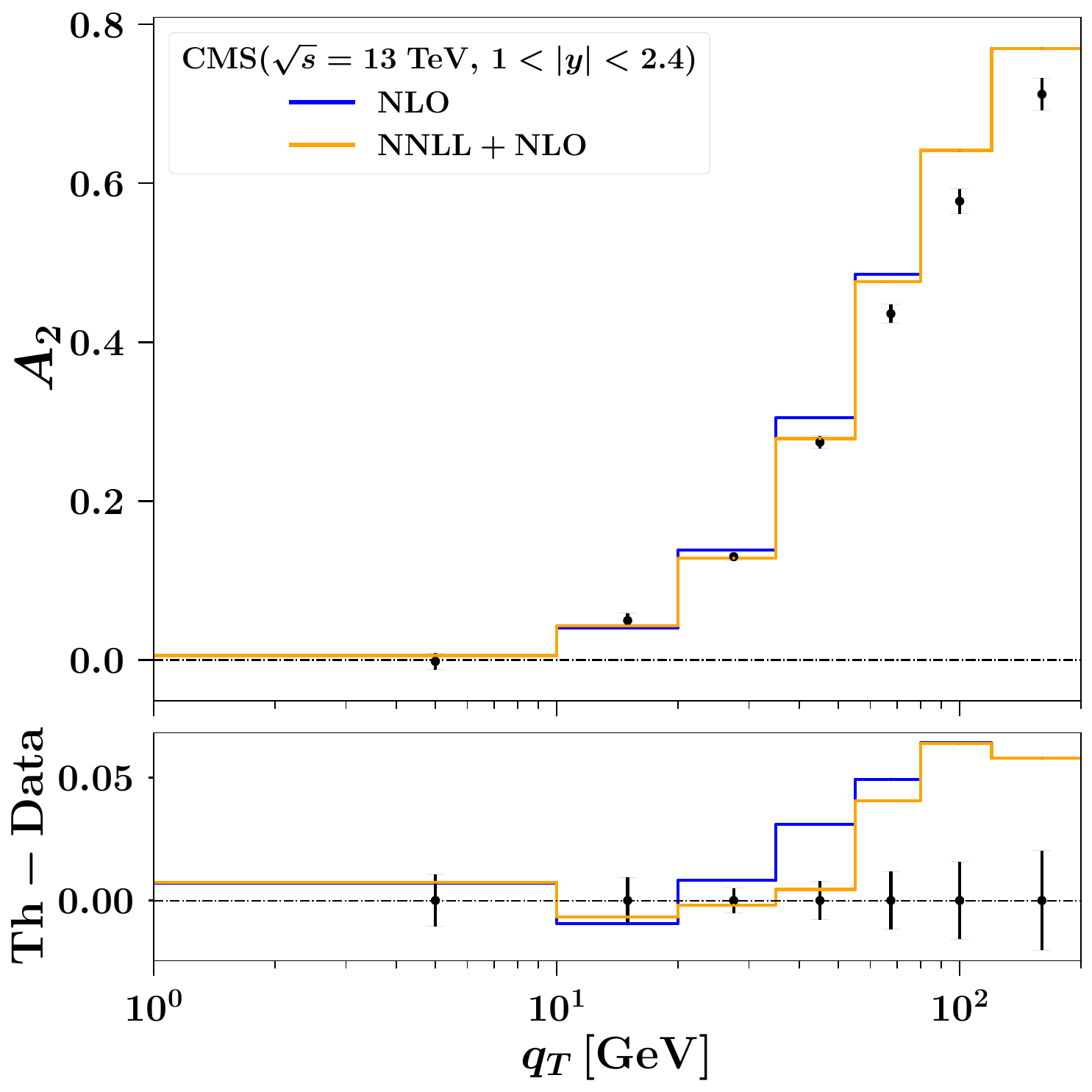}
\includegraphics[width=0.45\textwidth]{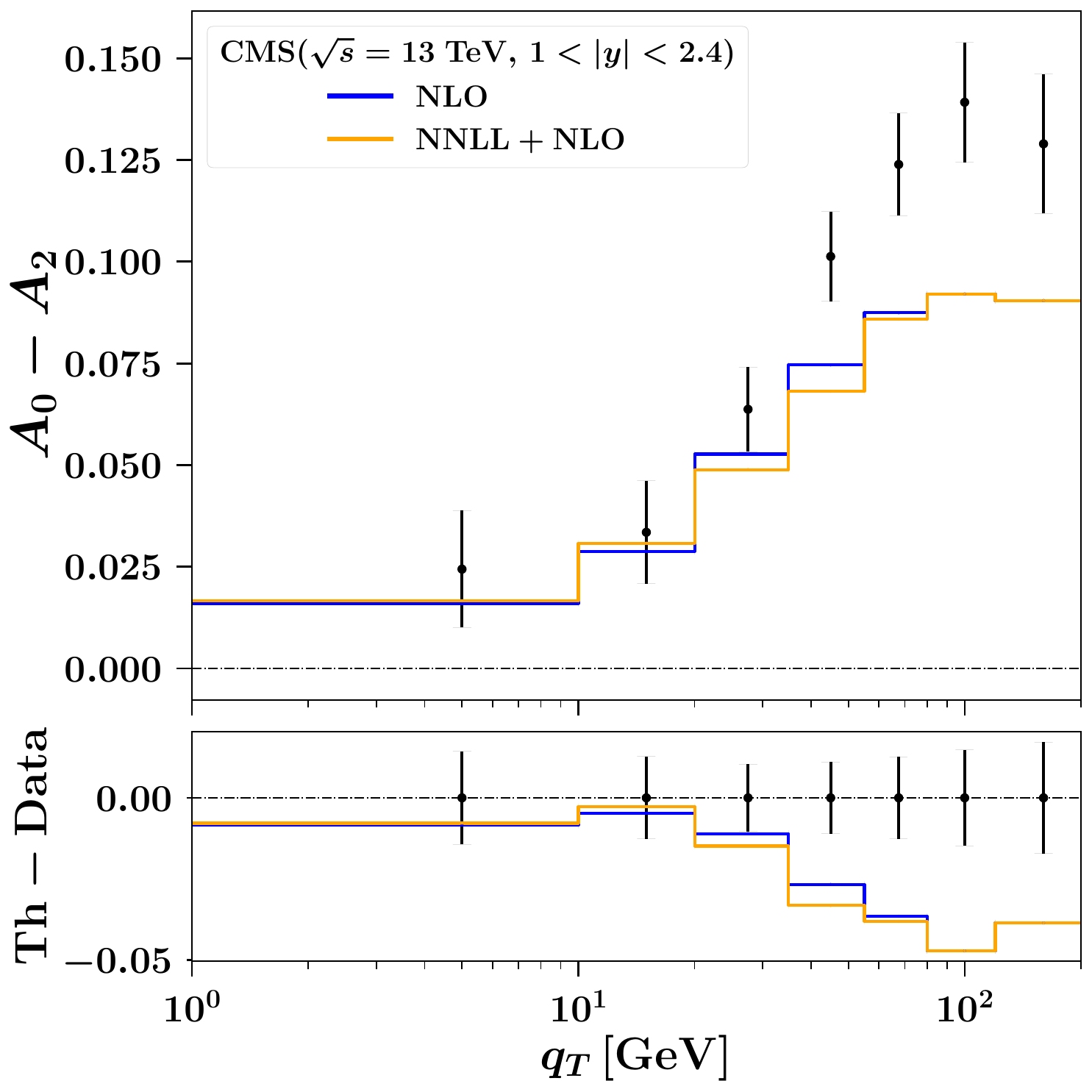}
\caption{Comparison between experimental data and theoretical predictions at NLO (blue) and NNLL+NLO (orange) for the angular coefficients $A_0$ (upper left), $A_2$ (upper right), and $A_0-A_2$ (lower) in $Z$-boson production at CMS at $\sqrt{s}=13$~TeV with $1<|y|<2.4$, shown as functions of $q_T$. The lower panels display the differences of the predictions with respect to the corresponding experimental results.}
\label{f:A_CMS13b}
\end{figure}

The comparison with the NLO and NNLL+NLO predictions is shown in Figs.~\ref{f:A_CMS13a} and~\ref{f:A_CMS13b}. The size of the resummation effect is similar to that observed at 8~TeV: in the region $20<q_T<55$~GeV, $A_0$ and $A_2$ are lowered by about $7\%$ for $|y|<1$ and by $8\%$ for $1<|y|<2.4$ (see Sec.~\ref{s:uncertainties}). Owing to the higher precision of the data, the shift corresponds in this case to $2$--$3$ experimental standard deviations, and it reaches $4.4$ standard deviations for $A_0$ in the bin $35<q_T<55$~GeV.

As for the other $Z$-boson measurements, resummation improves the description of $A_2$. The description of $A_0$, instead, worsens. In the region $20<q_T<55$~GeV the $A_0$ data agree with the NLO prediction within one standard deviation, and the resummed prediction lies $2$--$5$ experimental standard deviations below them, while the $A_2$ data lie $6$--$15\%$ below it. Since with the CS recoil prescription resummation rescales $A_0$ and $A_2$ by a common factor (see Eq.~\eqref{eq:Ai_res}), it cannot improve the description of both coefficients. For $q_T>80$~GeV, where the resummation effects vanish, both coefficients lie below the NLO prediction, by about $2\%$ for $A_0$ and by $7$--$16\%$ for $A_2$, which accounts for a large part of the $\chi^2$ of both predictions; in this region the $\mathcal{O}(\alpha_S^3)$ corrections are relevant\,\cite{Gauld:2017tww}.

The data for $A_0-A_2$ lie above the NLO prediction for $q_T>20$~GeV, by $45$--$95\%$ for $|y|<1$ and by $20$--$50\%$ for $1<|y|<2.4$: the measured violation of the Lam--Tung relation is larger than predicted at $\mathcal{O}(\alpha_S^2)$, as observed for the ATLAS data at 8~TeV. Resummation leaves $A_0-A_2$ essentially unchanged.

\subsection{\texorpdfstring{CDF $W$-boson}{CDF W-boson}}

In addition to being the first experiment to measure angular coefficients in
$Z$-boson production, the CDF collaboration was also the first to perform measurements
of angular coefficients in $W$-boson production, several years earlier. In
Ref.\,\cite{CDF:2005qwt}, CDF presented a measurement of the coefficients $A_2$ and $A_3$
for $W^-$ production.

Owing to the presence of an undetected neutrino in the final state, the full kinematics
of the $W$-boson decay could not be reconstructed. As a consequence, the analysis was
restricted to azimuthal angular modulations, obtained after integrating over the polar
angle $\theta$. This limitation prevents access to the complete set of angular
coefficients and explains why only $A_2$ and $A_3$ could be extracted in this
measurement.

\begin{figure}[h!]
\centering
  \includegraphics[width=0.44\textwidth]{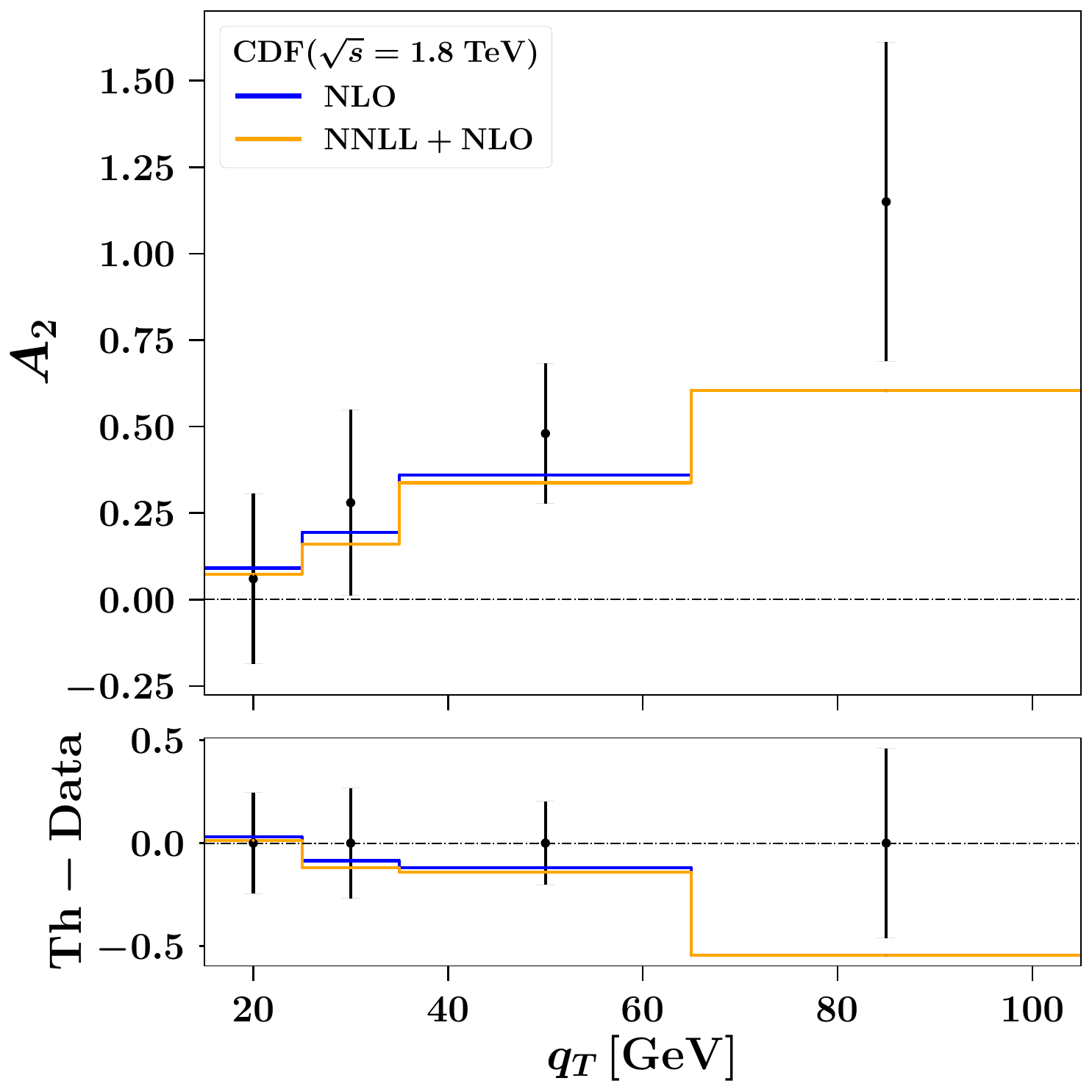}
\includegraphics[width=0.44\textwidth]{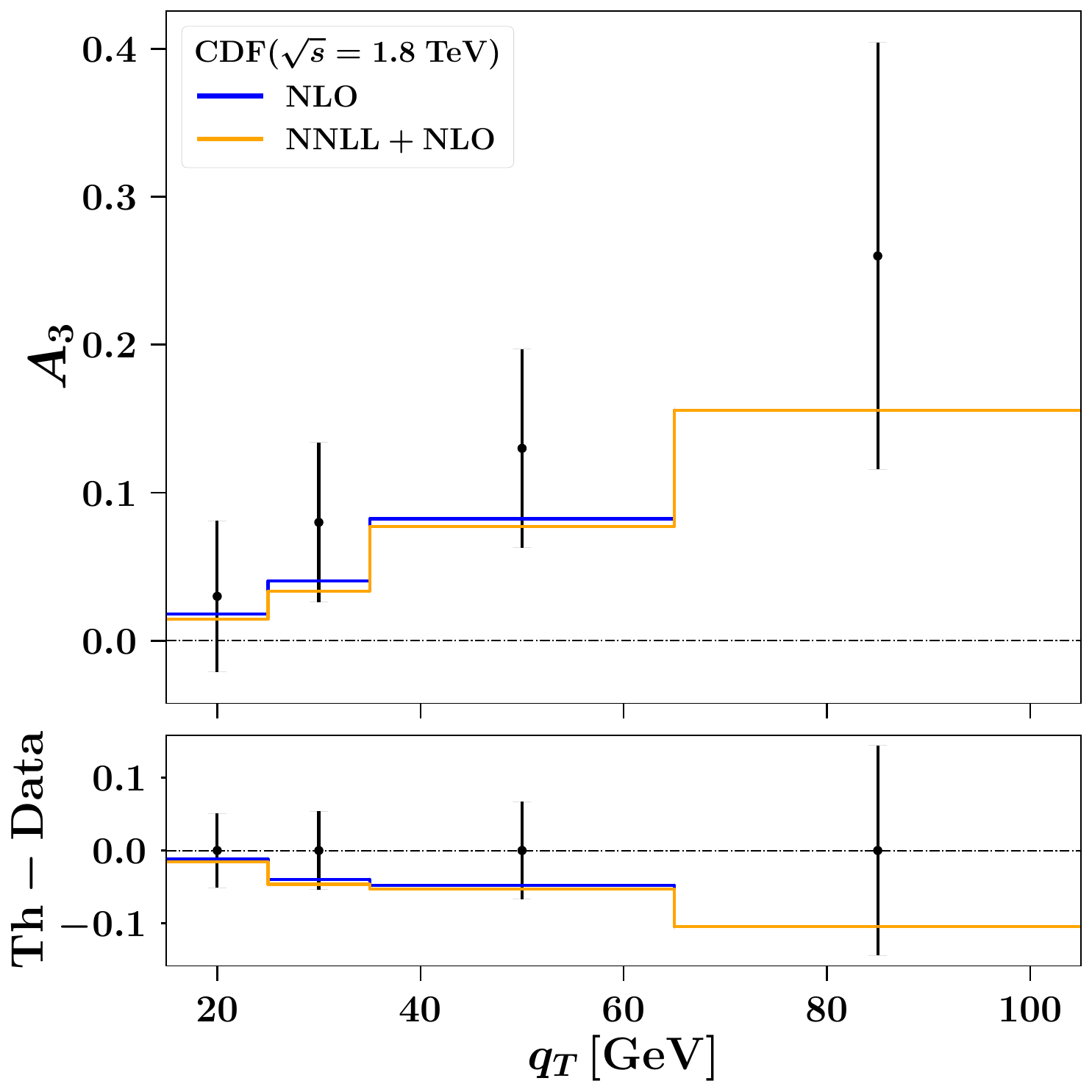}
 \caption{Comparison between experimental data and theoretical predictions at NLO (blue) and NNLL+NLO (orange) for the angular coefficients $A_2$ (left), $A_3$ (right), in $W^-$-boson production at CDF, shown as functions of $q_T$. The lower panels display the differences of the predictions with respect to the corresponding experimental results.}
\label{f:A_CDF_W}
\end{figure}

The $q_T$ distributions of the angular coefficients $A_2$ and $A_3$ are shown in
Fig.~\ref{f:A_CDF_W}, where the experimental data are compared to theoretical
predictions at NLO (blue) and NNLL+NLO (orange) accuracy. In contrast to the case of
$Z$-boson production, the coefficient $A_3$ plays a more prominent role for the charged
current process, as it is numerically sizeable over the measured $q_T$ range and exhibits a clearly visible dependence on
the transverse momentum. This feature makes $A_3$ sensitive to higher-order
QCD effects, as well as to the relative weight of the up- and down-type quark channels\,\cite{Gauld:2017tww}.

For this measurement we consider both the statistical and the systematic uncertainties.

As shown in Fig.~\ref{f:A_CDF_W}, the inclusion of resummation effects has a visible impact in the intermediate-$q_T$ region for both angular coefficients. In our setup resummation rescales $A_2$ and $A_3$ by the same factor (see Eq.~\eqref{eq:Ai_res}), so that the effect on the comparison with the data depends on the position of the NLO predictions with respect to the data, rather than on a different sensitivity of the two coefficients to resummation. We also note that $A_3$ is sensitive to linear power-suppressed
hadronic contributions, which are included at fixed order but are not
resummed (see Sec.~\ref{sec:sudakov}).

\subsection{\texorpdfstring{ATLAS $W$-boson}{ATLAS W-boson}}

The ATLAS collaboration has recently presented the first measurement of the complete set of angular coefficients for both charged $W$ bosons. This represents a major milestone in the study of the angular structure of charged-current Drell--Yan processes, as it constitutes the first experimental determination of the Lam--Tung relation for charged vector bosons. In this work, we investigate the impact of resummation effects on the angular coefficients for both $W^-$ and $W^+$ production.

\begin{figure}[h]
\centering \includegraphics[width=0.37\textwidth]{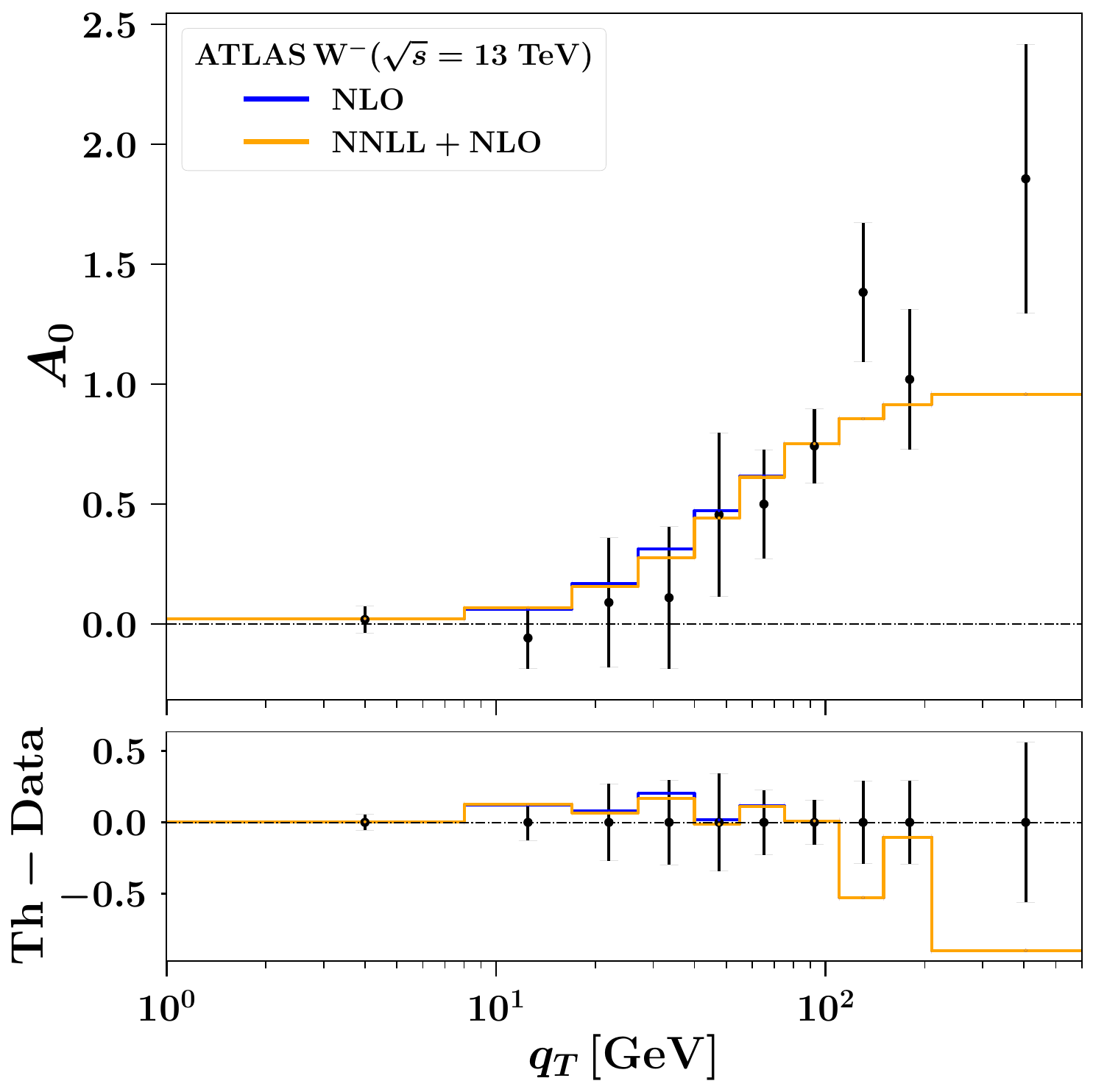}
\includegraphics[width=0.37\textwidth]{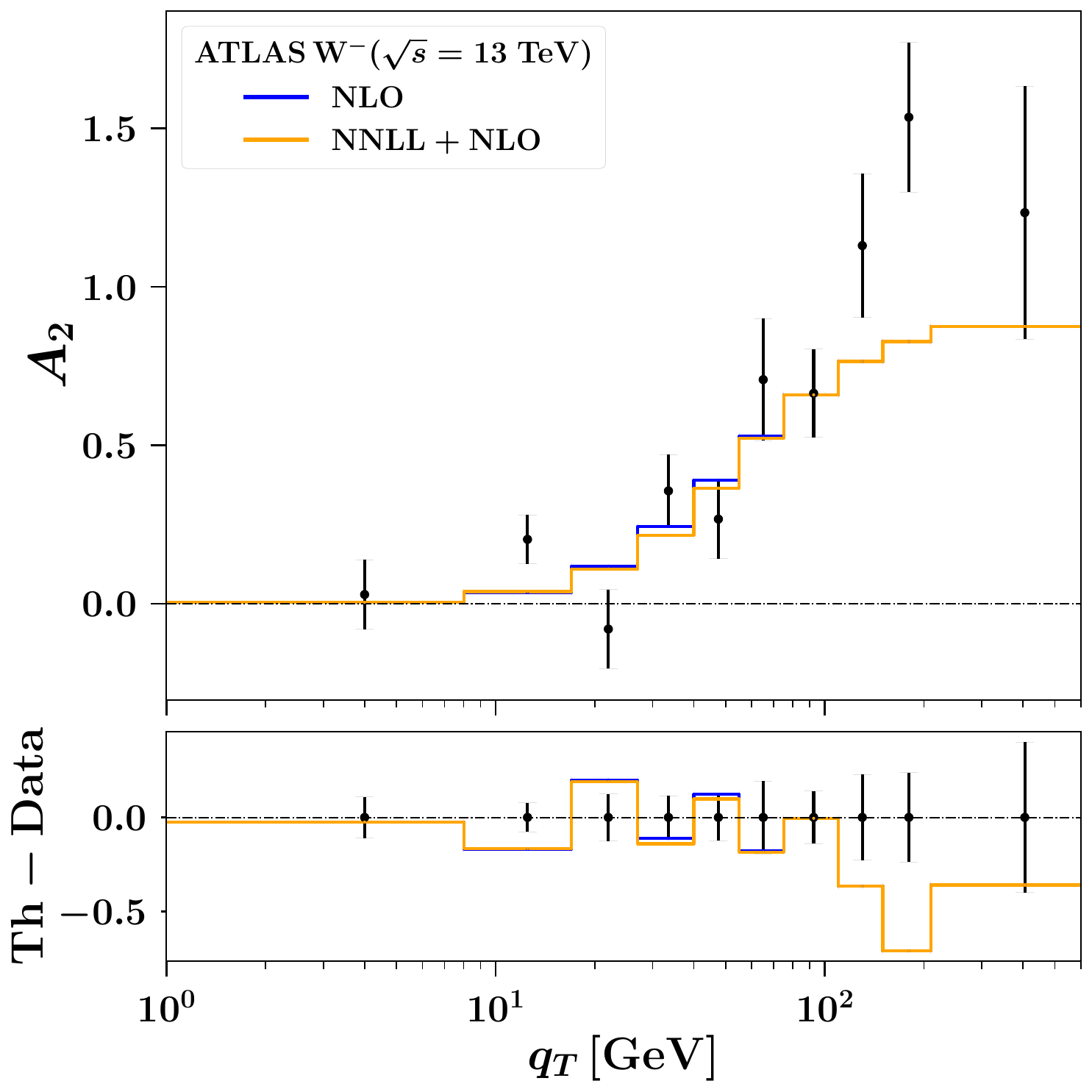}
\includegraphics[width=0.37\textwidth]{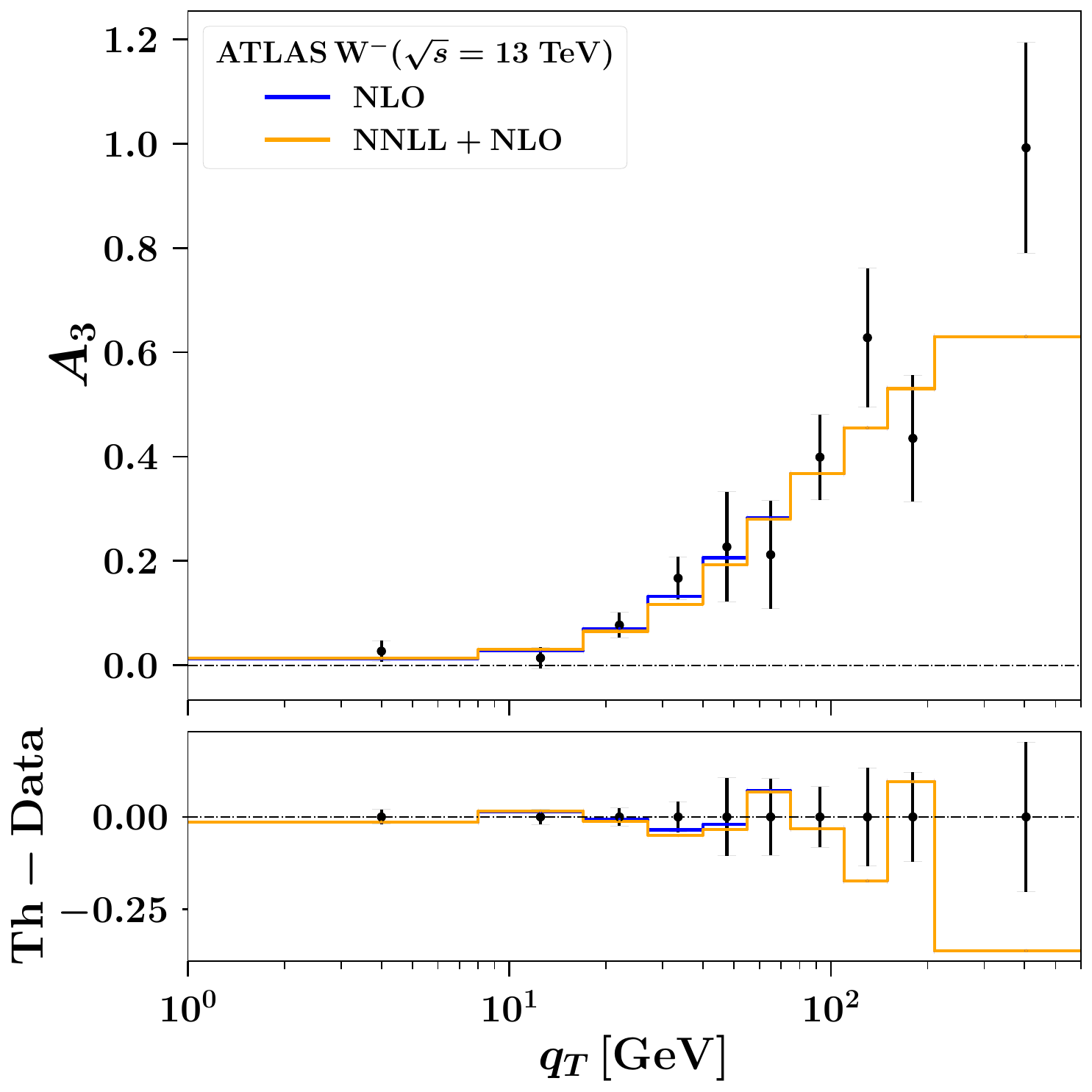}
\includegraphics[width=0.37\textwidth]{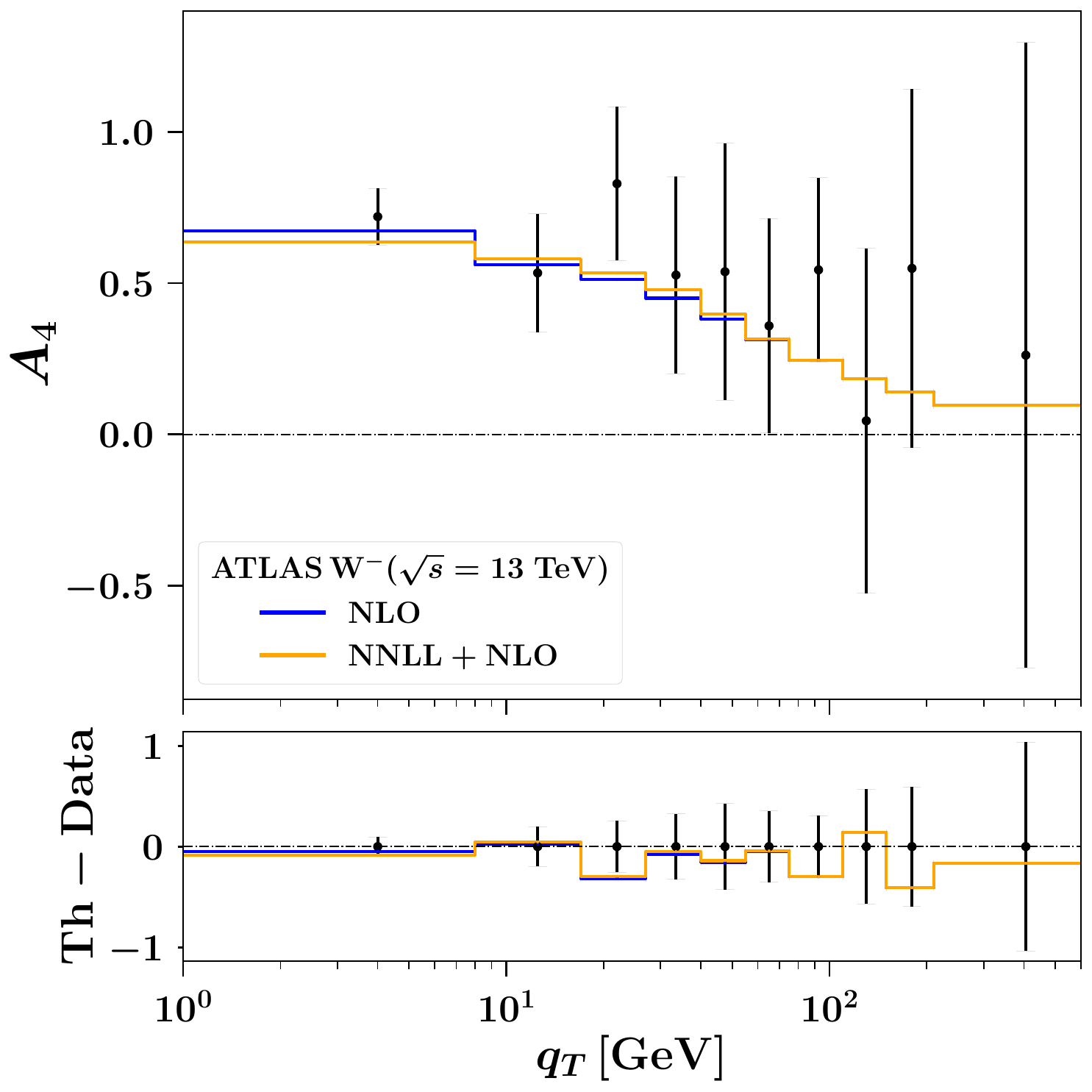}
\includegraphics[width=0.37\textwidth]{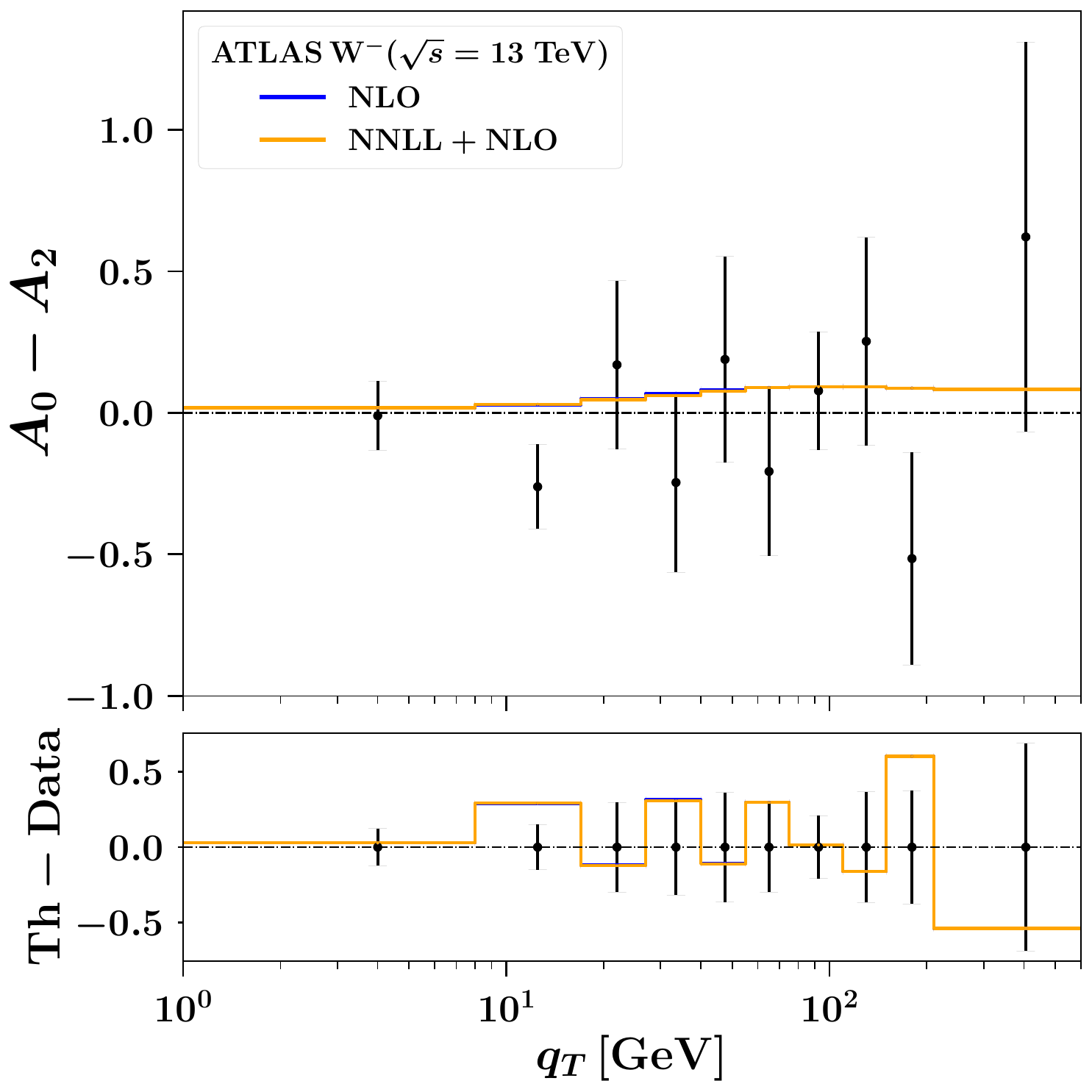}
 \caption{Comparison between experimental data and theoretical predictions at NLO (blue) and NNLL+NLO (orange) for the angular coefficients $A_0$ (upper left), $A_2$ (upper right), $A_3$ (central left), $A_4$ (central right), and the Lam--Tung combination $A_0-A_2$ (lower) in $W^-$ boson production at ATLAS, shown as functions of $q_T$. The lower panels display the differences between the predictions and the corresponding
experimental results.}
\label{f:A_ATLAS_Wm}
\end{figure}

\begin{figure}[h]
\centering
  \includegraphics[width=0.37\textwidth]{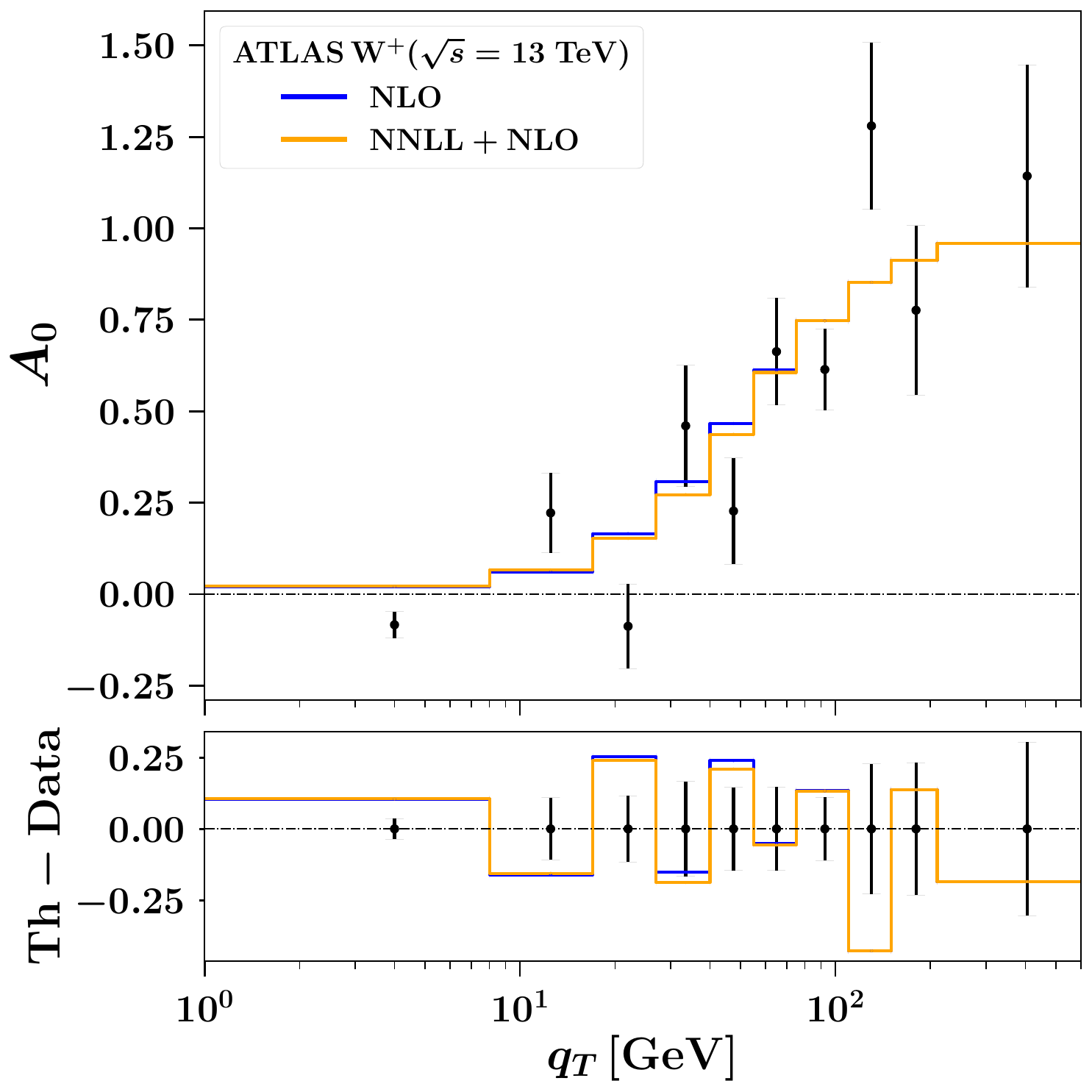}
\includegraphics[width=0.37\textwidth]{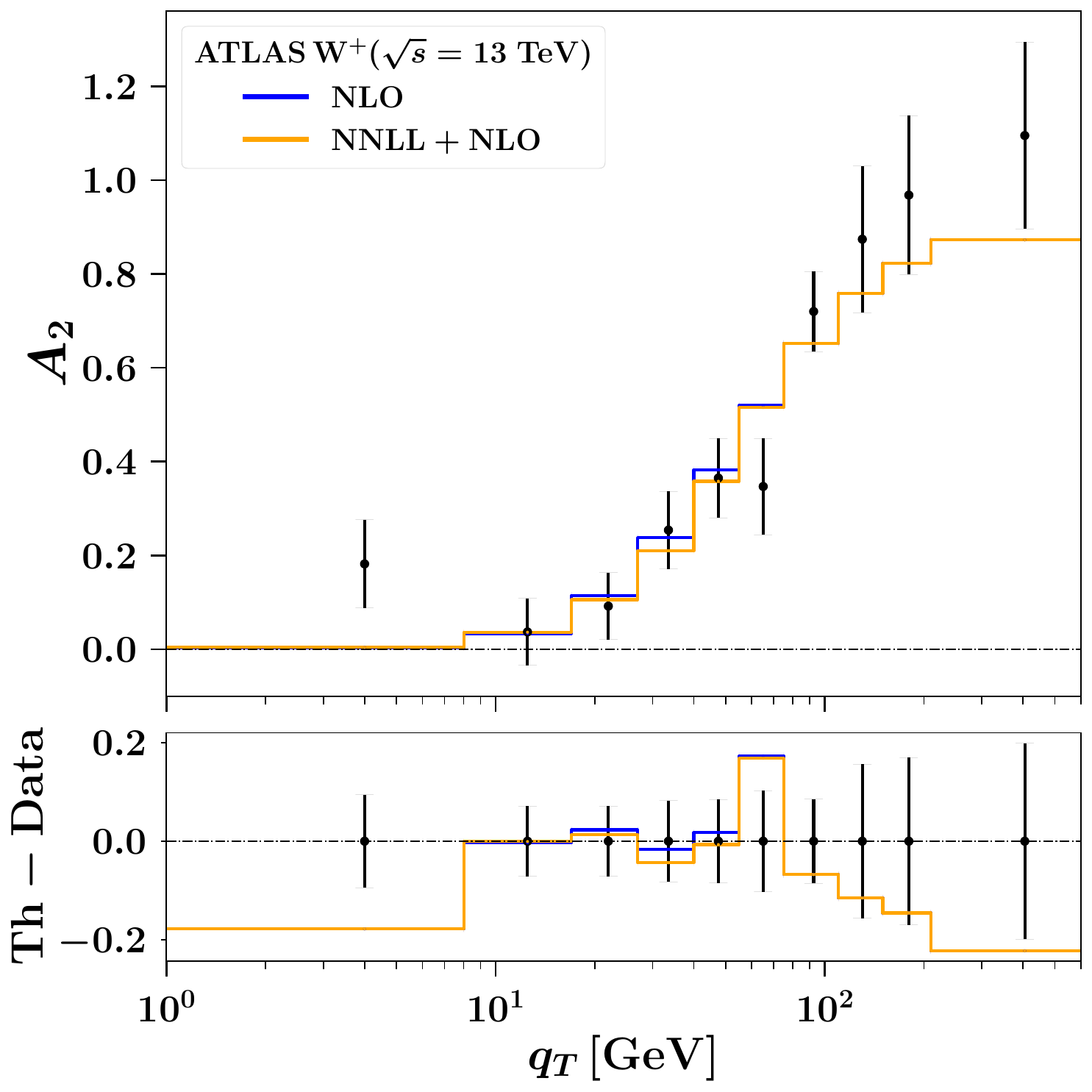}
\includegraphics[width=0.37\textwidth]{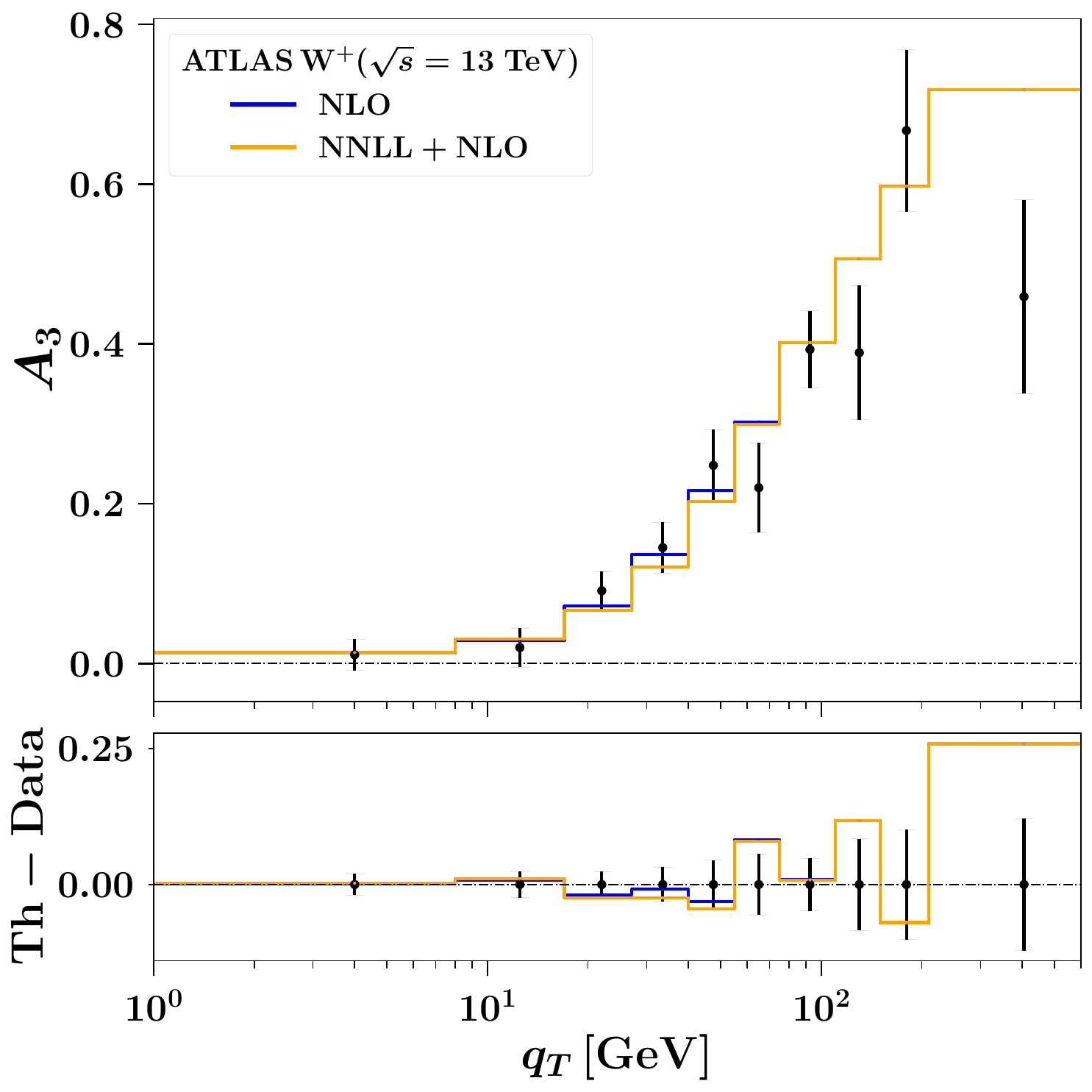}
\includegraphics[width=0.37\textwidth]{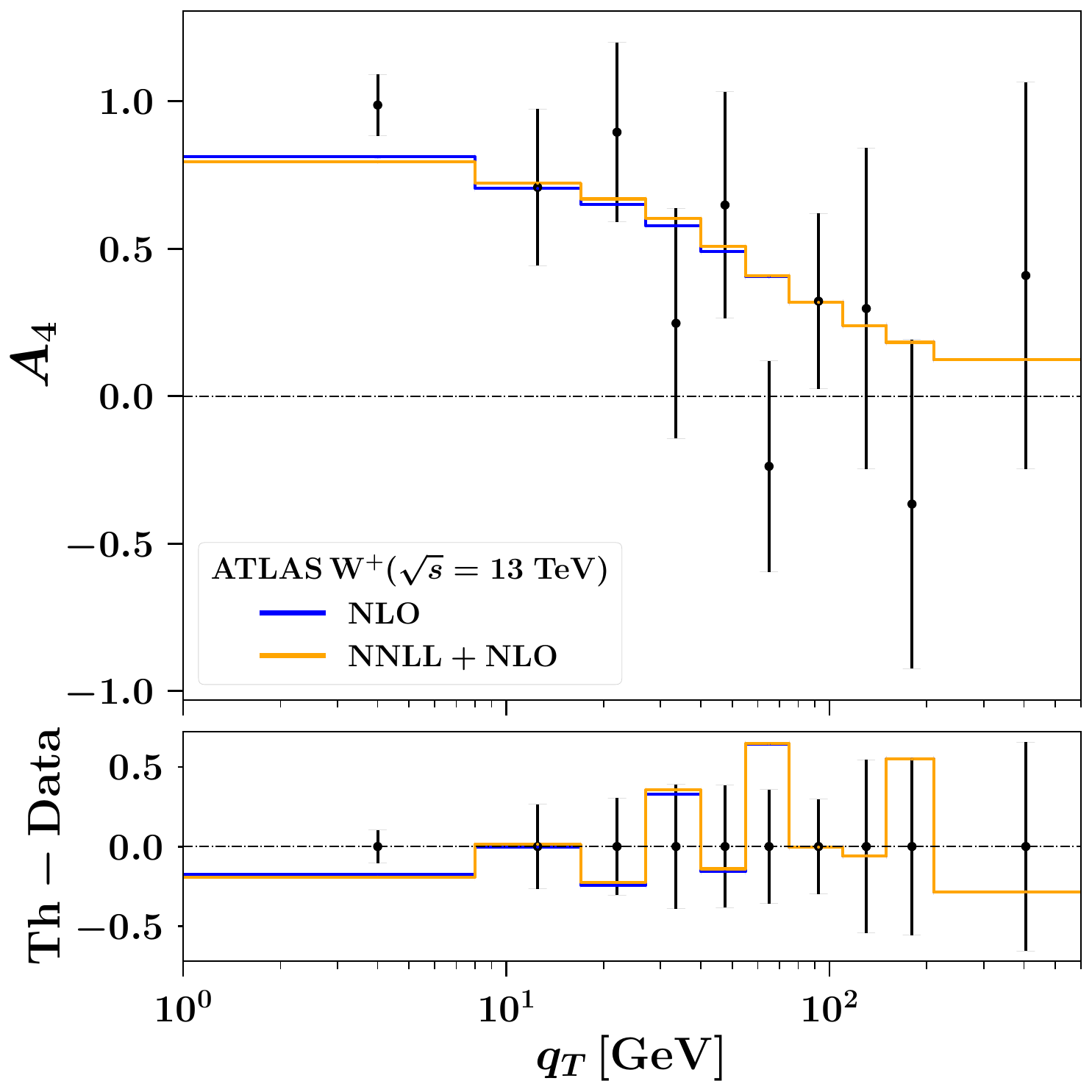}
\includegraphics[width=0.37\textwidth]{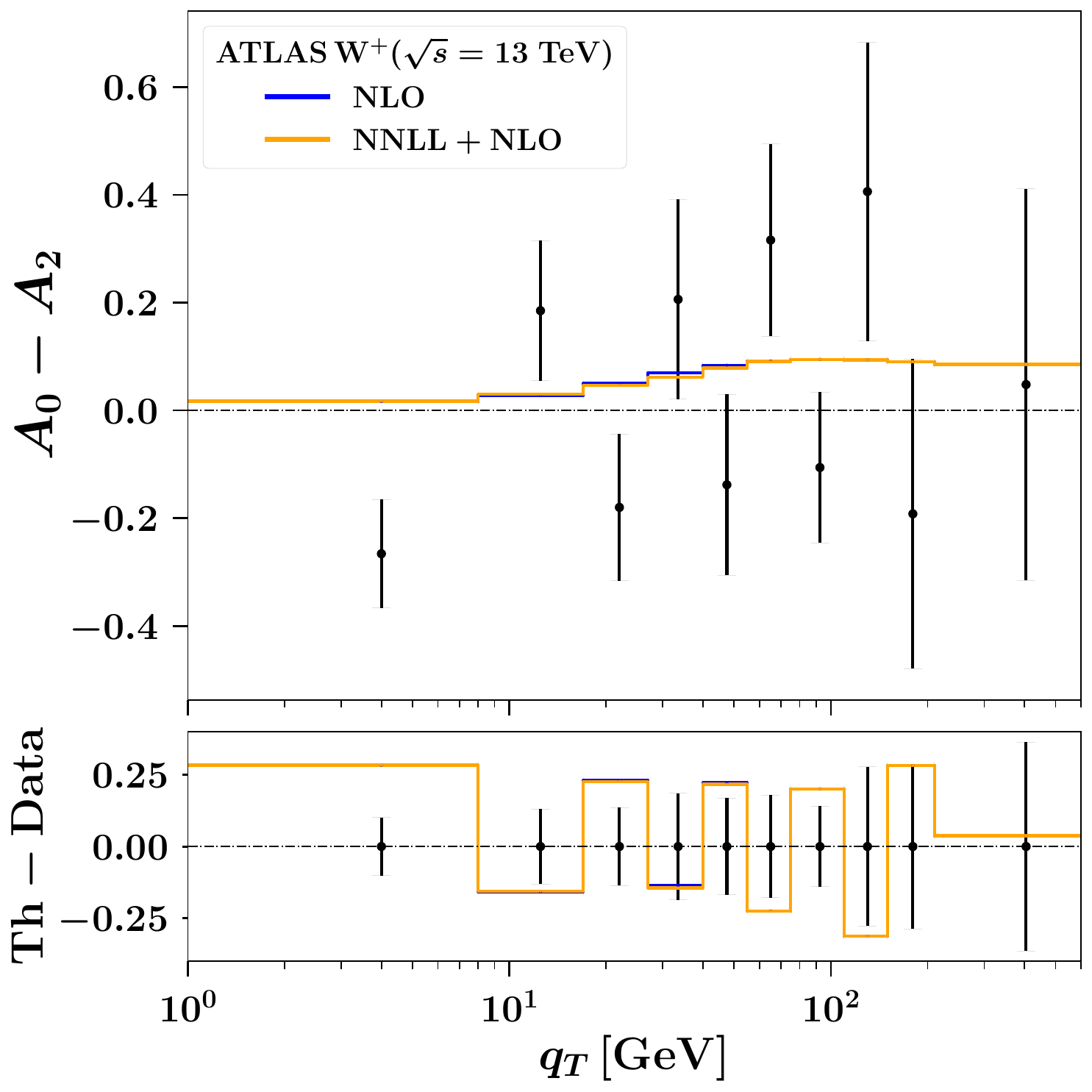}
 \caption{Comparison between experimental data and theoretical predictions at NLO (blue) and NNLL+NLO (orange) for the angular coefficients $A_0$ (upper left), $A_2$ (upper right), $A_3$ (central left), $A_4$ (central right), and the Lam--Tung combination $A_0-A_2$ (lower) in $W^+$ boson production at ATLAS, shown as functions of $q_T$. The lower panels display the differences between the predictions and the corresponding
experimental results.}
\label{f:A_ATLAS_Wp}
\end{figure}

Figures~\ref{f:A_ATLAS_Wm} and~\ref{f:A_ATLAS_Wp} show the $q_T$ distributions of the angular coefficients $A_0$, $A_2$, $A_3$, $A_4$, and the Lam--Tung combination $A_0-A_2$ for $W^-$ and $W^+$ production, respectively. The experimental measurements are compared to theoretical predictions obtained at NLO accuracy and at NNLL+NLO accuracy. The lower panels display the differences between the theoretical predictions and the corresponding experimental data.

As observed in the neutral-current case, the inclusion of resummed contributions modifies the theoretical predictions in the intermediate transverse-momentum region by a similar relative amount for both charge states of the $W$ boson: for $20<q_T<55$~GeV, $A_0$, $A_2$ and $A_3$ are lowered by about $8\%$ for both $W^+$ and $W^-$ (see Sec.~\ref{s:uncertainties}). A charge dependence appears in $A_4$, whose numerator receives a direct resummed contribution: for $8<q_T<55$~GeV the relative increase is $3.5$--$6\%$ for $W^-$ and $2.5$--$4.5\%$ for $W^+$.

In contrast to the case of $Z$-boson production, the resummation effects are small compared with the experimental uncertainties of the ATLAS $W$-boson measurement, which are larger than those of the $Z$-boson measurements and are given in wider $q_T$ bins: in the region $20<q_T<55$~GeV the shift induced by resummation amounts to $10$--$30\%$ of the experimental uncertainty (see Sec.~\ref{s:uncertainties}). For $A_4$, whose numerator receives a direct contribution from the resummed component and requires a separate resummation of numerator and denominator (see Sec.~\ref{sec:sudakov}), resummation increases the coefficient by $2.5$--$6\%$ ($3.5$--$6\%$ for $W^-$, $2.5$--$4.5\%$ for $W^+$) for $8<q_T<55$~GeV and lowers it by $0.04$ ($W^-$) and $0.02$ ($W^+$) in the first bin, where both predictions lie below the data; the shift is at most $0.04$, i.e.\ less than half of the experimental uncertainty in the first bin and less than $0.15$ of it elsewhere, well below the experimental uncertainties. For $A_3$, which is sensitive to hadronic linear power corrections and to the relative weight of the up- and down-type quark channels, a more refined theoretical treatment would be required.

To the best of our knowledge, this work is the first dedicated study of the impact of transverse-momentum
resummation on the angular coefficients of $W$-boson production. It is a
necessary step towards a complete theoretical description of
charged-current Drell--Yan processes at the level of angular observables,
and motivates further investigation on both the theoretical and
experimental sides.

\clearpage
 \section{Theoretical uncertainties and origin of the resummation effects}
\label{s:uncertainties}

In this section we verify the mechanism discussed in Sec.~\ref{sec:sudakov} and we quantify the theoretical uncertainties of the predictions presented in Sec.~\ref{s:results}. The analytic calculation allows us to carry out the large number of calculations required by the variations without statistical fluctuations. The settings and the experimental covariance matrices are those of Sec.~\ref{s:results}. For the $W$-boson measurements we also consider $A_4$, whose numerator receives a direct contribution from the resummed component; for the $Z$-boson measurements $A_4$ is not considered.

\subsection{Pointwise and bin-averaging effects}

In order to separate the two mechanisms discussed in Sec.~\ref{sec:sudakov}, we exploit the fine binning of the ATLAS measurement. For a set of wide bins, obtained by merging adjacent ATLAS bins, we compare the shift $\Delta A=A^{\rm NNLL+NLO}-A^{\rm NLO}$ of the bin-averaged coefficients with the shift obtained by weighting the NLO coefficients of the original bins with the NNLL+NLO unpolarised cross sections, which isolates the bin-averaging effect; the remainder is the pointwise contribution. The results, collected in Table~\ref{t:decomposition}, show that for $q_T>20$~GeV more than $95\%$ of the shift is pointwise, while the bin-averaging contribution is below $10^{-3}$. The bin-averaging effect is visible only below 20~GeV, where the shifts are small. Consistently, the size of the shift does not depend on the bin width: in the region $20<q_T<55$~GeV the relative shift of $A_0$ and $A_2$ is about $-9\%$ for CMS at 8~TeV and $-10\%$ for the more finely binned ATLAS measurement (Table~\ref{t:budget} and Fig.~\ref{f:shift}).

\begin{table}[h!]
\centering
\footnotesize
\renewcommand{\tabcolsep}{0.4pc}
\renewcommand{\arraystretch}{1.2}
\begin{tabular}{|c|c|c|c|c|c|c|}
\hline
 & \multicolumn{3}{c|}{$A_0$} & \multicolumn{3}{c|}{$A_2$} \\
\cline{2-7}
$q_T$ bin [GeV] & $\Delta A$ & bin averaging & pointwise & $\Delta A$ & bin averaging & pointwise \\
\hline\hline
[8, 22] & $+3.8$ & $+2.5$ & $+1.3$ & $+2.4$ & $+1.9$ & $+0.5$ \\ \hline
[22, 40.4] & $-22.1$ & $+0.7$ & $-22.8$ & $-17.3$ & $+0.6$ & $-17.9$ \\ \hline
[25.5, 44.9] & $-28.0$ & $+0.1$ & $-28.1$ & $-22.4$ & $+0.1$ & $-22.4$ \\ \hline
[40.4, 56.4] & $-31.8$ & $-0.8$ & $-31.0$ & $-26.7$ & $-0.8$ & $-25.9$ \\ \hline
\end{tabular}
\caption{Decomposition of the shift $\Delta A=A^{\rm NNLL+NLO}-A^{\rm NLO}$ of the bin-averaged coefficients $A_0$ and $A_2$, in units of $10^{-3}$, into a bin-averaging and a pointwise contribution, for wide bins obtained by merging adjacent bins of the ATLAS measurement ($|y|<1$). The bin-averaging contribution is obtained by weighting the NLO coefficients of the original bins with the NNLL+NLO unpolarised cross sections.}
\label{t:decomposition}
\end{table}

\subsection{Dependence on the recoil prescription}

We have computed the NNLL+NLO predictions for two alternative recoil prescriptions: the GJ prescription, in which the polar axis is chosen along the direction of the first incoming parton (from the beam moving along the positive $z$ axis, at all rapidities) and the transverse momentum is entirely assigned to the other parton, and the prescription in which the transverse momentum assigned to the first incoming parton is $\mathbf{k}_{T1}=[1+p_z/(M+E)]\,\mathbf{q}_T/2$, with $E$ and $p_z$ the energy and the longitudinal momentum of the vector boson (denoted as naive in the following), which coincides with the CS prescription at $y=0$ and departs from it with increasing $|y|$.

\begin{table}[h!]
\centering
\footnotesize
\renewcommand{\tabcolsep}{0.4pc}
\renewcommand{\arraystretch}{1.2}
\begin{tabular}{|c|c|c|c|c|c|c|c|}
\hline
 & & & \multicolumn{3}{c|}{NNLL+NLO} & \multicolumn{2}{c|}{shift reduction} \\
\cline{4-8}
Dataset & Coeff. & NLO & CS & GJ & naive & GJ & naive \\
\hline\hline
\multirow{2}{*}{CDF $Z$} & $A_0$ & 0.77 & 0.50 & 0.60 & 0.52 & 53\% & 13\% \\ \cline{2-8}
 & $A_2$ & 0.57 & 0.38 & 0.48 & 0.40 & 60\% & 15\% \\ \hline
\multirow{2}{*}{CMS $Z$ 8 TeV} & $A_0$ & 1.27 & 0.78 & 0.74 & 0.76 & 45\% & 4\% \\ \cline{2-8}
 & $A_2$ & 1.69 & 1.19 & 1.44 & 1.20 & 57\% & 5\% \\ \hline
\multirow{2}{*}{ATLAS $Z$} & $A_0$ & 2.13 & 2.80 & 1.60 & 2.65 & 44\% & 4\% \\ \cline{2-8}
 & $A_2$ & 5.07 & 3.04 & 3.97 & 3.10 & 56\% & 4\% \\ \hline
\multirow{2}{*}{LHCb $Z$} & $A_0$ & 13.33 (1.88) & 14.04 (2.18) & 13.24 (1.58) & 13.38 (1.64) & 47\% & 38\% \\ \cline{2-8}
 & $A_2$ & 3.34 (4.56) & 2.01 (2.59) & 2.57 (3.48) & 2.41 (3.24) & 61\% & 49\% \\ \hline
\multirow{2}{*}{\makecell{CMS $Z$ 13 TeV\\ $|y|<1$}} & $A_0$ & 3.20 & 6.12 & 3.83 & 5.86 & 45\% & 4\% \\ \cline{2-8}
 & $A_2$ & 12.96 & 9.12 & 10.79 & 9.23 & 59\% & 5\% \\ \hline
\multirow{2}{*}{\makecell{CMS $Z$ 13 TeV\\ $1<|y|<2.4$}} & $A_0$ & 3.17 & 4.69 & 3.08 & 3.69 & 46\% & 22\% \\ \cline{2-8}
 & $A_2$ & 8.86 & 5.43 & 6.88 & 5.95 & 60\% & 29\% \\ \hline
\multirow{2}{*}{CDF $W^-$} & $A_2$ & 0.47 & 0.52 & 0.49 & 0.51 & 60\% & 16\% \\ \cline{2-8}
 & $A_3$ & 0.41 & 0.50 & 0.50 & 0.34 & 1200\% & 197\% \\ \hline
\multirow{3}{*}{ATLAS $W^+$} & $A_0$ & 1.66 & 1.51 & 1.55 & 1.54 & 46\% & 27\% \\ \cline{2-8}
 & $A_2$ & 1.46 & 1.38 & 1.42 & 1.39 & 62\% & 36\% \\ \cline{2-8}
 & $A_3$ & 1.16 & 1.45 & 1.45 & 1.13 & 0\% & 173\% \\ \hline
\multirow{3}{*}{ATLAS $W^-$} & $A_0$ & 1.07 & 1.06 & 1.07 & 1.07 & 45\% & 23\% \\ \cline{2-8}
 & $A_2$ & 2.41 & 2.39 & 2.40 & 2.40 & 61\% & 30\% \\ \cline{2-8}
 & $A_3$ & 0.74 & 0.87 & 0.87 & 0.93 & 0\% & 161\% \\ \hline
\end{tabular}
\caption{Values of $\chi^2/N_{d.o.f.}$ for the NLO predictions and for the NNLL+NLO predictions with the Collins--Soper (CS), Gottfried--Jackson (GJ) and naive recoil prescriptions, without theoretical uncertainties, and median reduction of the shift $A^{\rm NNLL+NLO}-A^{\rm NLO}$ with respect to the CS prescription in the bins with $20<q_T<55$~GeV. For LHCb, the values in brackets are obtained without the first two bins. The combination $A_0-A_2$ is the same in the three prescriptions. For $A_3$, a reduction above $100\%$ means that the recoil contribution reverses the sign of the shift.}
\label{t:recoil}
\end{table}

Since $d\sigma^{(U+L)}$ does not depend on the prescription, we obtain the helicity cross sections from Eq.~\eqref{eq:rotation} as $d\sigma^{(0,2)}=d\sigma^{(0,2)}_{\rm NLO}+\sin^2\beta\,(d\sigma^{(U+L)}_{\rm NNLL+NLO}-d\sigma^{(U+L)}_{\rm NLO})$, evaluated in cells of $1$--$2$~GeV in $q_T$ and $0.25$--$0.5$ in rapidity, with $\sin^2\beta$ averaged over each cell, and summed over the cells of each bin of the measurements. Summing the cells reproduces the bin-averaged predictions of the CS prescription within $0.2\%$. In the alternative prescriptions $A_3$ also receives a contribution from the resummed $d\sigma^{(4)}$: according to Eq.~\eqref{eq:rotation}, the resummed correction $\Delta\sigma^{(4)}=d\sigma^{(4)}_{\rm NNLL+NLO}-d\sigma^{(4)}_{\rm NLO}$ enters $d\sigma^{(4)}$ multiplied by $\cos\beta$ and $d\sigma^{(3)}$ multiplied by $\sin\beta$, whose sign depends on the orientation of the rotation with respect to the polar axis of the Collins--Soper frame. In $pp$ collisions, where this axis follows the direction of the boson, the sign changes with $y$ in the GJ prescription and not in the naive one, so that in the GJ prescription the contribution to $A_3$ cancels in measurements integrated symmetrically over rapidity; in $p\bar p$ collisions, where the polar axis is always along the proton, the sign changes with $y$ in the naive prescription and not in the GJ one. We have checked these contributions against a direct computation of the lepton angular distributions.

The results are reported in Table~\ref{t:recoil}. In the GJ prescription the shift is reduced by $44$--$62\%$; the reduction is larger for $A_2$ than for $A_0$, since the additional term is the same for both coefficients and $A_2<A_0$, and the shift does not vanish because the quark--gluon channel yields values of $A_0$ well above $\sin^2\beta$. In the naive prescription the reduction increases with the rapidity: it amounts to $4$--$5\%$ for CMS and ATLAS ($|y|<1$), $13$--$15\%$ for CDF ($16\%$ for $A_2$ of the $W^-$), $22$--$29\%$ for CMS at 13~TeV with $1<|y|<2.4$, $23$--$36\%$ for the ATLAS $W$-boson measurements, integrated over rapidity, and $38$--$49\%$ for LHCb ($y>2$). In all the prescriptions, resummation improves the description of $A_0$ and $A_2$ for CDF and CMS at 8~TeV and of $A_2$ for ATLAS and LHCb, although in the GJ prescription the improvement of $A_2$ is roughly halved. For $A_0$ at ATLAS and at LHCb without the first two bins, instead, the $\chi^2$ increases with the CS prescription and decreases with the GJ one. For CMS at 13~TeV, resummation improves the description of $A_2$ and worsens that of $A_0$ in all the prescriptions, except for $A_0$ at $1<|y|<2.4$ in the GJ prescription, for which the $\chi^2$ slightly decreases. For $A_3$ of the ATLAS $W$-boson measurements the GJ contribution cancels, while in the naive prescription the contribution of $d\sigma^{(4)}$ has the opposite sign of the resummation shift and is larger than it. For $A_3$ of the CDF $W^-$ measurement the GJ contribution does not cancel and, since $A_4$ of the $W^-$ is large in $p\bar p$ collisions, is much larger than the resummation shift, while the naive one partially cancels between positive and negative rapidities. The corresponding values of $\chi^2/N_{d.o.f.}$ are reported in Table~\ref{t:recoil}.

\subsection{Theoretical uncertainties}

We consider the following sources of theoretical uncertainty.
\begin{itemize}
\item \emph{Scales}: the renormalisation and factorisation scales are varied by a factor of two around $M$ with $1/2\le\mu_R/\mu_F\le2$ (seven-point variation), for both the NLO and the NNLL+NLO predictions, and the resummation scale is varied to $Q=M/2$ and $Q=2M$ for the NNLL+NLO predictions. The same scales are used in the numerator and denominator of the coefficients, and the uncertainty is given by the envelope of the variations.
\item \emph{PDFs}: we use the 100 replicas of the NNPDF4.0 set\,\cite{NNPDF:2021njg}, with the same replica in the numerator and in the denominator.
\item \emph{Non-perturbative effects}: the parameters $g_1$, $\lambda$, $q$ and $g_0$ are varied within one standard deviation of the determination of Ref.\,\cite{Camarda:2025lbt}, and the uncertainty is given by the envelope of the 16 combinations of their extreme values. This is a conservative estimate, since it neglects the correlations among the parameters.
\item \emph{Transition function}: we vary $k$ to $0.25$ and $0.5$ and $\delta$ to $0.15$ and $0.35$, and we replace the Gaussian form with $w(q_T)=\{1+\cos[\pi(q_T-kM)/(\delta M)]\}/2$ for $kM<q_T<(k+\delta)M$ and $w(q_T)=0$ above.
\item \emph{Recoil prescription}: the envelope of the CS, GJ and naive prescriptions.
\end{itemize}

\begin{table}[h!]
\centering
\scriptsize
\renewcommand{\tabcolsep}{0.3pc}
\renewcommand{\arraystretch}{1.2}
\begin{tabular}{|c|c|c|c|c|c|c|c|c|c|}
\hline
 & & & & \multicolumn{6}{c|}{uncertainty [\% of $|\Delta A|$]} \\
\cline{5-10}
Dataset & Coeff. & \makecell{$\Delta A/A$\\ $[\%]$} & $\dfrac{|\Delta A|}{\sigma_{\rm exp}}$ & \makecell{scales\\ (NLO)} & \makecell{scales\\ (NNLL)} & PDF & NP & \makecell{transi-\\tion} & recoil \\
\hline\hline
\multirow{2}{*}{CDF $Z$} & $A_0$ & $-14$ & 0.6 & 3 & 57 & 1 & 0.3 & 16 & 26 \\ \cline{2-10}
 & $A_2$ & $-14$ & 0.4 & 18 & 38 & 1 & 0.3 & 16 & 30 \\ \hline
\multirow{2}{*}{CMS $Z$ 8 TeV} & $A_0$ & $-9$ & 1.8 & 3 & 51 & 0.6 & 0.3 & 20 & 23 \\ \cline{2-10}
 & $A_2$ & $-9$ & 1.0 & 46 & 17 & 0.7 & 0.3 & 20 & 28 \\ \hline
\multirow{2}{*}{ATLAS $Z$} & $A_0$ & $-10$ & 1.4 & 1.0 & 47 & 0.5 & 0.2 & 11 & 22 \\ \cline{2-10}
 & $A_2$ & $-10$ & 1.0 & 43 & 16 & 0.6 & 0.2 & 11 & 28 \\ \hline
\multirow{2}{*}{LHCb $Z$} & $A_0$ & $-12$ & 1.7 & 4 & 39 & 1 & 0.2 & 20 & 23 \\ \cline{2-10}
 & $A_2$ & $-12$ & 1.6 & 42 & 13 & 1 & 0.2 & 20 & 30 \\ \hline
\multirow{2}{*}{\makecell{CMS $Z$ 13 TeV\\ $|y|<1$}} & $A_0$ & $-7$ & 2.7 & 5 & 68 & 0.9 & 0.4 & 22 & 23 \\ \cline{2-10}
 & $A_2$ & $-7$ & 1.9 & 74 & 33 & 1.0 & 0.4 & 22 & 30 \\ \hline
\multirow{2}{*}{\makecell{CMS $Z$ 13 TeV\\ $1<|y|<2.4$}} & $A_0$ & $-8$ & 3.0 & 4 & 57 & 0.8 & 0.3 & 21 & 23 \\ \cline{2-10}
 & $A_2$ & $-8$ & 2.7 & 63 & 22 & 0.8 & 0.3 & 21 & 30 \\ \hline
\multirow{2}{*}{CDF $W^-$} & $A_2$ & $-17$ & 0.1 & 19 & 35 & 1.1 & 0.3 & 9 & 30 \\ \cline{2-10}
 & $A_3$ & $-17$ & 0.1 & 7 & 51 & 4 & 0.3 & 9 & 600 \\ \hline
\multirow{4}{*}{ATLAS $W^+$} & $A_0$ & $-8$ & 0.2 & 3 & 40 & 0.7 & 0.2 & 18 & 23 \\ \cline{2-10}
 & $A_2$ & $-8$ & 0.3 & 56 & 17 & 0.7 & 0.2 & 18 & 31 \\ \cline{2-10}
 & $A_3$ & $-8$ & 0.3 & 17 & 57 & 3 & 0.2 & 18 & 87 \\ \cline{2-10}
 & $A_4$ & $+3.7$ & 0.06 & 12 & 27 & 9 & 0.2 & 9 & 12 \\ \hline
\multirow{4}{*}{ATLAS $W^-$} & $A_0$ & $-8$ & 0.1 & 4 & 41 & 0.6 & 0.2 & 18 & 23 \\ \cline{2-10}
 & $A_2$ & $-8$ & 0.2 & 55 & 17 & 0.6 & 0.2 & 18 & 30 \\ \cline{2-10}
 & $A_3$ & $-8$ & 0.2 & 33 & 72 & 3 & 0.2 & 18 & 81 \\ \cline{2-10}
 & $A_4$ & $+4.7$ & 0.08 & 22 & 24 & 7 & 0.1 & 17 & 11 \\ \hline
\end{tabular}
\caption{Size of the resummation effect and of the theoretical uncertainties in the bins with $20\le q_T\le55$~GeV (median over the bins). $\Delta A/A$: relative shift $\Delta A=A^{\rm NNLL+NLO}-A^{\rm NLO}$ with the CS prescription; $|\Delta A|/\sigma_{\rm exp}$: shift in units of the experimental uncertainty. The other columns give the half-widths of the uncertainty envelopes in percent of $|\Delta A|$: scale variations of the NLO and of the NNLL+NLO predictions, PDF uncertainty of the shift, non-perturbative parameters (NP), transition function and recoil prescription. For $A_3$ of the ATLAS $W$-boson measurements the recoil uncertainty is given by the naive prescription, since the GJ contribution cancels after integration over rapidity; for CDF $W^-$ it is dominated by the GJ prescription (see the text).}
\label{t:budget}
\end{table}

We have also varied the treatment of the large-$b$ region, the
$b_\star$ prescription and the functional form of $\widetilde L$.
The large-$b$ variations mainly affect $q_T\lesssim10$~GeV and change
the reduced $\chi^2$ by at most $0.05$. Variations of the logarithmic
prescription act predominantly in the transition region and can change
the reduced $\chi^2$ by up to about one unit for $A_0$ at ATLAS, although
they do not change the direction of the resummation shift. Since the
non-perturbative parameters of Ref.\,\cite{Camarda:2025lbt} were
determined with the default choices, a consistent variation of the
large-$b$ treatment and of the logarithmic prescription would require a
new determination of these parameters. We therefore regard
this dependence as an additional resummation-scheme uncertainty, which is not
included in Table~\ref{t:budget} and in $V^{\rm th}$.

\begin{figure}[h!]
\centering
\includegraphics[width=0.9\textwidth]{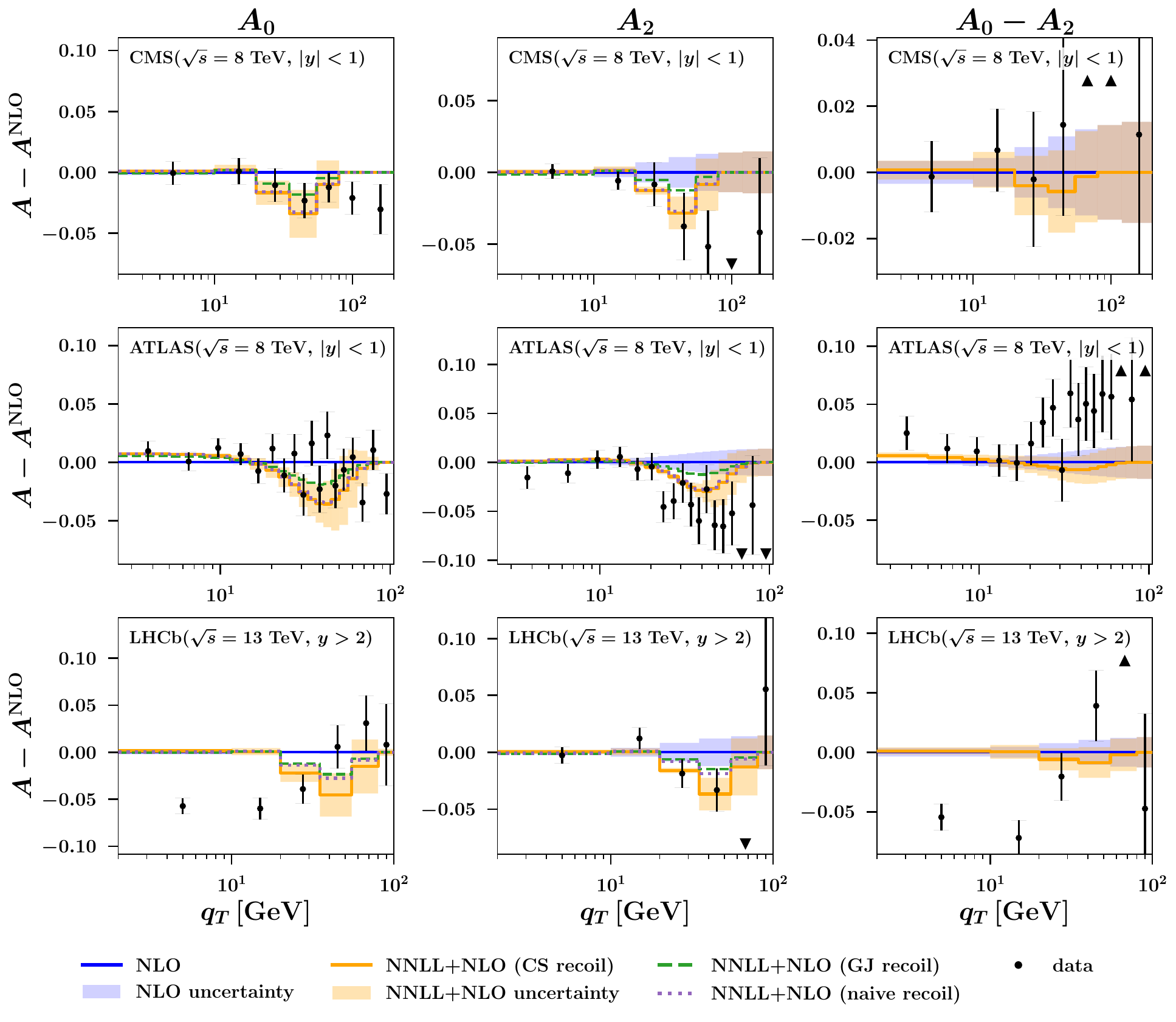}
\caption{Difference between the theoretical predictions and the NLO prediction, $A-A^{\rm NLO}$, for $A_0$ (left), $A_2$ (centre) and $A_0-A_2$ (right) in $Z$-boson production at CMS ($\sqrt{s}=8$~TeV, $|y|<1$, upper), ATLAS ($|y|<1$, middle) and LHCb ($y>2$, lower). Solid orange line: NNLL+NLO with the CS recoil prescription, with its theoretical uncertainty (scales, PDFs, non-perturbative parameters and transition function, added in quadrature); dashed green and dotted purple lines: NNLL+NLO with the GJ and naive recoil prescriptions, which coincide with the CS one for $A_0-A_2$; blue line and band: NLO prediction and its uncertainty (scales and PDFs). The points show the difference between the data and the NLO prediction; data points outside the vertical range are indicated by triangles at the edge of the panel.}
\label{f:shift}
\end{figure}

\begin{figure}[h!]
\centering
\includegraphics[width=\textwidth]{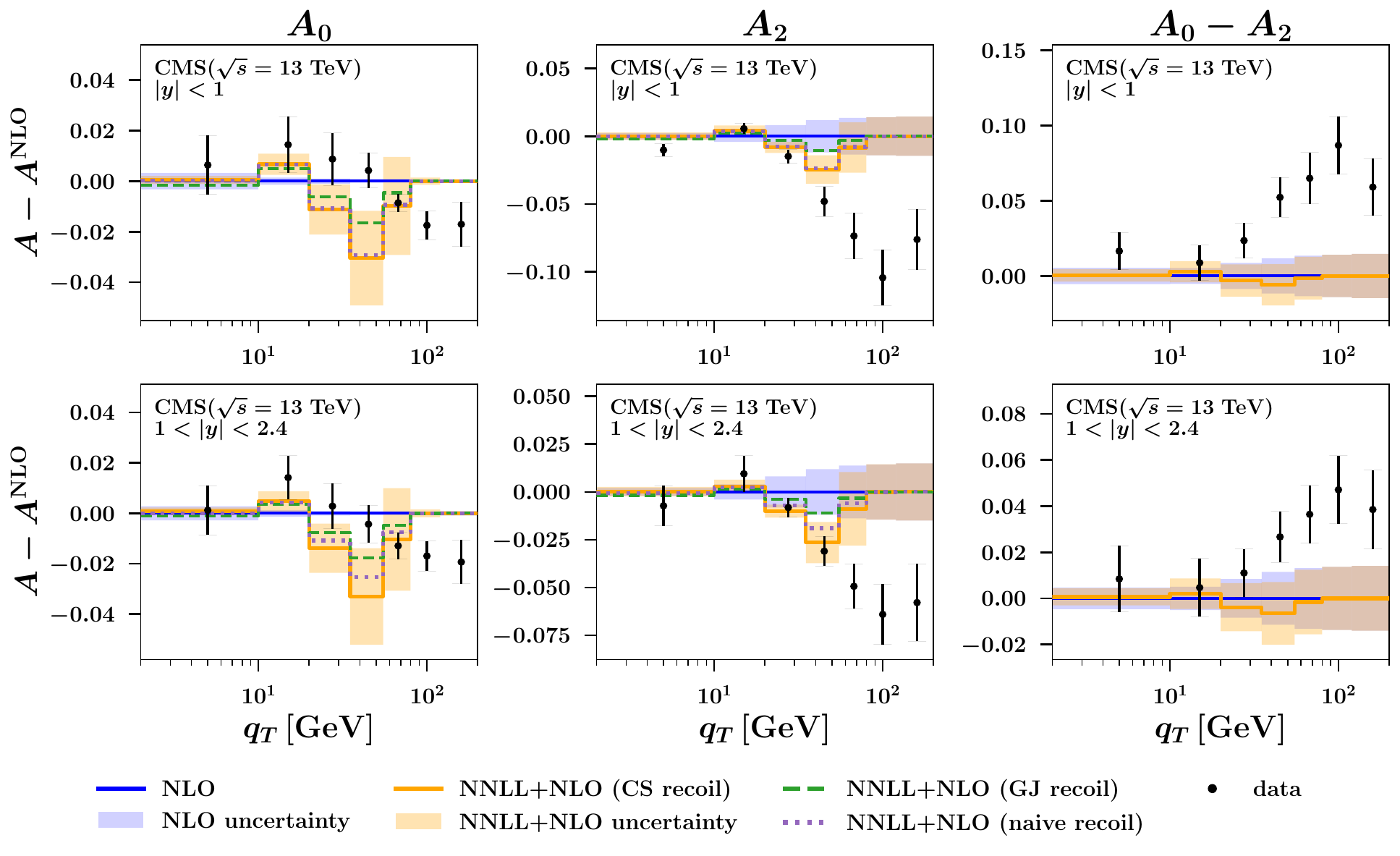}
\caption{As in Fig.~\ref{f:shift}, for the CMS measurement at $\sqrt{s}=13$~TeV with $|y|<1$ (upper) and $1<|y|<2.4$ (lower).}
\label{f:shift13}
\end{figure}

The size of the resummation effect and of the uncertainties in the intermediate region is summarised in Table~\ref{t:budget}, and the predictions for the CMS, ATLAS and LHCb measurements are shown in Fig.~\ref{f:shift} (and for the CMS measurement at 13~TeV in Fig.~\ref{f:shift13}) as differences with respect to the NLO prediction. In the region $20\le q_T\le55$~GeV resummation lowers $A_0$, $A_2$ and $A_3$ by $7$--$17\%$ (and raises $A_4$ of the $W$ bosons by $4$--$5\%$), which corresponds to $1$--$2$ experimental standard deviations for CMS at 8~TeV, ATLAS and LHCb, to $2$--$3$ standard deviations for CMS at 13~TeV, to about half a standard deviation for CDF and to $10$--$30\%$ of a standard deviation for the $W$-boson measurements. The PDF and non-perturbative uncertainties of the shift are negligible, about $1\%$ (up to $3\%$ for $A_3$, and $7$--$9\%$ for $A_4$ of the $W$ bosons, whose shift is small) and below $0.5\%$ of the shift, respectively: the PDF dependence largely cancels in the ratios, and the non-perturbative effects are concentrated at $q_T\lesssim10$~GeV. The dominant uncertainties are those from the scales, the transition function and the recoil prescription. For $A_0$ and $A_2$, they typically amount to between about $10\%$ and $75\%$ of the shift. For $A_3$, the relative recoil uncertainty can be much larger because the Collins--Soper shift is small and can change sign. The scale uncertainty of the NNLL+NLO prediction for $A_0$ is particularly large, $39$--$68\%$ of the shift: the fixed-order numerator is proportional to $\alpha_S(\mu_R^2)$, while the resummed denominator is much less sensitive to $\mu_R$, and the dependence does not cancel in the ratio. At NLO, instead, the scale uncertainty of $A_0$ is a few percent of the shift, while that of $A_2$ is about half of it at the LHC.

\clearpage

\subsection{Impact on the comparison with data}
\label{s:impact}

We include the theoretical uncertainties in the $\chi^2$ by adding to the covariance matrix the theoretical covariance
\begin{equation}
V^{\rm th}
=
{\rm diag}\left(
s_{\rm scales}^2+s_{\rm NP}^2+s_{\rm tr}^2+s_{\rm rec}^2
\right)
+C_{\rm PDF}\,,
\label{e:Vth}
\end{equation}
where the $s$ denote the half-widths of the scale, non-perturbative,
transition-function and recoil envelopes, while $C_{\rm PDF}$ is the
covariance over the PDF replicas. For the NLO predictions,
$s_{\rm NP}=s_{\rm tr}=s_{\rm rec}=0$; we also set $s_{\rm rec}=0$
when quoting NNLL+NLO results without the recoil uncertainty. We treat
the envelope uncertainties as uncorrelated between bins. Assuming
instead full bin-to-bin correlation for the transition and recoil
variations changes the reduced $\chi^2$ by at most $0.35$ and does not
affect our conclusions.

The results are reported in Table~\ref{t:chi2th}. After including the
theoretical uncertainties, NNLL+NLO gives a lower $\chi^2$ than NLO for
$A_0$ and $A_2$ in all the $Z$-boson datasets and recoil prescriptions.
For $A_0$ at ATLAS, CMS at 13~TeV and LHCb, however, this improvement is due to the
larger NNLL+NLO uncertainty rather than to the central prediction. The
description of $A_0-A_2$ slightly worsens, while the changes for the
$W$-boson measurements are small.

\begin{table}[h!]
\centering
\scriptsize
\renewcommand{\tabcolsep}{0.18pc}
\renewcommand{\arraystretch}{1.2}
\begin{tabular}{|c|c|c|c|c|c|c|c|c|c|}
\hline
 & & & \multicolumn{2}{c|}{NLO} & \multicolumn{5}{c|}{NNLL+NLO} \\
\cline{4-10}
Dataset & Coeff. & $N_{\rm dat}$ & & $+V^{\rm th}$ & CS & \makecell{CS\\ $+V^{\rm th}$} & \makecell{CS $+V^{\rm th}$\\ incl. recoil} & \makecell{GJ\\ $+V^{\rm th}$} & \makecell{naive\\ $+V^{\rm th}$} \\
\hline\hline
\multirow{3}{*}{CDF $Z$} & $A_0$ & 5 & 0.77 & 0.77 & 0.50 & 0.48 & 0.48 & 0.58 & 0.50 \\ \cline{2-10}
 & $A_2$ & 5 & 0.57 & 0.56 & 0.38 & 0.37 & 0.37 & 0.48 & 0.40 \\ \cline{2-10}
 & $A_0-A_2$ & 5 & 0.13 & 0.13 & 0.14 & 0.14 & 0.14 & 0.14 & 0.14 \\ \hline
\multirow{3}{*}{CMS $Z$ 8 TeV} & $A_0$ & 7 & 1.27 & 1.26 & 0.78 & 0.71 & 0.71 & 0.68 & 0.70 \\ \cline{2-10}
 & $A_2$ & 7 & 1.69 & 1.41 & 1.19 & 0.95 & 0.95 & 1.14 & 0.96 \\ \cline{2-10}
 & $A_0-A_2$ & 7 & 0.74 & 0.61 & 0.82 & 0.66 & 0.66 & 0.66 & 0.66 \\ \hline
\multirow{3}{*}{ATLAS $Z$} & $A_0$ & 18 & 2.13 & 2.06 & 2.80 & 1.48 & 1.39 & 1.02 & 1.43 \\ \cline{2-10}
 & $A_2$ & 18 & 5.07 & 4.23 & 3.04 & 2.64 & 2.58 & 3.39 & 2.68 \\ \cline{2-10}
 & $A_0-A_2$ & 18 & 2.71 & 2.27 & 3.06 & 2.41 & 2.41 & 2.41 & 2.41 \\ \hline
\multirow{3}{*}{LHCb $Z$} & $A_0$ & 6 (4) & 13.33 (1.88) & 12.83 (1.87) & 14.04 (2.18) & 12.75 (1.16) & 12.62 (1.08) & 12.22 (0.97) & 12.32 (0.97) \\ \cline{2-10}
 & $A_2$ & 6 (4) & 3.34 (4.56) & 2.70 (3.66) & 2.01 (2.59) & 1.37 (1.66) & 1.36 (1.64) & 1.76 (2.30) & 1.65 (2.13) \\ \cline{2-10}
 & $A_0-A_2$ & 6 (4) & 10.25 (3.50) & 9.36 (3.22) & 10.56 (3.67) & 9.50 (3.29) & 9.50 (3.29) & 9.50 (3.29) & 9.50 (3.29) \\ \hline
\multirow{3}{*}{\makecell{CMS $Z$ 13 TeV\\ $|y|<1$}} & $A_0$ & 7 & 3.20 & 3.17 & 6.12 & 2.63 & 2.58 & 2.30 & 2.60 \\ \cline{2-10}
 & $A_2$ & 7 & 12.96 & 7.61 & 9.12 & 5.68 & 5.57 & 6.51 & 5.73 \\ \cline{2-10}
 & $A_0-A_2$ & 7 & 9.56 & 5.97 & 10.30 & 6.05 & 6.05 & 6.05 & 6.05 \\ \hline
\multirow{3}{*}{\makecell{CMS $Z$ 13 TeV\\ $1<|y|<2.4$}} & $A_0$ & 7 & 3.17 & 3.11 & 4.69 & 2.44 & 2.39 & 2.15 & 2.26 \\ \cline{2-10}
 & $A_2$ & 7 & 8.86 & 4.06 & 5.43 & 2.66 & 2.64 & 3.15 & 2.81 \\ \cline{2-10}
 & $A_0-A_2$ & 7 & 4.44 & 2.35 & 5.11 & 2.50 & 2.50 & 2.50 & 2.50 \\ \hline
\multirow{2}{*}{CDF $W^-$} & $A_2$ & 4 & 0.47 & 0.47 & 0.52 & 0.52 & 0.52 & 0.49 & 0.51 \\ \cline{2-10}
 & $A_3$ & 4 & 0.41 & 0.40 & 0.50 & 0.49 & 0.40 & 0.50 & 0.34 \\ \hline
\multirow{5}{*}{ATLAS $W^+$} & $A_0$ & 10 & 1.66 & 1.65 & 1.51 & 1.50 & 1.50 & 1.53 & 1.52 \\ \cline{2-10}
 & $A_2$ & 10 & 1.46 & 1.42 & 1.38 & 1.35 & 1.35 & 1.37 & 1.35 \\ \cline{2-10}
 & $A_3$ & 10 & 1.16 & 1.10 & 1.45 & 1.35 & 1.25 & 1.35 & 1.05 \\ \cline{2-10}
 & $A_4$ & 10 & 1.37 & 1.21 & 1.48 & 1.30 & 1.30 & 1.28 & 1.29 \\ \cline{2-10}
 & $A_0-A_2$ & 10 & 1.32 & 1.32 & 1.30 & 1.30 & 1.30 & 1.30 & 1.30 \\ \hline
\multirow{5}{*}{ATLAS $W^-$} & $A_0$ & 10 & 1.07 & 1.07 & 1.06 & 1.06 & 1.06 & 1.06 & 1.06 \\ \cline{2-10}
 & $A_2$ & 10 & 2.41 & 2.40 & 2.39 & 2.38 & 2.38 & 2.39 & 2.38 \\ \cline{2-10}
 & $A_3$ & 10 & 0.74 & 0.72 & 0.87 & 0.84 & 0.82 & 0.84 & 0.91 \\ \cline{2-10}
 & $A_4$ & 10 & 0.53 & 0.51 & 0.47 & 0.45 & 0.45 & 0.46 & 0.46 \\ \cline{2-10}
 & $A_0-A_2$ & 10 & 0.85 & 0.85 & 0.86 & 0.86 & 0.86 & 0.86 & 0.86 \\ \hline
\end{tabular}
\caption{Values of $\chi^2/N_{d.o.f.}$ without and with the theoretical covariance of Eq.~\eqref{e:Vth}, for the NLO and NNLL+NLO predictions. For the ATLAS $W$-boson measurements the experimental covariance includes the statistical correlations between the bins. ``CS $+V^{\rm th}$ incl. recoil'': CS prescription, including the recoil uncertainty in $V^{\rm th}$; ``GJ $+V^{\rm th}$'', ``naive $+V^{\rm th}$'': predictions in the corresponding prescription, with $V^{\rm th}$ without the recoil uncertainty. For LHCb, the values in brackets are obtained without the first two bins.}
\label{t:chi2th}
\end{table}

A $\chi^2$ that includes the theoretical uncertainties measures the compatibility of the predictions with the data, but it does not establish whether resummation improves their description. To address this question, we compare the $\chi^2/N_{d.o.f.}$ of the NNLL+NLO and NLO predictions for each variation separately: at the same values of $\mu_R$ and $\mu_F$ for the scale variations, and with respect to the central NLO prediction for the other variations, which do not affect the NLO calculation. The results are reported in Table~\ref{t:robust}, where the outcome is defined as ``improves'' (``worsens'') if the difference $\Delta\chi^2/N_{d.o.f.}$ is negative (positive) for the central predictions and for all the variations, ``unchanged'' if $|\Delta\chi^2/N_{d.o.f.}|<0.05$ for all of them, and ``undetermined'' otherwise.

\begin{table}[h!]
\centering
\scriptsize
\renewcommand{\tabcolsep}{0.3pc}
\renewcommand{\arraystretch}{1.2}
\begin{tabular}{|c|c|c|c|c|c|c|}
\hline
Dataset & Coeff. & central & scales & transition & recoil & outcome \\
\hline\hline
\multirow{3}{*}{CDF $Z$} & $A_0$ & $-$0.27 & [$-$0.28, $-$0.18] & [$-$0.28, $-$0.25] & [$-$0.25, $-$0.17] & improves \\ \cline{2-7}
 & $A_2$ & $-$0.19 & [$-$0.25, $-$0.10] & [$-$0.20, $-$0.17] & [$-$0.16, $-$0.08] & improves \\ \cline{2-7}
 & $A_0-A_2$ & +0.01 & [+0.01, +0.02] & [+0.01, +0.01] & [+0.01, +0.01] & unchanged \\ \hline
\multirow{3}{*}{CMS $Z$ 8 TeV} & $A_0$ & $-$0.49 & [$-$0.56, $-$0.05] & [$-$0.51, +0.35] & [$-$0.53, $-$0.51] & undetermined \\ \cline{2-7}
 & $A_2$ & $-$0.50 & [$-$0.98, $-$0.13] & [$-$0.89, $-$0.23] & [$-$0.49, $-$0.25] & improves \\ \cline{2-7}
 & $A_0-A_2$ & +0.08 & [$-$0.01, +0.19] & [+0.03, +0.18] & [+0.08, +0.08] & undetermined \\ \hline
\multirow{3}{*}{ATLAS $Z$} & $A_0$ & +0.67 & [$-$0.28, +2.06] & [+0.02, +2.53] & [$-$0.53, +0.52] & undetermined \\ \cline{2-7}
 & $A_2$ & $-$2.02 & [$-$3.23, $-$0.85] & [$-$2.65, $-$1.41] & [$-$1.97, $-$1.10] & improves \\ \cline{2-7}
 & $A_0-A_2$ & +0.35 & [+0.17, +0.52] & [+0.25, +0.52] & [+0.35, +0.35] & worsens \\ \hline
\multirow{6}{*}{LHCb $Z$} & $A_0$ & +0.71 & [+0.17, +1.36] & [+0.01, +2.38] & [$-$0.09, +0.05] & undetermined \\ \cline{2-7}
 & $A_0^\dagger$ & +0.30 & [$-$0.30, +1.12] & [$-$0.74, +2.81] & [$-$0.30, $-$0.24] & undetermined \\ \cline{2-7}
 & $A_2$ & $-$1.33 & [$-$2.33, $-$0.54] & [$-$2.11, $-$0.84] & [$-$0.93, $-$0.77] & improves \\ \cline{2-7}
 & $A_2^\dagger$ & $-$1.98 & [$-$3.52, $-$0.73] & [$-$3.15, $-$1.24] & [$-$1.33, $-$1.08] & improves \\ \cline{2-7}
 & $A_0-A_2$ & +0.30 & [+0.14, +0.44] & [+0.19, +0.52] & [+0.30, +0.30] & worsens \\ \cline{2-7}
 & $(A_0-A_2)^\dagger$ & +0.17 & [$-$0.01, +0.34] & [0.00, +0.48] & [+0.17, +0.17] & undetermined \\ \hline
\multirow{3}{*}{\makecell{CMS $Z$ 13 TeV\\ $|y|<1$}} & $A_0$ & +2.92 & [+0.07, +6.69] & [+1.35, +15.20] & [+0.63, +2.66] & worsens \\ \cline{2-7}
 & $A_2$ & $-$3.84 & [$-$6.56, $-$1.72] & [$-$5.67, $-$2.59] & [$-$3.73, $-$2.16] & improves \\ \cline{2-7}
 & $A_0-A_2$ & +0.74 & [+0.28, +1.15] & [+0.41, +1.26] & [+0.74, +0.74] & worsens \\ \hline
\multirow{3}{*}{\makecell{CMS $Z$ 13 TeV\\ $1<|y|<2.4$}} & $A_0$ & +1.52 & [$-$0.40, +4.29] & [+0.54, +7.03] & [$-$0.09, +0.52] & undetermined \\ \cline{2-7}
 & $A_2$ & $-$3.43 & [$-$6.73, $-$1.08] & [$-$5.01, $-$2.12] & [$-$2.91, $-$1.98] & improves \\ \cline{2-7}
 & $A_0-A_2$ & +0.67 & [+0.21, +1.16] & [+0.37, +1.22] & [+0.67, +0.67] & worsens \\ \hline
\multirow{2}{*}{CDF $W^-$} & $A_2$ & +0.06 & [+0.02, +0.10] & [+0.04, +0.08] & [+0.02, +0.05] & worsens \\ \cline{2-7}
 & $A_3$ & +0.09 & [+0.04, +0.15] & [+0.07, +0.11] & [$-$0.06, +0.09] & undetermined \\ \hline
\multirow{5}{*}{ATLAS $W^+$} & $A_0$ & $-$0.15 & [$-$0.19, $-$0.09] & [$-$0.24, $-$0.07] & [$-$0.12, $-$0.11] & improves \\ \cline{2-7}
 & $A_2$ & $-$0.07 & [$-$0.19, $-$0.00] & [$-$0.18, $-$0.00] & [$-$0.07, $-$0.04] & improves \\ \cline{2-7}
 & $A_3$ & +0.29 & [+0.11, +0.46] & [+0.23, +0.34] & [$-$0.03, +0.29] & undetermined \\ \cline{2-7}
 & $A_4$ & +0.11 & [+0.07, +0.18] & [+0.10, +0.13] & [+0.09, +0.10] & worsens \\ \cline{2-7}
 & $A_0-A_2$ & $-$0.02 & [$-$0.02, $-$0.02] & [$-$0.03, $-$0.01] & [$-$0.02, $-$0.02] & unchanged \\ \hline
\multirow{5}{*}{ATLAS $W^-$} & $A_0$ & $-$0.01 & [$-$0.02, 0.00] & [$-$0.03, 0.00] & [$-$0.01, $-$0.01] & unchanged \\ \cline{2-7}
 & $A_2$ & $-$0.02 & [$-$0.04, $-$0.01] & [$-$0.03, +0.01] & [$-$0.02, $-$0.01] & unchanged \\ \cline{2-7}
 & $A_3$ & +0.13 & [+0.08, +0.19] & [+0.10, +0.15] & [+0.13, +0.19] & worsens \\ \cline{2-7}
 & $A_4$ & $-$0.05 & [$-$0.08, $-$0.03] & [$-$0.06, $-$0.05] & [$-$0.05, $-$0.05] & improves \\ \cline{2-7}
 & $A_0-A_2$ & 0.00 & [0.00, +0.01] & [0.00, +0.01] & [0.00, 0.00] & unchanged \\ \hline
\end{tabular}
\caption{Difference $\Delta\chi^2/N_{d.o.f.}=\chi^2/N_{d.o.f.}({\rm NNLL{+}NLO})-\chi^2/N_{d.o.f.}({\rm NLO})$ for the central predictions and range over the variations of each source of uncertainty (scales: six $(\mu_R,\mu_F)$ combinations and $Q=M/2,\,2M$; recoil: GJ and naive prescriptions). The variations of the non-perturbative parameters change $\Delta\chi^2/N_{d.o.f.}$ by at most $0.01$ and are not shown. $^\dagger$\,Without the first two bins.}
\label{t:robust}
\end{table}

In summary, the improvement in the description of $A_2$ for the CMS (at both energies), ATLAS and LHCb measurements and of $A_0$ and $A_2$ for the CDF measurement is stable against all the variations considered. The improvement of $A_0$ for CMS at 8~TeV holds for all the variations except the transition function with $k=0.5$. For $A_0$ at ATLAS and LHCb, the sign of the change depends on the recoil prescription and on the scales. For the CMS data at 13~TeV, $A_0$ is worse described for all the variations at $|y|<1$, while at $1<|y|<2.4$ the sign of the change depends on the scales and on the recoil prescription. The combination $A_0-A_2$ is never improved by resummation and is slightly worse described for ATLAS, CMS at 13~TeV and LHCb, consistently with the preservation of the Lam--Tung relation by the resummed contributions; its description requires higher-order corrections to the fixed-order component. For CDF $W^-$, the values of $\chi^2/N_{d.o.f.}$ for $A_2$ and $A_3$ are similar for all the variations and below one, i.e.\ the measurement is well described with all the theoretical choices. For the ATLAS $W$-boson measurements, the description of $A_0$ and $A_2$ slightly improves for $W^+$, for all the variations, and is unchanged for $W^-$, while
$A_3$ is slightly worse described for all the variations, except for the naive prescription for $W^+$, and the description of $A_4$ slightly improves for $W^-$ and slightly worsens for $W^+$. These contrasting indications may also be due to the large experimental uncertainties of the $W$-boson measurements.

\clearpage
 \section{Conclusions}
\label{s:conclusions}

We have presented a study of the impact of transverse-momentum ($q_T$) resummation on the angular coefficients of $Z$ and $W$ boson production at hadron colliders. Theoretical predictions have been obtained using the {\ttfamily DYTurbo} \cite{Camarda:2019zyx} numerical program by combining resummation of logarithmically enhanced QCD corrections at small $q_T$ up to NNLL accuracy with fixed-order NLO (i.e.\ $\mathcal{O}(\alpha_S^2)$) calculations. The predictions have been compared to measurements from the CDF, CMS, ATLAS, and LHCb collaborations, spanning a wide range of centre-of-mass energies and rapidity intervals.

For $Z$-boson production, the inclusion of resummed contributions lowers the coefficients $A_0$ and $A_2$ by $5$--$20\%$ in the intermediate transverse-momentum region, $q_T \simeq 20$--$50$~GeV. We have shown that this effect arises pointwise in $q_T$ from the matching of the resummed and fixed-order components: with the Collins--Soper recoil prescription, the resummed component contributes only to the unpolarised cross section and to $A_4$, and the other coefficients are rescaled by a common factor, while the averaging over finite bins gives a subdominant contribution. The size of the effect depends on the recoil prescription, and it is reduced by about $45$--$60\%$ if the polar axis of the frame in which the Born-level distribution is generated is chosen along one of the incoming partons. We have estimated the theoretical uncertainties of the predictions, including scale variations, PDFs, non-perturbative effects, the transition function and the recoil prescription. The PDF and non-perturbative uncertainties are negligible, while the other sources amount each to between about $10\%$ and $75\%$ of the effect for $A_0$ and $A_2$. Within these uncertainties, resummation improves the description of $A_2$ in the CMS (at 8 and 13~TeV), ATLAS and LHCb data and of $A_0$ and $A_2$ in the CDF data, and the improvement of $A_0$ in the CMS data at 8~TeV persists for all the variations but one. For $A_0$ at ATLAS and LHCb, the direction of the change depends on the theoretical choices. In the more precise CMS data at 13~TeV, $A_0$ agrees with the NLO prediction in the transition region while $A_2$ lies below it, i.e.\ the data show a larger violation of the Lam--Tung relation than predicted at $\mathcal{O}(\alpha_S^2)$; since resummation lowers both coefficients by a common factor, it worsens the description of $A_0$, for all the variations at $|y|<1$. The Lam--Tung combination $A_0-A_2$ is essentially unaffected by resummation, consistently with the preservation of the Lam--Tung relation by the resummed contributions, and its description requires higher-order corrections to the fixed-order component.

For $W$-boson production, we have presented, to the best of our knowledge, the first dedicated study of resummation effects on the angular coefficients. The resummed contributions modify the theoretical predictions in the intermediate $q_T$ region by a relative amount similar to the neutral-current case, but the effects are small compared with the experimental uncertainties. The CDF $W^-$ data are well described by all the predictions, with changes that depend on the recoil prescription. For the ATLAS measurements, resummation slightly improves $A_0$ and $A_2$ for $W^+$, leaves them essentially unchanged for $W^-$, and slightly worsens the description of $A_3$. For $A_4$, whose numerator receives a direct contribution from the resummed component, resummation increases the coefficient by $2.5$--$6\%$ for $8<q_T<55$~GeV and lowers it in the first bin; the shift is at most $0.04$, well below the experimental uncertainties.

Taken together, these results show that transverse-momentum resummation has a non-negligible impact on the power-suppressed angular coefficients in the transition region, comparable to the precision of the most accurate measurements, and that its size is controlled by the treatment of the power corrections and of the transition between the resummed and fixed-order regimes. A more precise assessment requires the inclusion of the $\mathcal{O}(\alpha_S^3)$ corrections to the fixed-order component, a systematic treatment of the power-suppressed hadronic contributions and a matching procedure with a reduced dependence on the transition function. This work motivates further investigations in these directions, as well as the extension to additional angular coefficients and kinematic regimes.

\section*{Acknowledgments}
We thank Dani\"el Boer and Werner Vogelsang for useful discussions.
\bibliographystyle{spphys_arxiv}
  \bibliography{biblio}

@article{Camarda:2026kdu,
    author = "Camarda, Stefano and Cieri, Leandro and Ferrera, Giancarlo and Rossi, Lorenzo",
    title = "{Analytic next-to-leading-order helicity cross sections for vector-boson production at finite transverse momentum}",
    eprint = "2609.37238",
    archivePrefix = "arXiv",
    primaryClass = "hep-ph",
    month = "9",
    year = "2026"
}

@article{Monni:2016ktx,
    author = "Monni, Pier Francesco and Re, Emanuele and Torrielli, Paolo",
    title = "{Higgs Transverse-Momentum Resummation in Direct Space}",
    eprint = "1604.02191",
    archivePrefix = "arXiv",
    primaryClass = "hep-ph",
    reportNumber = "LAPTH-017-16, OUTP-16-07P",
    doi = "10.1103/PhysRevLett.116.242001",
    journal = "Phys. Rev. Lett.",
    volume = "116",
    number = "24",
    pages = "242001",
    year = "2016"
}

@article{ATLAS:2016rnf,
    author = "Aad, Georges and others",
    collaboration = "ATLAS",
    title = "{Measurement of the angular coefficients in $Z$-boson events using electron and muon pairs from data taken at $\sqrt{s}=8$ TeV with the ATLAS detector}",
    eprint = "1606.00689",
    archivePrefix = "arXiv",
    primaryClass = "hep-ex",
    reportNumber = "CERN-EP-2016-087",
    doi = "10.1007/JHEP08(2016)159",
    journal = "JHEP",
    volume = "08",
    pages = "159",
    year = "2016"
}

@article{ATLAS:2025mlt,
    author = "Aad, Georges and others",
    collaboration = "ATLAS",
    title = "{Measurement of the $W$-boson angular coefficients and transverse momentum in $pp$ collisions at $\sqrt{s}=13$ TeV with the ATLAS detector}",
    eprint = "2509.13759",
    archivePrefix = "arXiv",
    primaryClass = "hep-ex",
    reportNumber = "CERN-EP-2025-188",
    month = "9",
    year = "2025"
}

@article{CDF:2011ksg,
    author = "Aaltonen, T. and others",
    collaboration = "CDF",
    title = "{First Measurement of the Angular Coefficients of Drell-Yan $e^{+}e^{-}$ pairs in the Z Mass Region from $p\bar{p}$ Collisions at $\sqrt{s}$ = 1.96 TeV}",
    eprint = "1103.5699",
    archivePrefix = "arXiv",
    primaryClass = "hep-ex",
    reportNumber = "FERMILAB-PUB-11-153-E",
    doi = "10.1103/PhysRevLett.106.241801",
    journal = "Phys. Rev. Lett.",
    volume = "106",
    pages = "241801",
    year = "2011"
}

@article{CMS:2015cyj,
    author = "Khachatryan, Vardan and others",
    collaboration = "CMS",
    title = "{Angular coefficients of Z bosons produced in pp collisions at $\sqrt{s}$ = 8 TeV and decaying to $\mu^+ \mu^-$ as a function of transverse momentum and rapidity}",
    eprint = "1504.03512",
    archivePrefix = "arXiv",
    primaryClass = "hep-ex",
    reportNumber = "CMS-SMP-13-010, CERN-PH-EP-2015-046",
    doi = "10.1016/j.physletb.2015.08.061",
    journal = "Phys. Lett. B",
    volume = "750",
    pages = "154--175",
    year = "2015"
}

@article{LHCb:2022tbc,
    author = "Aaij, R. and others",
    collaboration = "LHCb",
    title = "{First Measurement of the Z{\textrightarrow}{\ensuremath{\mu}}+{\ensuremath{\mu}}- Angular Coefficients in the Forward Region of pp Collisions at s=13{\,}{\,}TeV}",
    eprint = "2203.01602",
    archivePrefix = "arXiv",
    primaryClass = "hep-ex",
    reportNumber = "LHCb-PAPER-2021-048, CERN-EP-2022-030",
    doi = "10.1103/PhysRevLett.129.091801",
    journal = "Phys. Rev. Lett.",
    volume = "129",
    number = "9",
    pages = "091801",
    year = "2022"
}

@article{Affolder:1999jh,
    author = "Affolder, T. and others",
    collaboration = "CDF",
    title = "{The transverse momentum and total cross section of $e^+e^-$ pairs in the $Z$ boson region from $p\bar{p}$ collisions at $\sqrt{s} = 1.8$ TeV}",
    eprint = "hep-ex/0001021",
    archivePrefix = "arXiv",
    reportNumber = "FERMILAB-PUB-99-220-E",
    doi = "10.1103/PhysRevLett.84.845",
    journal = "Phys. Rev. Lett.",
    volume = "84",
    pages = "845--850",
    year = "2000"
}

@article{Aaltonen:2012fi,
    author = "Aaltonen, T. and others",
    collaboration = "CDF",
    title = "{Transverse momentum cross section of $e^+e^-$ pairs in the $Z$-boson region from $p\bar{p}$ collisions at $\sqrt{s}=1.96$ TeV}",
    eprint = "1207.7138",
    archivePrefix = "arXiv",
    primaryClass = "hep-ex",
    reportNumber = "FERMILAB-PUB-12-421-E",
    doi = "10.1103/PhysRevD.86.052010",
    journal = "Phys. Rev. D",
    volume = "86",
    pages = "052010",
    year = "2012"
}

@article{Abbott:1999wk,
    author = "Abbott, B. and others",
    collaboration = "D0",
    title = "{Measurement of the inclusive differential cross section for $Z$ bosons as a function of transverse momentum in $\bar{p}p$ collisions at $\sqrt{s} = 1.8$ TeV}",
    eprint = "hep-ex/9907009",
    archivePrefix = "arXiv",
    reportNumber = "FERMILAB-PUB-99-197-E",
    doi = "10.1103/PhysRevD.61.032004",
    journal = "Phys. Rev. D",
    volume = "61",
    pages = "032004",
    year = "2000"
}

@article{Abazov:2007ac,
    author = "Abazov, V. M. and others",
    collaboration = "D0",
    title = "{Measurement of the shape of the boson transverse momentum distribution in $p \bar{p} \to Z / \gamma^{*} \to e^+ e^- + X$ events produced at $\sqrt{s}$=1.96-TeV}",
    eprint = "0712.0803",
    archivePrefix = "arXiv",
    primaryClass = "hep-ex",
    reportNumber = "FERMILAB-PUB-07-642-E",
    doi = "10.1103/PhysRevLett.100.102002",
    journal = "Phys. Rev. Lett.",
    volume = "100",
    pages = "102002",
    year = "2008"
}

@article{Abazov:2010kn,
    author = "Abazov, Victor Mukhamedovich and others",
    collaboration = "D0",
    title = "{Measurement of the Normalized $Z/\gamma^* -> \mu^+\mu^-$ Transverse Momentum Distribution in $p\bar{p}$ Collisions at $\sqrt{s}=1.96$ TeV}",
    eprint = "1006.0618",
    archivePrefix = "arXiv",
    primaryClass = "hep-ex",
    reportNumber = "FERMILAB-PUB-10-183-E",
    doi = "10.1016/j.physletb.2010.09.012",
    journal = "Phys. Lett. B",
    volume = "693",
    pages = "522--530",
    year = "2010"
}

@article{Aad:2014xaa,
    author = "Aad, Georges and others",
    collaboration = "ATLAS",
    title = "{Measurement of the $Z/\gamma^*$ boson transverse momentum distribution in $pp$ collisions at $\sqrt{s}$ = 7 TeV with the ATLAS detector}",
    eprint = "1406.3660",
    archivePrefix = "arXiv",
    primaryClass = "hep-ex",
    reportNumber = "CERN-PH-EP-2014-075",
    doi = "10.1007/JHEP09(2014)145",
    journal = "JHEP",
    volume = "09",
    pages = "145",
    year = "2014"
}

@article{Aad:2015auj,
    author = "Aad, Georges and others",
    collaboration = "ATLAS",
    title = "{Measurement of the transverse momentum and $\phi ^*_{\eta }$ distributions of Drell\textendash{}Yan lepton pairs in proton\textendash{}proton collisions at $\sqrt{s}=8$  TeV with the ATLAS detector}",
    eprint = "1512.02192",
    archivePrefix = "arXiv",
    primaryClass = "hep-ex",
    reportNumber = "CERN-PH-EP-2015-275",
    doi = "10.1140/epjc/s10052-016-4070-4",
    journal = "Eur. Phys. J. C",
    volume = "76",
    number = "5",
    pages = "291",
    year = "2016"
}

@article{ATLAS:2019zci,
    author = "Aad, Georges and others",
    collaboration = "ATLAS",
    title = "{Measurement of the transverse momentum distribution of Drell\textendash{}Yan lepton pairs in proton\textendash{}proton collisions at $\sqrt{s}=13$ TeV with the ATLAS detector}",
    eprint = "1912.02844",
    archivePrefix = "arXiv",
    primaryClass = "hep-ex",
    reportNumber = "CERN-EP-2019-223",
    doi = "10.1140/epjc/s10052-020-8001-z",
    journal = "Eur. Phys. J. C",
    volume = "80",
    number = "7",
    pages = "616",
    year = "2020"
}

@article{CMS:2019raw,
    author = "Sirunyan, Albert M and others",
    collaboration = "CMS",
    title = "{Measurements of differential Z boson production cross sections in proton-proton collisions at $ \sqrt{s} $ = 13 TeV}",
    eprint = "1909.04133",
    archivePrefix = "arXiv",
    primaryClass = "hep-ex",
    reportNumber = "CMS-SMP-17-010, CERN-EP-2019-175",
    doi = "10.1007/JHEP12(2019)061",
    journal = "JHEP",
    volume = "12",
    pages = "061",
    year = "2019"
}

@article{Chatrchyan:2011wt,
    author = "Chatrchyan, Serguei and others",
    collaboration = "CMS",
    title = "{Measurement of the Rapidity and Transverse Momentum Distributions of $Z$ Bosons in $pp$ Collisions at $\sqrt{s}=7$ TeV}",
    eprint = "1110.4973",
    archivePrefix = "arXiv",
    primaryClass = "hep-ex",
    reportNumber = "CMS-EWK-10-010, CERN-PH-EP-2011-169",
    doi = "10.1103/PhysRevD.85.032002",
    journal = "Phys. Rev. D",
    volume = "85",
    pages = "032002",
    year = "2012"
}

@article{Khachatryan:2016nbe,
    author = "Khachatryan, Vardan and others",
    collaboration = "CMS",
    title = "{Measurement of the transverse momentum spectra of weak vector bosons produced in proton-proton collisions at $ \sqrt{s}=8 $ TeV}",
    eprint = "1606.05864",
    archivePrefix = "arXiv",
    primaryClass = "hep-ex",
    reportNumber = "CMS-SMP-14-012, CERN-EP-2016-152",
    doi = "10.1007/JHEP02(2017)096",
    journal = "JHEP",
    volume = "02",
    pages = "096",
    year = "2017"
}

@article{Aaij:2015gna,
    author = "Aaij, Roel and others",
    collaboration = "LHCb",
    title = "{Measurement of the forward $Z$ boson production cross-section in $pp$ collisions at $\sqrt{s}=7$ TeV}",
    eprint = "1505.07024",
    archivePrefix = "arXiv",
    primaryClass = "hep-ex",
    reportNumber = "LHCB-PAPER-2015-001, CERN-PH-EP-2015-102",
    doi = "10.1007/JHEP08(2015)039",
    journal = "JHEP",
    volume = "08",
    pages = "039",
    year = "2015"
}

@article{Aaij:2015zlq,
    author = "Aaij, Roel and others",
    collaboration = "LHCb",
    title = "{Measurement of forward W and Z boson production in $pp$ collisions at $ \sqrt{s}=8 $ TeV}",
    eprint = "1511.08039",
    archivePrefix = "arXiv",
    primaryClass = "hep-ex",
    reportNumber = "LHCB-PAPER-2015-049, CERN-PH-EP-2015-301",
    doi = "10.1007/JHEP01(2016)155",
    journal = "JHEP",
    volume = "01",
    pages = "155",
    year = "2016"
}

@article{Aaij:2016mgv,
    author = "Aaij, Roel and others",
    collaboration = "LHCb",
    title = "{Measurement of the forward Z boson production cross-section in pp collisions at $\sqrt{s} = 13$ TeV}",
    eprint = "1607.06495",
    archivePrefix = "arXiv",
    primaryClass = "hep-ex",
    reportNumber = "LHCB-PAPER-2016-021, CERN-EP-2016-170",
    doi = "10.1007/JHEP09(2016)136",
    journal = "JHEP",
    volume = "09",
    pages = "136",
    year = "2016"
}

@article{ATLAS:2023lsr,
    author = "Aad, Georges and others",
    collaboration = "ATLAS",
    title = "{A precise measurement of the Z-boson double-differential transverse momentum and rapidity distributions in the full phase space of the decay leptons with the ATLAS experiment at~$\sqrt{s}=8$~TeV}",
    eprint = "2309.09318",
    archivePrefix = "arXiv",
    primaryClass = "hep-ex",
    reportNumber = "CERN-EP-2023-171",
    doi = "10.1140/epjc/s10052-024-12438-w",
    journal = "Eur. Phys. J. C",
    volume = "84",
    number = "3",
    pages = "315",
    year = "2024"
}

@article{ATLAS:2023lhg,
    author = "Aad, Georges and others",
    collaboration = "ATLAS",
    title = "{A precise determination of the strong-coupling constant from the recoil of $Z$ bosons with the ATLAS experiment at $\sqrt{s} = 8$ TeV}",
    eprint = "2309.12986",
    archivePrefix = "arXiv",
    primaryClass = "hep-ex",
    month = "9",
    year = "2023"
}

@article{Collins:1977iv,
    author = "Collins, John C. and Soper, Davison E.",
    title = "{Angular Distribution of Dileptons in High-Energy Hadron Collisions}",
    reportNumber = "Print-77-0288 (PRINCETON)",
    doi = "10.1103/PhysRevD.16.2219",
    journal = "Phys. Rev. D",
    volume = "16",
    pages = "2219",
    year = "1977"
}

@article{Lam:1978pu,
    author = "Lam, C. S. and Tung, Wu-Ki",
    title = "{A Systematic Approach to Inclusive Lepton Pair Production in Hadronic Collisions}",
    reportNumber = "Print-78-0553 (MCGILL)",
    doi = "10.1103/PhysRevD.18.2447",
    journal = "Phys. Rev. D",
    volume = "18",
    pages = "2447",
    year = "1978"
}

@article{Lam:1978zr,
    author = "Lam, C. S. and Tung, Wu-Ki",
    title = "{Structure Function Relations at Large Transverse Momenta in Lepton Pair Production Processes}",
    reportNumber = "RL-78-054",
    doi = "10.1016/0370-2693(79)90204-1",
    journal = "Phys. Lett. B",
    volume = "80",
    pages = "228--231",
    year = "1979"
}

@article{Mirkes:1992hu,
    author = "Mirkes, E.",
    title = "{Angular decay distribution of leptons from W bosons at NLO in hadronic collisions}",
    reportNumber = "TTP-92-12",
    doi = "10.1016/0550-3213(92)90046-E",
    journal = "Nucl. Phys. B",
    volume = "387",
    pages = "3--85",
    year = "1992"
}

@article{Mirkes:1994eb,
    author = "Mirkes, E. and Ohnemus, J.",
    title = "{$W$ and $Z$ polarization effects in hadronic collisions}",
    eprint = "hep-ph/9406381",
    archivePrefix = "arXiv",
    reportNumber = "MAD-PH-834, UCD-94-23",
    doi = "10.1103/PhysRevD.50.5692",
    journal = "Phys. Rev. D",
    volume = "50",
    pages = "5692--5703",
    year = "1994"
}

@article{Mirkes:1994dp,
    author = "Mirkes, E. and Ohnemus, J.",
    title = "{Angular distributions of Drell-Yan lepton pairs at the Tevatron: Order $\alpha-s^{2}$ corrections and Monte Carlo studies}",
    eprint = "hep-ph/9412289",
    archivePrefix = "arXiv",
    reportNumber = "MAD-PH-857, UCD-94-39",
    doi = "10.1103/PhysRevD.51.4891",
    journal = "Phys. Rev. D",
    volume = "51",
    pages = "4891--4904",
    year = "1995"
}

@article{ATLAS:2012au,
    author = "Aad, Georges and others",
    collaboration = "ATLAS",
    title = "{Measurement of the polarisation of $W$ bosons produced with large transverse momentum in $pp$ collisions at $\sqrt{s}=7$ TeV with the ATLAS experiment}",
    eprint = "1203.2165",
    archivePrefix = "arXiv",
    primaryClass = "hep-ex",
    reportNumber = "CERN-PH-EP-2012-016",
    doi = "10.1140/epjc/s10052-012-2001-6",
    journal = "Eur. Phys. J. C",
    volume = "72",
    pages = "2001",
    year = "2012"
}

@article{CMS:2011kaj,
    author = "Chatrchyan, Serguei and others",
    collaboration = "CMS",
    title = "{Measurement of the Polarization of W Bosons with Large Transverse Momenta in W+Jets Events at the LHC}",
    eprint = "1104.3829",
    archivePrefix = "arXiv",
    primaryClass = "hep-ex",
    reportNumber = "CERN-PH-EP-2011-043, CMS-EWK-10-014",
    doi = "10.1103/PhysRevLett.107.021802",
    journal = "Phys. Rev. Lett.",
    volume = "107",
    pages = "021802",
    year = "2011"
}

@article{Pellen:2022fom,
    author = "Pellen, Mathieu and Poncelet, Rene and Popescu, Andrei and Vitos, Timea",
    title = "{Angular coefficients in $\hbox {W}+\hbox {j}$ production at the LHC with high precision}",
    eprint = "2204.12394",
    archivePrefix = "arXiv",
    primaryClass = "hep-ph",
    reportNumber = "CAVENDISH--HEP--22/04, FR-PHENO-2022-04, Lund-22-24",
    doi = "10.1140/epjc/s10052-022-10641-1",
    journal = "Eur. Phys. J. C",
    volume = "82",
    number = "8",
    pages = "693",
    year = "2022"
}

@article{Gauld:2017tww,
    author = "Gauld, R. and Gehrmann-De Ridder, A. and Gehrmann, T. and Glover, E. W. N. and Huss, A.",
    title = "{Precise predictions for the angular coefficients in Z-boson production at the LHC}",
    eprint = "1708.00008",
    archivePrefix = "arXiv",
    primaryClass = "hep-ph",
    reportNumber = "IPPP-17-58, ZU-TH-21-17",
    doi = "10.1007/JHEP11(2017)003",
    journal = "JHEP",
    volume = "11",
    pages = "003",
    year = "2017"
}

@article{Dokshitzer:1978yd,
    author = "Dokshitzer, Yuri L. and Diakonov, Dmitri and Troian, S. I.",
    title = "{On the Transverse Momentum Distribution of Massive Lepton Pairs}",
    doi = "10.1016/0370-2693(78)90240-X",
    journal = "Phys. Lett. B",
    volume = "79",
    pages = "269--272",
    year = "1978"
}

@article{Parisi:1979se,
    author = "Parisi, G. and Petronzio, R.",
    title = "{Small Transverse Momentum Distributions in Hard Processes}",
    reportNumber = "CERN-TH-2627",
    doi = "10.1016/0550-3213(79)90040-3",
    journal = "Nucl. Phys. B",
    volume = "154",
    pages = "427--440",
    year = "1979"
}

@article{Collins:1984kg,
    author = "Collins, John C. and Soper, Davison E. and Sterman, George F.",
    title = "{Transverse Momentum Distribution in Drell-Yan Pair and W and Z Boson Production}",
    reportNumber = "CERN-TH-3923",
    doi = "10.1016/0550-3213(85)90479-1",
    journal = "Nucl. Phys. B",
    volume = "250",
    pages = "199--224",
    year = "1985"
}

@article{Catani:2010pd,
    author = "Catani, Stefano and Grazzini, Massimiliano",
    title = "{QCD transverse-momentum resummation in gluon fusion processes}",
    eprint = "1011.3918",
    archivePrefix = "arXiv",
    primaryClass = "hep-ph",
    doi = "10.1016/j.nuclphysb.2010.12.007",
    journal = "Nucl. Phys. B",
    volume = "845",
    pages = "297--323",
    year = "2011"
}

@article{Becher:2010tm,
    author = "Becher, Thomas and Neubert, Matthias",
    title = "{Drell-Yan Production at Small $q_T$, Transverse Parton Distributions and the Collinear Anomaly}",
    eprint = "1007.4005",
    archivePrefix = "arXiv",
    primaryClass = "hep-ph",
    reportNumber = "HD-THEP-10-13, MZ-TH-10-26",
    doi = "10.1140/epjc/s10052-011-1665-7",
    journal = "Eur. Phys. J. C",
    volume = "71",
    pages = "1665",
    year = "2011"
}

@article{Scimemi:2017etj,
    author = "Scimemi, Ignazio and Vladimirov, Alexey",
    title = "{Analysis of vector boson production within TMD factorization}",
    eprint = "1706.01473",
    archivePrefix = "arXiv",
    primaryClass = "hep-ph",
    doi = "10.1140/epjc/s10052-018-5557-y",
    journal = "Eur. Phys. J. C",
    volume = "78",
    number = "2",
    pages = "89",
    year = "2018"
}

@article{Bizon:2018foh,
    author = "Bizo{\'n}, Wojciech and Chen, Xuan and Gehrmann-De Ridder, Aude and Gehrmann, Thomas and Glover, Nigel and Huss, Alexander and Monni, Pier Francesco and Re, Emanuele and Rottoli, Luca and Torrielli, Paolo",
    title = "{Fiducial distributions in Higgs and Drell-Yan production at N$^{3}$LL+NNLO}",
    eprint = "1805.05916",
    archivePrefix = "arXiv",
    primaryClass = "hep-ph",
    reportNumber = "CERN-TH-2018-105, IPPP/18/34, LAPTH-015/18, OUTP-17-19P, ZU-TH 17/18, IPPP-18-34, LAPTH-015-18, ZU-TH-17-18",
    doi = "10.1007/JHEP12(2018)132",
    journal = "JHEP",
    volume = "12",
    pages = "132",
    year = "2018"
}

@article{Becher:2019bnm,
    author = "Becher, Thomas and Hager, Monika",
    title = "{Event-Based Transverse Momentum Resummation}",
    eprint = "1904.08325",
    archivePrefix = "arXiv",
    primaryClass = "hep-ph",
    doi = "10.1140/epjc/s10052-019-7136-2",
    journal = "Eur. Phys. J. C",
    volume = "79",
    number = "8",
    pages = "665",
    year = "2019"
}

@article{Alioli:2021qbf,
    author = "Alioli, Simone and Bauer, Christian W. and Broggio, Alessandro and Gavardi, Alessandro and Kallweit, Stefan and Lim, Matthew A. and Nagar, Riccardo and Napoletano, Davide and Rottoli, Luca",
    title = "{Matching NNLO predictions to parton showers using N3LL color-singlet transverse momentum resummation in geneva}",
    eprint = "2102.08390",
    archivePrefix = "arXiv",
    primaryClass = "hep-ph",
    reportNumber = "ZU-TH 8/21",
    doi = "10.1103/PhysRevD.104.094020",
    journal = "Phys. Rev. D",
    volume = "104",
    number = "9",
    pages = "094020",
    year = "2021"
}

@article{Ebert:2016gcn,
    author = "Ebert, Markus A. and Tackmann, Frank J.",
    title = "{Resummation of Transverse Momentum Distributions in Distribution Space}",
    eprint = "1611.08610",
    archivePrefix = "arXiv",
    primaryClass = "hep-ph",
    reportNumber = "DESY-16-215",
    doi = "10.1007/JHEP02(2017)110",
    journal = "JHEP",
    volume = "02",
    pages = "110",
    year = "2017"
}

@article{Bizon:2019zgf,
    author = "Bizon, Wojciech and Gehrmann-De Ridder, Aude and Gehrmann, Thomas and Glover, Nigel and Huss, Alexander and Monni, Pier Francesco and Re, Emanuele and Rottoli, Luca and Walker, Duncan M.",
    title = "{The transverse momentum spectrum of weak gauge bosons at N ${}^3$ LL + NNLO}",
    eprint = "1905.05171",
    archivePrefix = "arXiv",
    primaryClass = "hep-ph",
    reportNumber = "ZU-TH 21/1, CERN-TH-2019-050, LAPTH-026/19, IPPP/19/38",
    doi = "10.1140/epjc/s10052-019-7324-0",
    journal = "Eur. Phys. J. C",
    volume = "79",
    number = "10",
    pages = "868",
    year = "2019"
}

@article{Ebert:2020dfc,
    author = "Ebert, Markus A. and Michel, Johannes K. L. and Stewart, Iain W. and Tackmann, Frank J.",
    title = "{Drell-Yan $q_{T}$ resummation of fiducial power corrections at N$^{3}$LL}",
    eprint = "2006.11382",
    archivePrefix = "arXiv",
    primaryClass = "hep-ph",
    reportNumber = "DESY-20-016, DESY 20-016, MIT-CTP 5205",
    doi = "10.1007/JHEP04(2021)102",
    journal = "JHEP",
    volume = "04",
    pages = "102",
    year = "2021"
}

@article{Billis:2024dqq,
    author = "Billis, Georgios and Michel, Johannes K. L. and Tackmann, Frank J.",
    title = "{Drell-Yan transverse-momentum spectra at N$^{3}$LL$^{'}$ and approximate N$^{4}$LL with SCETlib}",
    eprint = "2411.16004",
    archivePrefix = "arXiv",
    primaryClass = "hep-ph",
    reportNumber = "DESY-23-081, MIT-CTP 5572, Nikhef 2024-007",
    doi = "10.1007/JHEP02(2025)170",
    journal = "JHEP",
    volume = "02",
    pages = "170",
    year = "2025"
}

@article{Catani:2015vma,
    author = "Catani, Stefano and de Florian, Daniel and Ferrera, Giancarlo and Grazzini, Massimiliano",
    title = "{Vector boson production at hadron colliders: transverse-momentum resummation and leptonic decay}",
    eprint = "1507.06937",
    archivePrefix = "arXiv",
    primaryClass = "hep-ph",
    reportNumber = "ICAS-02-15, TIF-UNIMI-2015-10, ZU-TH-24-15",
    doi = "10.1007/JHEP12(2015)047",
    journal = "JHEP",
    volume = "12",
    pages = "047",
    year = "2015"
}

@article{Camarda:2019zyx,
    author = "Camarda, Stefano and others",
    title = "{DYTurbo: Fast predictions for Drell-Yan processes}",
    eprint = "1910.07049",
    archivePrefix = "arXiv",
    primaryClass = "hep-ph",
    doi = "10.1140/epjc/s10052-020-7757-5",
    journal = "Eur. Phys. J. C",
    volume = "80",
    number = "3",
    pages = "251",
    year = "2020",
    note = "[Erratum: Eur.Phys.J.C 80, 440 (2020)]"
}

@article{Camarda:2025lbt,
    author = "Camarda, Stefano and Ferrera, Giancarlo and Rossi, Lorenzo",
    title = "{Drell-Yan lepton pair production at low invariant masses: transverse-momentum resummation and non-perturbative effects in QCD}",
    eprint = "2508.06201",
    archivePrefix = "arXiv",
    primaryClass = "hep-ph",
    doi = "10.1007/JHEP01(2026)150",
    journal = "JHEP",
    volume = "01",
    pages = "150",
    year = "2026"
}

@article{Bozzi:2005wk,
    author = "Bozzi, Giuseppe and Catani, Stefano and de Florian, Daniel and Grazzini, Massimiliano",
    title = "{Transverse-momentum resummation and the spectrum of the Higgs boson at the LHC}",
    eprint = "hep-ph/0508068",
archivePrefix = "arXiv",
    doi = "10.1016/j.nuclphysb.2005.12.022",
    journal = "Nucl. Phys. B",
    volume = "737",
    pages = "73--120",
    year = "2006"
}

@article{Bozzi:2010xn,
    author = "Bozzi, Giuseppe and Catani, Stefano and Ferrera, Giancarlo and de Florian, Daniel and Grazzini, Massimiliano",
    title = "{Production of Drell-Yan lepton pairs in hadron collisions: Transverse-momentum resummation at next-to-next-to-leading logarithmic accuracy}",
    eprint = "1007.2351",
    archivePrefix = "arXiv",
    primaryClass = "hep-ph",
    doi = "10.1016/j.physletb.2010.12.024",
    journal = "Phys. Lett. B",
    volume = "696",
    pages = "207--213",
    year = "2011"
}

@article{Lambertsen:2016wgj,
    author = "Lambertsen, Martin and Vogelsang, Werner",
    title = "{Drell-Yan lepton angular distributions in perturbative QCD}",
    eprint = "1605.02625",
    archivePrefix = "arXiv",
    primaryClass = "hep-ph",
    doi = "10.1103/PhysRevD.93.114013",
    journal = "Phys. Rev. D",
    volume = "93",
    number = "11",
    pages = "114013",
    year = "2016"
}

@article{CMS:2011utm,
    author = "Chatrchyan, Serguei and others",
    collaboration = "CMS",
    title = "{Measurement of the weak mixing angle with the Drell-Yan process in proton-proton collisions at the LHC}",
    eprint = "1110.2682",
    archivePrefix = "arXiv",
    primaryClass = "hep-ex",
    reportNumber = "CMS-EWK-11-003, CERN-PH-EP-2011-159",
    doi = "10.1103/PhysRevD.84.112002",
    journal = "Phys. Rev. D",
    volume = "84",
    pages = "112002",
    year = "2011"
}

@article{NNPDF:2021njg,
    author = "Ball, Richard D. and others",
    collaboration = "NNPDF",
    title = "{The path to proton structure at 1{\%} accuracy}",
    eprint = "2109.02653",
    archivePrefix = "arXiv",
    primaryClass = "hep-ph",
    reportNumber = "Edinburgh 2021/12, Nikhef-2021-013, TIF-UNIMI-2021-11",
    doi = "10.1140/epjc/s10052-022-10328-7",
    journal = "Eur. Phys. J. C",
    volume = "82",
    number = "5",
    pages = "428",
    year = "2022"
}

@article{CDF:2005qwt,
    author = "Acosta, D. and others",
    collaboration = "CDF",
    title = "{Measurement of the azimuthal angle distribution of leptons from $W$ boson decays as a function of the $W$ transverse momentum in $p \bar{p}$ collisions at $\sqrt{s} = 1.8$ TeV}",
    eprint = "hep-ex/0504020",
    archivePrefix = "arXiv",
    reportNumber = "FERMILAB-PUB-05-063-E",
    doi = "10.1103/PhysRevD.73.052002",
    journal = "Phys. Rev. D",
    volume = "73",
    pages = "052002",
    year = "2006"
}

@article{CMS:2026amb,
    author = "Gevorgyan, Arzunik and others",
    collaboration = "CMS",
    title = "{Measurement of the Z $\to$$\mu^+\mu^-$ angular coefficients in pp collisions at $\sqrt{s}$ = 13 TeV as functions of transverse momentum and rapidity}",
    eprint = "2604.25678",
    archivePrefix = "arXiv",
    primaryClass = "hep-ex",
    reportNumber = "CMS-SMP-23-007, CERN-EP-2026-099",
    doi = "10.1016/j.physletb.2026.140988",
    journal = "Phys. Lett. B",
    volume = "882",
    pages = "140988",
    year = "2026"
}

@article{Bacchetta:2025ara,
    author = "Bacchetta, Alessandro and Bertone, Valerio and Bissolotti, Chiara and Cerutti, Matteo and Radici, Marco and Rodini, Simone and Rossi, Lorenzo",
    collaboration = "MAP (Multi-dimensional Analyses of Partonic distributions)",
    title = "{Neural-Network Extraction of Unpolarized Transverse-Momentum-Dependent Distributions}",
    eprint = "2502.04166",
    archivePrefix = "arXiv",
    primaryClass = "hep-ph",
    reportNumber = "DESY-25-022, JLAB-THY-25-4221",
    doi = "10.1103/csc2-bj91",
    journal = "Phys. Rev. Lett.",
    volume = "135",
    number = "2",
    pages = "021904",
    year = "2025"
}

@article{Bacchetta:2024qre,
    author = "Bacchetta, Alessandro and Bertone, Valerio and Bissolotti, Chiara and Bozzi, Giuseppe and Cerutti, Matteo and Delcarro, Filippo and Radici, Marco and Rossi, Lorenzo and Signori, Andrea",
    collaboration = "MAP (Multi-dimensional Analyses of Partonic distributions)",
    title = "{Flavor dependence of unpolarized quark transverse momentum distributions from a global fit}",
    eprint = "2405.13833",
    archivePrefix = "arXiv",
    primaryClass = "hep-ph",
    reportNumber = "JLAB-THY-24-4066",
    doi = "10.1007/JHEP08(2024)232",
    journal = "JHEP",
    volume = "08",
    pages = "232",
    year = "2024"
}

@article{Qiu:2000hf,
    author = "Qiu, Jian-wei and Zhang, Xiao-fei",
    title = "{Role of the nonperturbative input in QCD resummed Drell-Yan $Q_{T}$ distributions}",
    eprint = "hep-ph/0012348",
    archivePrefix = "arXiv",
    doi = "10.1103/PhysRevD.63.114011",
    journal = "Phys. Rev. D",
    volume = "63",
    pages = "114011",
    year = "2001"
}

@article{Catani:1996yz,
    author = "Catani, Stefano and Mangano, Michelangelo L. and Nason, Paolo and Trentadue, Luca",
    title = "{The Resummation of soft gluons in hadronic collisions}",
    eprint = "hep-ph/9604351",
    archivePrefix = "arXiv",
    reportNumber = "CERN-TH-96-86",
    doi = "10.1016/0550-3213(96)00399-9",
    journal = "Nucl. Phys. B",
    volume = "478",
    pages = "273--310",
    year = "1996"
}

@article{Laenen:2000de,
    author = "Laenen, Eric and Sterman, George F. and Vogelsang, Werner",
    title = "{Higher order QCD corrections in prompt photon production}",
    eprint = "hep-ph/0002078",
    archivePrefix = "arXiv",
    reportNumber = "YITP-99-69, NIKHEF-00-002",
    doi = "10.1103/PhysRevLett.84.4296",
    journal = "Phys. Rev. Lett.",
    volume = "84",
    pages = "4296--4299",
    year = "2000"
}

@article{Lustermans:2019plv,
    author = "Lustermans, Gillian and Michel, Johannes K. L. and Tackmann, Frank J. and Waalewijn, Wouter J.",
    title = "{Joint two-dimensional resummation in $q_{T}$ and $0$-jettiness at NNLL}",
    eprint = "1901.03331",
    archivePrefix = "arXiv",
    primaryClass = "hep-ph",
    reportNumber = "DESY-19-004, DESY 19-004, NIKHEF 2018-053",
    doi = "10.1007/JHEP03(2019)124",
    journal = "JHEP",
    volume = "03",
    pages = "124",
    year = "2019"
}

@article{Bozzi:2003jy,
    author = "Bozzi, G. and Catani, S. and de Florian, D. and Grazzini, M.",
    title = "{The q(T) spectrum of the Higgs boson at the LHC in QCD perturbation theory}",
    eprint = "hep-ph/0302104",
    archivePrefix = "arXiv",
    reportNumber = "CERN-TH-2003-026",
    doi = "10.1016/S0370-2693(03)00656-7",
    journal = "Phys. Lett. B",
    volume = "564",
    pages = "65--72",
    year = "2003"
}

@article{Catani:2000vq,
    author = "Catani, Stefano and de Florian, Daniel and Grazzini, Massimiliano",
    title = "{Universality of nonleading logarithmic contributions in transverse momentum distributions}",
    eprint = "hep-ph/0008184",
    archivePrefix = "arXiv",
    reportNumber = "CERN-TH-2000-244",
    doi = "10.1016/S0550-3213(00)00617-9",
    journal = "Nucl. Phys. B",
    volume = "596",
    pages = "299--312",
    year = "2001"
}

@article{Catani:2013tia,
    author = "Catani, Stefano and Cieri, Leandro and de Florian, Daniel and Ferrera, Giancarlo and Grazzini, Massimiliano",
    title = "{Universality of transverse-momentum resummation and hard factors at the NNLO}",
    eprint = "1311.1654",
    archivePrefix = "arXiv",
    primaryClass = "hep-ph",
    reportNumber = "ZU-TH-25-13",
    doi = "10.1016/j.nuclphysb.2014.02.011",
    journal = "Nucl. Phys. B",
    volume = "881",
    pages = "414--443",
    year = "2014"
}

@article{Catani:2012qa,
    author = "Catani, Stefano and Cieri, Leandro and de Florian, Daniel and Ferrera, Giancarlo and Grazzini, Massimiliano",
    title = "{Vector boson production at hadron colliders: hard-collinear coefficients at the NNLO}",
    eprint = "1209.0158",
    archivePrefix = "arXiv",
    primaryClass = "hep-ph",
    reportNumber = "ZU-TH-16-12",
    doi = "10.1140/epjc/s10052-012-2195-7",
    journal = "Eur. Phys. J. C",
    volume = "72",
    pages = "2195",
    year = "2012"
}

@article{Kulesza:2002rh,
    author = "Kulesza, Anna and Sterman, George F. and Vogelsang, Werner",
    title = "{Joint resummation in electroweak boson production}",
    eprint = "hep-ph/0202251",
    archivePrefix = "arXiv",
    reportNumber = "BNL-HET-01-43, BNL-NT-01-32, RBRC-232, YITP-SB-01-75",
    doi = "10.1103/PhysRevD.66.014011",
    journal = "Phys. Rev. D",
    volume = "66",
    pages = "014011",
    year = "2002"
}

@article{Kulesza:2003wn,
    author = "Kulesza, Anna and Sterman, George F. and Vogelsang, Werner",
    title = "{Joint resummation for Higgs production}",
    eprint = "hep-ph/0309264",
    archivePrefix = "arXiv",
    reportNumber = "BNL-HET-03-20, BNL-NT-03-26, RBRC-335, YITP-SB-03-47",
    doi = "10.1103/PhysRevD.69.014012",
    journal = "Phys. Rev. D",
    volume = "69",
    pages = "014012",
    year = "2004"}

@article{Gehrmann:2012ze,
    author = "Gehrmann, Thomas and Lubbert, Thomas and Yang, Li Lin",
    title = "{Transverse parton distribution functions at next-to-next-to-leading order: the quark-to-quark case}",
    eprint = "1209.0682",
    archivePrefix = "arXiv",
    primaryClass = "hep-ph",
    doi = "10.1103/PhysRevLett.109.242003",
    journal = "Phys. Rev. Lett.",
    volume = "109",
    pages = "242003",
    year = "2012"
}

@article{Boer:1999mm,
    author = "Boer, Daniel",
    title = "{Investigating the origins of transverse spin asymmetries at RHIC}",
    eprint = "hep-ph/9902255",
    archivePrefix = "arXiv",
    reportNumber = "RIKEN-BNL-PREPRINT",
    doi = "10.1103/PhysRevD.60.014012",
    journal = "Phys. Rev. D",
    volume = "60",
    pages = "014012",
    year = "1999"
}

@article{Boer:1997nt,
    author = "Boer, Daniel and Mulders, P. J.",
    title = "{Time reversal odd distribution functions in leptoproduction}",
    eprint = "hep-ph/9711485",
    archivePrefix = "arXiv",
    reportNumber = "NIKHEF-97-049, VUTH-97-20",
    doi = "10.1103/PhysRevD.57.5780",
    journal = "Phys. Rev. D",
    volume = "57",
    pages = "5780--5786",
    year = "1998"
}

@article{Camarda:2022qdg,
    author = "Camarda, Stefano and Ferrera, Giancarlo and Schott, Matthias",
    title = "{Determination of the strong-coupling constant from the Z-boson transverse-momentum distribution}",
    eprint = "2203.05394",
    archivePrefix = "arXiv",
    primaryClass = "hep-ph",
    doi = "10.1140/epjc/s10052-023-12373-2",
    journal = "Eur. Phys. J. C",
    volume = "84",
    number = "1",
    pages = "39",
    year = "2024"
}

@article{ATLAS:2024erm,
    author = "Aad, Georges and others",
    collaboration = "ATLAS",
    title = "{Measurement of the W-boson mass and width with the ATLAS detector using proton{\textendash}proton collisions at $\sqrt{s}=7$ TeV}",
    eprint = "2403.15085",
    archivePrefix = "arXiv",
    primaryClass = "hep-ex",
    reportNumber = "CERN-EP-2024-074",
    doi = "10.1140/epjc/s10052-024-13190-x",
    journal = "Eur. Phys. J. C",
    volume = "84",
    number = "12",
    pages = "1309",
    year = "2024"
}

@article{LHCb:2021bjt,
    author = "Aaij, Roel and others",
    collaboration = "LHCb",
    title = "{Measurement of the W boson mass}",
    eprint = "2109.01113",
    archivePrefix = "arXiv",
    primaryClass = "hep-ex",
    reportNumber = "LHCb-PAPER-2021-024, CERN-EP-2021-170",
    doi = "10.1007/JHEP01(2022)036",
    journal = "JHEP",
    volume = "01",
    pages = "036",
    year = "2022"
}

@article{CMS:2024lrd,
    author = "Chekhovsky, Vladimir and others",
    collaboration = "CMS",
    title = "{High-precision measurement of the W boson mass with the CMS experiment}",
    eprint = "2412.13872",
    archivePrefix = "arXiv",
    primaryClass = "hep-ex",
    reportNumber = "CMS-SMP-23-002, CERN-EP-2024-308",
    doi = "10.1038/s41586-026-10168-5",
    journal = "Nature",
    volume = "652",
    number = "8109",
    pages = "321--327",
    year = "2026"
}

@article{CDF:2022hxs,
    author = "Aaltonen, T. and others",
    collaboration = "CDF",
    title = "{High-precision measurement of the $W$          boson mass with the CDF II detector}",
    reportNumber = "FERMILAB-PUB-22-254-PPD",
    doi = "10.1126/science.abk1781",
    journal = "Science",
    volume = "376",
    number = "6589",
    pages = "170--176",
    year = "2022"
}

@article{D0:2013jba,
    author = "Abazov, Victor Mukhamedovich and others",
    collaboration = "D0",
    title = "{Measurement of the $W$ boson mass with the D0 detector}",
    eprint = "1310.8628",
    archivePrefix = "arXiv",
    primaryClass = "hep-ex",
    reportNumber = "FERMILAB-PUB-13-489-E",
    doi = "10.1103/PhysRevD.89.012005",
    journal = "Phys. Rev. D",
    volume = "89",
    number = "1",
    pages = "012005",
    year = "2014"
}

@article{D0:2017ekd,
    author = "Abazov, Victor Mukhamedovich and others",
    collaboration = "D0",
    title = "{Measurement of the Effective Weak Mixing Angle in $p\bar{p}\rightarrow Z/\gamma^* \rightarrow \ell^+\ell^-$ Events}",
    eprint = "1710.03951",
    archivePrefix = "arXiv",
    primaryClass = "hep-ex",
    reportNumber = "FERMILAB-PUB-17-434-E",
    doi = "10.1103/PhysRevLett.120.241802",
    journal = "Phys. Rev. Lett.",
    volume = "120",
    number = "24",
    pages = "241802",
    year = "2018"
}

@article{Rottoli:2023xdc,
    author = "Rottoli, Luca and Torrielli, Paolo and Vicini, Alessandro",
    title = "{Determination of the W-boson mass at hadron colliders}",
    eprint = "2301.04059",
    archivePrefix = "arXiv",
    primaryClass = "hep-ph",
    reportNumber = "ZU-TH 01/23, TIF-UNIMI-2023-1",
    doi = "10.1140/epjc/s10052-023-12128-z",
    journal = "Eur. Phys. J. C",
    volume = "83",
    number = "10",
    pages = "948",
    year = "2023"
}

@article{Boer:2006eq,
    author = "Boer, Daniel and Vogelsang, Werner",
    title = "{Drell-Yan lepton angular distribution at small transverse momentum}",
    eprint = "hep-ph/0604177",
    archivePrefix = "arXiv",
    journal = "Phys. Rev. D",
    volume = "74",
    pages = "014004",
    year = "2006"
}

@article{Berger:2007jw,
    author = "Berger, Edmond L. and Qiu, Jian-Wei and Rodriguez-Pedraza, Ricardo A.",
    title = "{Transverse momentum dependence of the angular distribution of the Drell-Yan process}",
    eprint = "0707.3150",
    archivePrefix = "arXiv",
    primaryClass = "hep-ph",
    journal = "Phys. Rev. D",
    volume = "76",
    pages = "074006",
    year = "2007"
}

@article{Berger:2007si,
    author = "Berger, Edmond L. and Qiu, Jian-Wei and Rodriguez-Pedraza, Ricardo A.",
    title = "{Angular distribution of leptons from the decay of massive vector bosons}",
    eprint = "0708.0578",
    archivePrefix = "arXiv",
    primaryClass = "hep-ph",
    journal = "Phys. Lett. B",
    volume = "656",
    pages = "74--80",
    year = "2007"
}
\end{document}